\documentclass[twocolumn,prl,floatfix,superscriptaddress,nobibnotes]{revtex4-1}
\usepackage{amsmath}
\usepackage{amssymb}
\usepackage{graphicx}
\usepackage{color}
\usepackage[dvipsnames]{xcolor}
\usepackage[colorlinks=true, citecolor={blue!80!black}, urlcolor={blue!50!black}, linkcolor = {blue!80!black}]{hyperref}

\begin{document}

\title{Flux-induced topological superconductivity in full-shell nanowires}

\author{S.~Vaitiek\.{e}nas}
\affiliation{Center for Quantum Devices and Microsoft Quantum Lab--Copenhagen, Niels Bohr Institute, University of Copenhagen, 2100 Copenhagen, Denmark}

\author{G.~W.~Winkler} \affiliation{Microsoft Quantum, Microsoft Station Q, University of California, Santa Barbara, California 93106-6105 USA}

\author{B.~van~Heck} \affiliation{Microsoft Quantum, Microsoft Station Q, University of California, Santa Barbara, California 93106-6105 USA}

\author{T.~Karzig}
\affiliation{Microsoft Quantum, Microsoft Station Q, University of California, Santa Barbara, California 93106-6105 USA}

\author{M.-T.~Deng}
\affiliation{Center for Quantum Devices and Microsoft Quantum Lab--Copenhagen, Niels Bohr Institute, University of Copenhagen, 2100 Copenhagen, Denmark}

\author{K.~Flensberg}
\affiliation{Center for Quantum Devices and Microsoft Quantum Lab--Copenhagen, Niels Bohr Institute, University of Copenhagen, 2100 Copenhagen, Denmark}

\author{L.~I.~Glazman}
\affiliation{Departments of Physics and Applied Physics, Yale University, New Haven, CT 06520, USA}

\author{C.~Nayak}
\affiliation{Microsoft Quantum, Microsoft Station Q, University of California, Santa Barbara, California 93106-6105 USA}

\author{P.~Krogstrup}
\affiliation{Center for Quantum Devices and Microsoft Quantum Lab--Copenhagen, Niels Bohr Institute, University of Copenhagen, 2100 Copenhagen, Denmark}

\author{R.~M.~Lutchyn} \affiliation{Microsoft Quantum, Microsoft Station Q, University of California, Santa Barbara, California 93106-6105 USA}

\author{C.~M.~Marcus}
\affiliation{Center for Quantum Devices and Microsoft Quantum Lab--Copenhagen, Niels Bohr Institute, University of Copenhagen, 2100 Copenhagen, Denmark}

\date{\today}

\begin{abstract}
We present a novel route to realizing topological superconductivity using magnetic flux applied to a full superconducting shell surrounding a semiconducting nanowire core.
In the destructive Little-Parks regime, reentrant regions of superconductivity are associated with integer number of phase windings in the shell. Tunneling into the core reveals a hard induced gap near zero applied flux, corresponding to zero phase winding, and a gapped region with a discrete zero-energy state around one applied flux quantum, $\Phi_{0} = h/2e$, corresponding to $2 \pi$ phase winding.
Theoretical analysis indicates that in the presence of radial spin-orbit coupling in the semiconductor, the winding of the superconducting phase can induce a transition to a topological phase supporting Majorana zero modes.
Realistic modeling shows a topological phase persisting over a wide range of parameters, and reproduces experimental tunneling conductance data.
Further measurements of Coulomb blockade peak spacing around one flux quantum in full-shell nanowire islands shows exponentially decreasing deviation from $1e$ periodicity with device length, consistent with Majorana modes at the ends of the nanowire.
\end{abstract}

\maketitle

\section{Introduction}

Majorana zero modes (MZMs) at the ends of one-dimensional topological superconductors are expected to exhibit non-trivial braiding statistics \cite{Read2000,Kitaev01,Alicea2011}, opening a path toward topologically protected quantum computing \cite{Nayak2008,DasSarma2015}.
Among the proposals to realize MZMs, one approach \cite{Lutchyn2010,Oreg2010}  based on semiconducting nanowires with strong spin-orbit coupling subject to a Zeeman field and superconducting proximity effect has received particular attention, yielding numerous compelling experimental signatures \cite{Mourik2012,Deng2016,Albrecht2016,Zhang2018,Lutchyn2018}.
An alternative route to MZMs aims to create vortices in spinless superconductors, by various means, for instance by coupling a vortex in a conventional superconductor to a topological insulator \cite{Fu2008,Chiu11,Cook11,deJuan14, Xu2015} or conventional semiconductor \cite{Sau2010, Alicea2010}, using doped topological insulators \cite{Hosur2011}, iron-based superconductors \cite{Wang2018}, or using vortices in exotic quantum Hall analogs of spinless superconductors \cite{DasSarma2005}.

In this Article, we demonstrate both experimentally and theoretically that a hybrid nanowire consisting of a full superconducting (Al) shell surrounding a semiconducting (InAs) core can be driven into a topological phase that supports MZMs at the wire ends by a flux-induced winding of the superconducting phase.
This conceptually new approach contains elements of both proximitized-wire schemes \cite{Lutchyn2010,Oreg2010}  and vortex-based schemes \cite{Read2000,Fu2008} for creating MZMs.
The topological phase sets in at relatively low magnetic fields, is controlled discretely by moving from zero to one phase twists around the superconducting shell, and does not require a large $g$ factor in the semiconductor, broadening the landscape of candidate materials.

While it is known that well-chosen superconducting phase differences can be used to break time-reversal symmetry and localize MZMs in semiconducting heterostructures \cite{Romito2012,vanHeck2014, Kotetes2015,Hell2017, Pientka2017,Stanescu2018}, the corresponding realizations typically require careful tuning of the fluxes. In contrast, vortices in a full-shell wire provide a naturally quantized means of controlling superconducting phase. In the destructive Little-Parks regime \cite{Little1962,deGennes1981}, the modulation of critical current and temperature with flux applied along the hybrid nanowire results in a sequence of lobes with reentrant superconductivity \cite{Liu2001,Sternfeld2011}. Each lobe is associated with a quantized number of twists of the superconducting phase \cite{Tinkham1966}, determined by the external field so that the free energy of the superconducting shell is minimized.
The result is a series of topologically locked boundary conditions for the proximity effect in the semiconducting core, with a dramatic effect on the subgap density of states.

Our measurements reveal that tunneling into the core in the zeroth superconducting lobe, around zero flux, yields a hard proximity-induced gap with no subgap features.
In the superconducting regions around one quantum of applied flux, corresponding to phase twists of $\pm 2\pi$ in the shell, tunneling spectra into the core shows stable zero-bias peaks, indicating a discrete subgap state fixed at zero energy.

Theoretically, we find that a Rashba field arising from the breaking of local radial inversion symmetry at the semiconductor-superconductor interface~\cite{Antipov2018,Mikkelsen2018, Woods2019}, along with $2\pi$-phase twists in the boundary condition, can induce a topological state supporting MZMs.
We calculate the topological phase diagram of the system as a function of various parameters such as Rashba spin-orbit coupling, radius of the semiconducting core and band bending at the superconductor-semiconductor interface~\cite{Antipov2018,Mikkelsen2018,Woods2019}.
Our analysis shows that topological superconductivity extends in a reasonably large portion of the parameter space.
Transport simulations of the tunneling conductance in the presence of Majorana zero modes qualitatively reproduce the experimental data in the entire voltage-bias range.

We obtain further experimental evidence that the zero-energy states are localized at wire ends by investigating Coulomb blockade conductance peaks in full-shell wire islands of various lengths. In the zeroth lobe, Coulomb blockade peaks show $2e$ spacing, indicating Cooper-pair tunneling and an induced gap exceeding the island charging energy.
In the first lobe, peak spacings are roughly $1e$-periodic, with slight even-odd alternation that vanishes exponentially with island length consistent with overlapping Majorana modes at the two ends of the Coulomb island, as investigated previously \cite{Albrecht2016,vanHeck2016}. The exponential dependence on length, and incompatibility with a power-law dependence, strongly suggests that MZMs reside at the ends of the hybrid islands.

\section{Device description}

\begin{figure}[h!t]
\includegraphics[width=\linewidth]{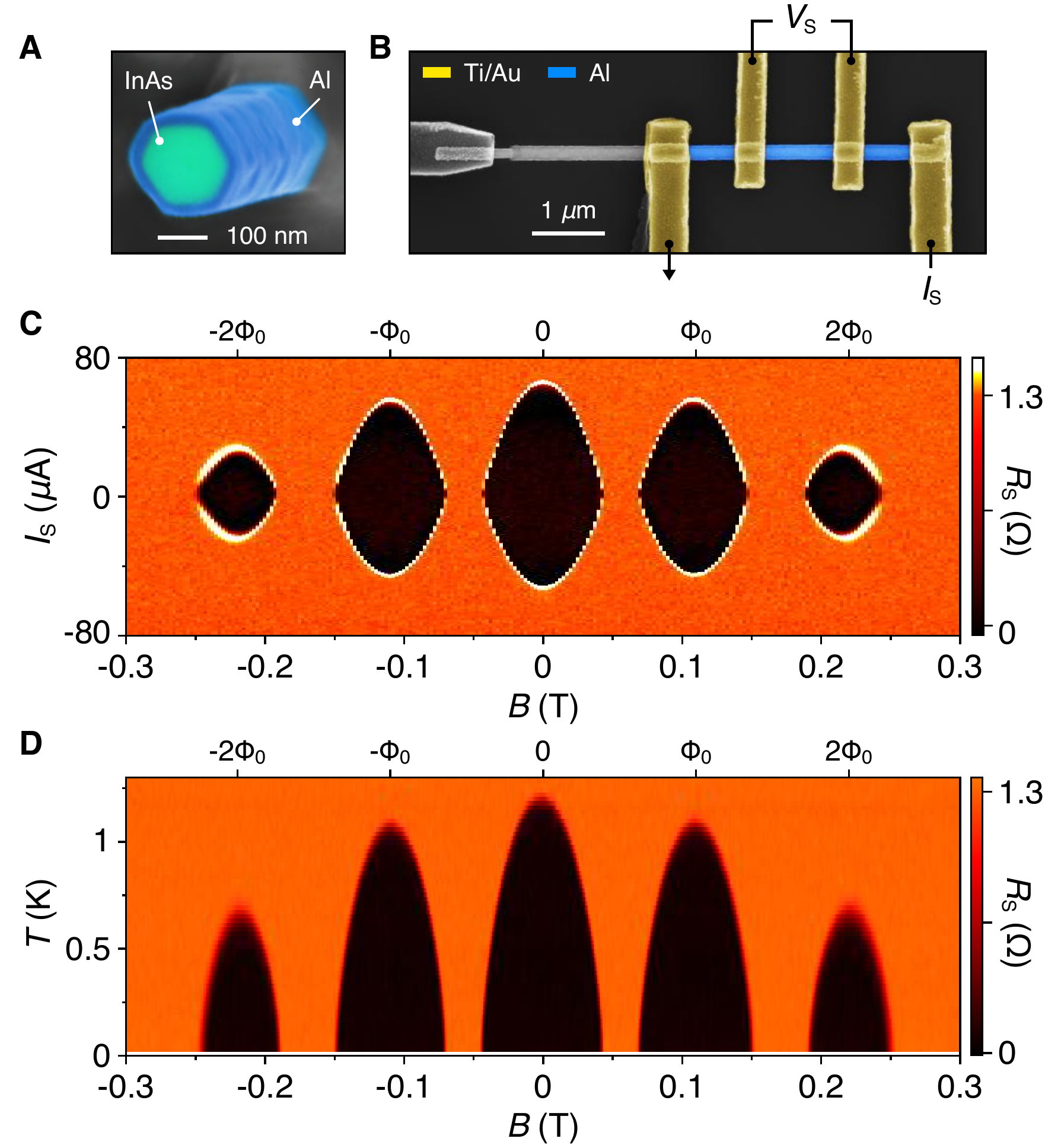}
\caption{\label{fig:1} \textbf{Destructive Little-Parks regime in full-shell nanowire device.} (\textbf{A}) Colorized material-sensitive  electron micrograph of InAs-Al hybrid nanowire. Hexagonal InAs core (maximum diameter 130~nm) with 30~nm full-shell epitaxial Al. (\textbf{B}) Micrograph of device 1, colorized to highlight 4-probe measurement setup. (\textbf{C}) Differential resistance of the Al shell, $R_{\rm S}$, as a function of current bias, $I_{\rm S}$, and axial magnetic field, $B$, measured at $20$~mK. Top axis shows flux, $BA_{\rm wire}$, in units of the flux quantum $\Phi_{0} = h/2e$. Superconducting lobes are separated by destructive regions near odd half-integer flux quanta. (\textbf{D}) Temperature evolution of $R_{\rm S}$ as a function of $B$ measured around $I_{\rm S}=0$. Note that $R_{\rm S}$ equals the normal-state resistance in all destructive regimes.}
\end{figure}

InAs nanowires were grown by the vapor-liquid-solid method using molecular beam epitaxy (MBE). The nanowires had a hexagonal cross section with maximum diameter ${\rm D}=130$~nm. A 30~nm epitaxial Al layer was grown while rotating the sample, yielding a fully enclosing shell (Fig.~\ref{fig:1}A) \cite{Krogstrup2015}. Devices were fabricated using electron-beam lithography. Standard ac lock-in measurements were carried out in a dilution refrigerator with a base temperature of 20~mK. Magnetic field was applied parallel to the nanowire using three-axis vector magnet. Two device geometries, measured in three devices each, showed similar results. Data from two representative devices are reported in the main text: device 1 was used for 4-probe measurements of the shell (Fig.~\ref{fig:1}B) and tunneling spectroscopy of the core (Fig.~\ref{fig:2}A); device 2 comprised six Coulomb islands of different lengths fabricated on a single nanowire, each with separate ohmic contacts, two side gates to trim tunnel barriers, and a plunger gate to change occupancy (Fig.~\ref{fig:6}A). Supporting data from additional three tunneling devices, one of which has thinner shell, and two Coulomb-blockaded devices are presented in \cite{Supplementary}. For more detailed description of the wire growth, device fabrication and measurement techniques see \cite{Supplementary}.

%%%%%%%%%%%%%%%%%%%%%%%%%%%%%%%%%%%%%%%%%%%%%%%%%%%%%%
%%%%%%%%%%%%%%%%%%%%%% FIG. 1 %%%%%%%%%%%%%%%%%%%%%%%%
%%%%%%%%%%%%%%%%%%%%%%%%%%%%%%%%%%%%%%%%%%%%%%%%%%%%%%
%%%%%%%%%%%%%%%%%%%%%%%%%%%%%%%%%%%%%%%%%%%%%%%%%%%%%%
%%%%%%%%%%%%%%%%%%%%%%%%%%%%%%%%%%%%%%%%%%%%%%%%%%%%%%

Differential resistance of the shell, $R_{\rm S} = dV_{\rm S}/dI_{\rm S}$, measured for device 1 as a function of bias current, $I_{\rm S}$, and axial magnetic field, $B$, showed a lobe pattern characteristic of the destructive regime (Fig.~\ref{fig:1}C) with maximum switching current of $70~\mu$A at $B = 0$, the center of the zeroth lobe. Between the zeroth and first lobes, supercurrent vanished at $\vert B\vert=45$~mT, re-emerged at $70$~mT, and had a maximum near the center of the first lobe, at $\vert B\vert =110$ mT. A second lobe with smaller critical current was also observed, but no third lobe.

Temperature dependence of $R_{\rm S}$ around zero bias yielded a reentrant phase diagram with superconducting regions separated by destructive regions with temperature-independent normal-state resistance $R_{\rm S}^{\rm (N)} = 1.3~\Omega$ (Fig.~\ref{fig:1}D). $R_{\rm S}^{\rm (N)}$ and shell dimensions from Fig.~\ref{fig:1}A yield a Drude mean free path of $l=19$~nm. The dirty-limit shell coherence length \cite{Tinkham1966,Gordon1984} 
\begin{align}\label{eq:xis}
\xi_{\rm S}=\sqrt{\frac{\pi\hbar v_{\rm F} l}{24 k_{\rm B} T_{\rm C}}}
\end{align}
can then be found using the zero-field critical temperature $T_{\rm C}=1.2$~K from Fig.~\ref{fig:1}D and Fermi velocity of Al, $v_{\rm F}=2\times 10^6$~m/s \cite{Kittel2005}, with Planck constant $\hbar$ and Boltzmann constant $k_{\rm B}$, yielding $\xi_{\rm S} = 180$~nm. The same values for $\xi_{\rm S}$ is found using the onset of the first destructive regime \cite{Schwiete2009}.

\section{Tunneling spectroscopy}

\begin{figure}[h!t]
\includegraphics[width=0.975\linewidth]{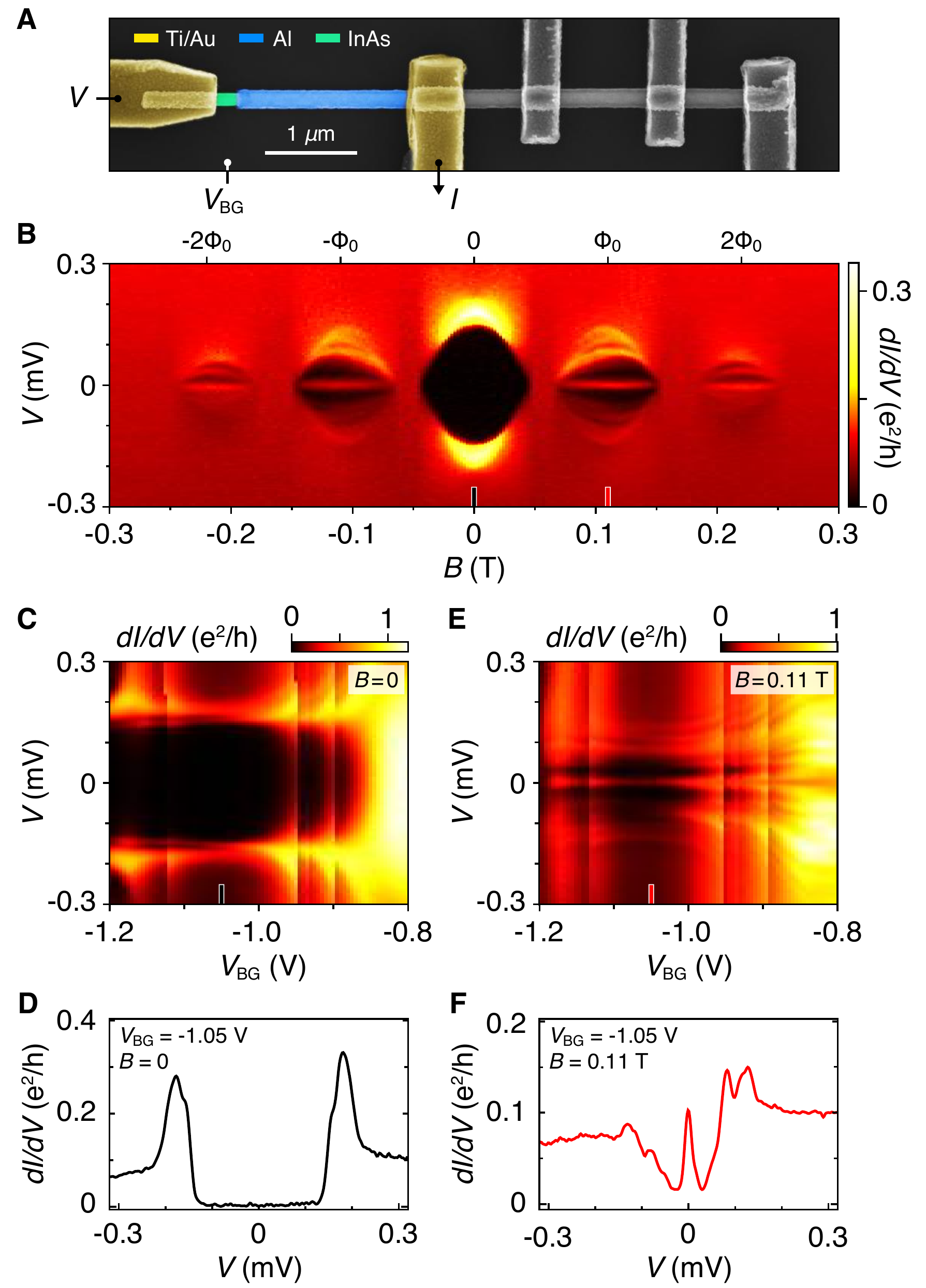}
\caption{\label{fig:2} \textbf{Experimental tunneling spectrum: hard gap in the zeroth lobe, zero-bias peak in the first lobe.} (\textbf{A}) Micrograph of device 1 colorized to highlight tunneling spectroscopy set-up. (\textbf{B}) Differential conductance, $dI/dV$, as a function of source-drain bias voltage, $V$, and axial field, $B$. The zeroth lobe shows a hard superconducting gap, the first lobes show zero-bias peak, the second lobes show non-zero subgap states. The lobes are separated by featureless normal-state spectra. (\textbf{C}) Zero-field conductance as a function of $V$ and back-gate voltage, $V_{\rm BG}$. (\textbf{D}) Line-cut of the conductance taken at $B=0$ and $V_{\rm BG}=-1.05$~V. (\textbf{E} and \textbf{F}) Similar to (C) and (D), measured in the first lobe at $B=110$~mT. Data shown are from two-terminal measurements, which include line resistances \cite{Supplementary}.
}
\end{figure}

Differential conductance, $dI/dV$, as a function of source-drain voltage, $V$, measured in the tunneling regime as a probe of the local density of states at the end of the nanowire is shown in Fig.~\ref{fig:2}. The Al shell was removed at the end of the wire and the tunnel barrier was controlled by the global back-gate at voltage $V_{\rm BG}$. At zero field, a hard superconducting gap was observed throughout the zeroth superconducting lobe (Fig.~\ref{fig:2},~B~and~D). Similar to the supercurrent measurements presented above, the superconducting gap in the core closed at $|B|=45$~mT and reopened at $70$~mT, separated by a gapless destructive regime. Upon reopening, a narrow zero-bias conductance peak was observed throughout the first gapped lobe (Fig.~\ref{fig:2},~B~and~F). Several flux-dependent subgap states are also visible, separated from the zero-bias peak in the first lobe. These nonzero subgap states are analogs of Caroli-de Gennes-Matricon bound states \cite{Caroli1964}, in this case confined at the metal-semiconductor interface rather than around a vortex core. 

The first lobe persists to $150$~mT, above which a second gapless destructive regime was observed.  A second gapped lobe centered around $\vert B\vert = 220$~mT then appeared, containing several subgap states away from zero energy, as shown in greater detail in \cite{Supplementary}. The second lobe closes at $250$~mT, above which only normal-state behavior was observed.

%%%%%%%%%%%%%%%%%%%%%%%%%%%%%%%%%%%%%%%%%%%%%%%%%%%%%%
%%%%%%%%%%%%%%%%%%%%%% FIG. 2 %%%%%%%%%%%%%%%%%%%%%%%%
%%%%%%%%%%%%%%%%%%%%%%%%%%%%%%%%%%%%%%%%%%%%%%%%%%%%%%
%%%%%%%%%%%%%%%%%%%%%%%%%%%%%%%%%%%%%%%%%%%%%%%%%%%%%%
%%%%%%%%%%%%%%%%%%%%%%%%%%%%%%%%%%%%%%%%%%%%%%%%%%%%%%

The dependence of tunneling spectra on back-gate voltage in the zeroth lobe is shown in Fig.~\ref{fig:2}C. In weak tunneling regime, for $V_{\rm BG} < -1$~V a hard gap was observed, with $\Delta = 180~\mu$eV (Fig.~\ref{fig:2},~C~and~D). As the device is opened, for $V_{\rm BG} \sim -0.8$~V subgap conductance is enhanced due to Andreev processes. The increase  in conductance at $V_{\rm BG} \sim -1.2$~V is likely due to a resonance in the barrier. In the first lobe, at $B=110$~mT, the sweep of $V_{\rm BG}$ showed a zero-energy state throughout the tunneling regime (Fig.~\ref{fig:2}E). The cut displayed in Fig.~\ref{fig:2}F shows a discrete zero-bias peak well separated from other states.
As the tunnel barrier is opened, the zero-bias peak gradually evolves into a zero-bias dip.
This behavior is in qualitative agreement with theory~\cite{Vuik2018}, although the crossover occurs at lower conductance than expected.
Additional line-cuts as well as the tunneling spectroscopy for the second lobe are provided in~\cite{Supplementary}. Several switches in data occurred at the same gate voltages in Fig.~\ref{fig:2},~C~and~E, presumably due to gate-dependent charge motion in the barrier.

\section{Modeling of topological phases}

\begin{figure}[h!t]
  \begin{center}
    \includegraphics[width=0.975\columnwidth]{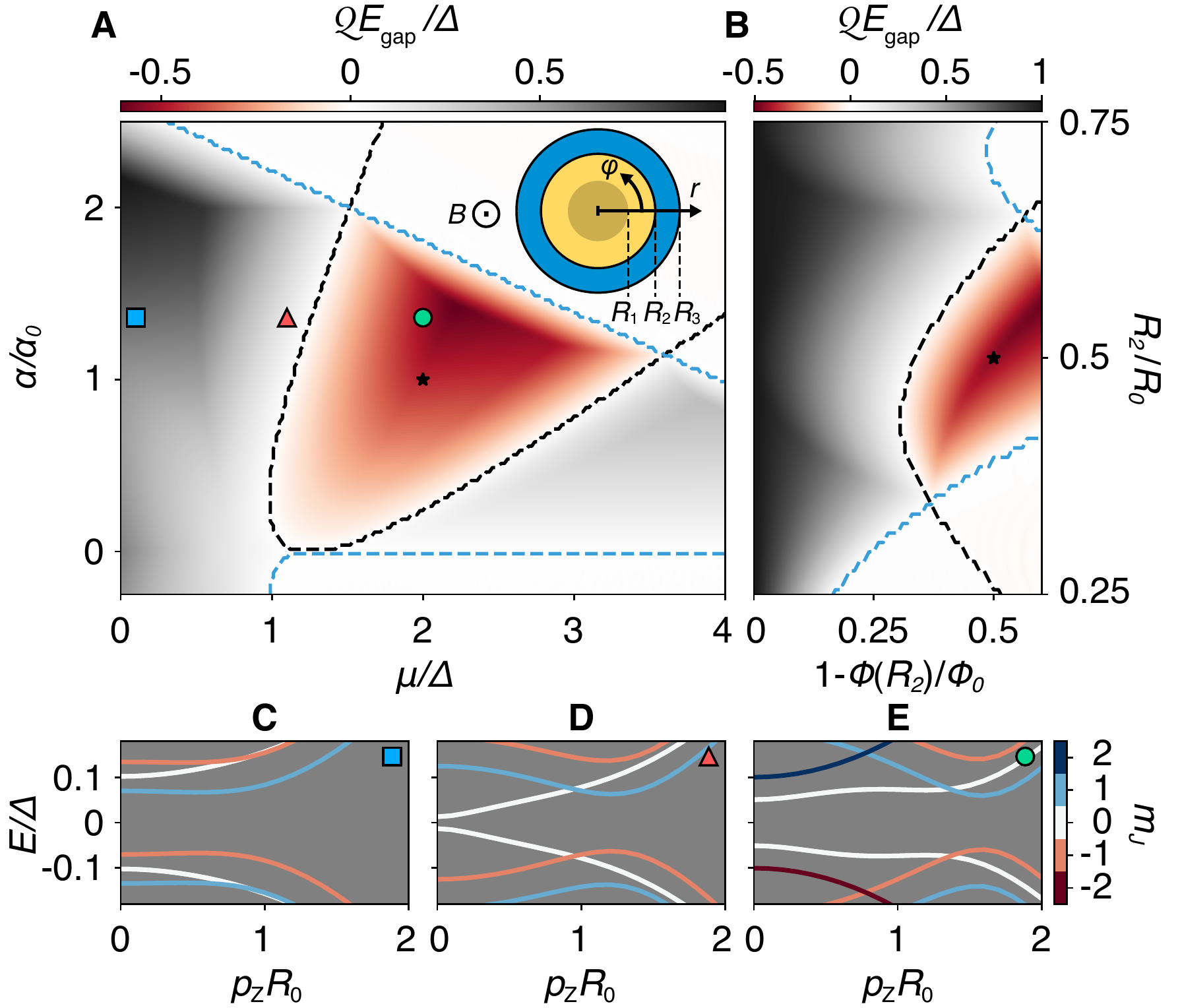}
\caption{\textbf{Topological phase diagram in hollow-cylinder model.} (\textbf{A}) Bulk energy gap, $E_\textrm{g}$, as a function of chemical potential and spin-orbit coupling. The energy gap is multiplied by the topological index $\mathcal{Q}=\pm 1$, so that red regions correspond to the gapped topological phase. The black dashed line denotes the boundary of the topological phase in the $m_J=0$ sector, according to Eq.~\eqref{eq:topological_condition}, while the blue dashed lines denote the boundaries at which higher $m_J$ sectors become gapless~\cite{Supplementary}. Here $\Phi(R_2)/\Phi_0=\tfrac{1}{2}$, $R/R_0=\tfrac{1}{2}$. We define $\alpha_0=\sqrt{\Delta/2m}$ and $R_0=1/\sqrt{2m\Delta}$. For reference, using realistic parameters $m^*=0.026\;m_\textrm{e}$ and $\Delta=0.2$~meV, one obtains $\alpha_0\approx 17$~meV$\cdot$nm and $R_0\approx 85$~nm. The inset shows a cross-section of a semiconducting nanowire (yellow) with a full superconducting shell (blue), subject to a weak axial magnetic field $B$. The shaded yellow region with $r<R_1$ indicates the possible presence of an insulating core in the semiconductor. (\textbf{B}) Bulk energy gap at fixed $\mu/\Delta=2$ and $\alpha/\alpha_0=1$, as indicated by a black marker in (A), as a function of flux and $R$. (\textbf{C}-\textbf{E}) Band-structures at the points indicated in (A), illustrating the closing and re-opening of the bulk gap in the $m_J=0$ sector.
  \label{fig:3}}
\end{center}
\end{figure}

\begin{figure}[t!]
  \begin{center}
    \includegraphics[width=\columnwidth]{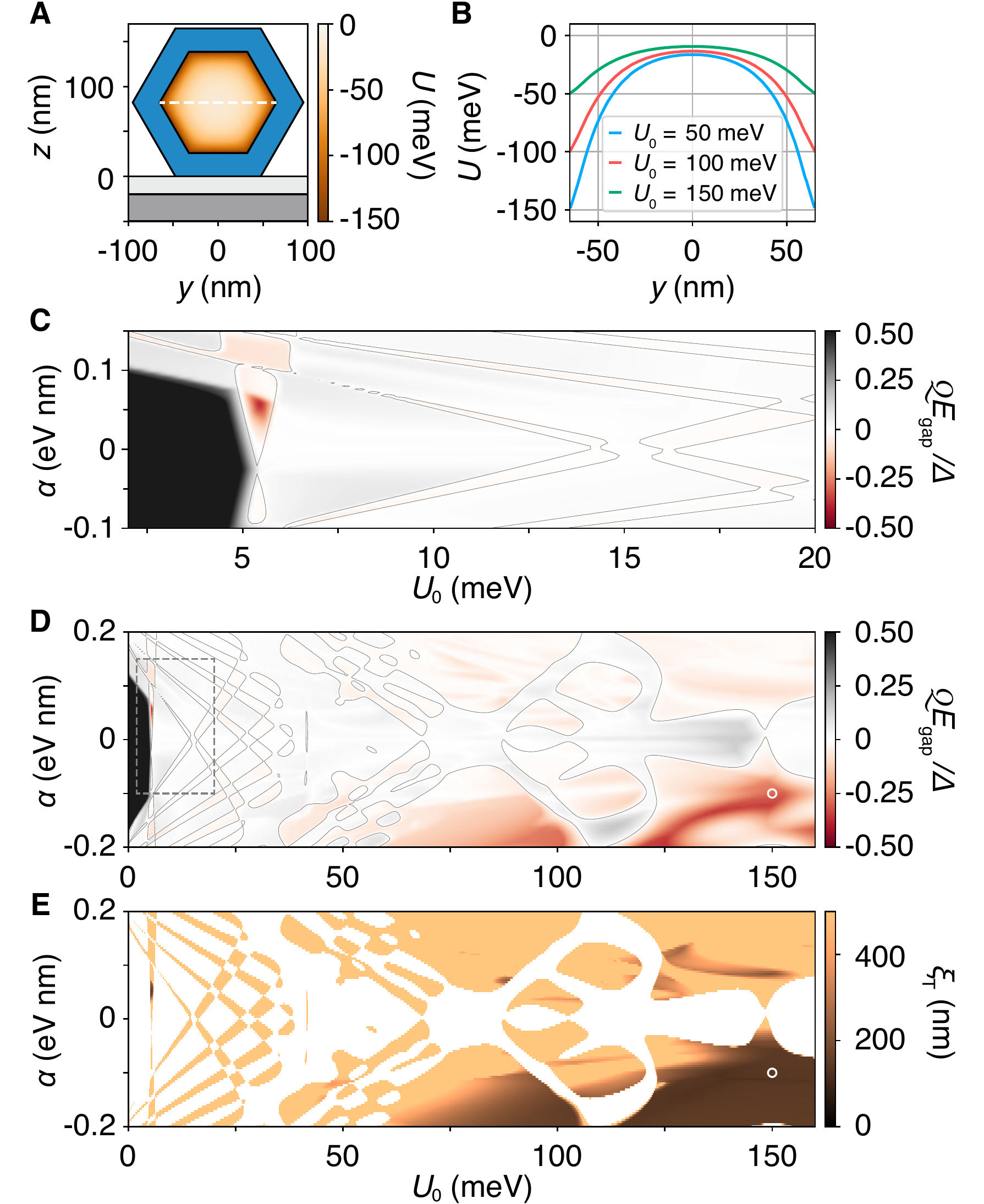}
\caption{\textbf{Modeling the electrostatic potential and topological phase diagram.}
(\textbf{A}) Schematic cross-section of the wire superimposed with the simulated potential energy, $U$, in the semiconductor for band offset $U_0=150$\,meV.
(\textbf{B}) Horizontal cuts of the  potential in the wire for different band offsets.
(\textbf{C}) Topological phase diagram of the full-shell nanowire in the first lobe at $B=0.124$\,T as a function $U_0$ and spin-orbit coupling, $\alpha$, close to the $m_J=0$ topological phase. The gray lines indicate a change of the sign of the Pfaffian, ${\cal Q}$.
(\textbf{D}) Topological phase diagram for the same set of parameters as in (B) over a large range of band offsets.
(\textbf{E})~Topological coherence length, $\xi_{\rm T}$, computed for the same $U_0$ and $\alpha$ ranges as in (D).
  \label{fig:4}}
\end{center}
\end{figure}

To better understand the origin of the zero-energy modes in the first lobe we analyze theoretically a semiconducting nanowire covered by a superconducting shell.
First, we present a toy model of a cylindically  symmetric full-shell wire (Fig.~\ref{fig:3}), highlighting the underlying mechanism of the topological phase appearance. Thereafter we move on to simulations of realistic geometries (Fig.~\ref{fig:4} and Fig.~\ref{fig:5}).

We assume that the semiconductor (InAs) has a large Rashba spin-orbit coupling due to the local inversion symmetry breaking in the radial direction at the semiconductor-superconductor interface (corresponding to an electric field pointing along the radial direction at the superconductor-semiconductor interface).
The system is subject to a magnetic field along the direction of the nanowire, $\vec{B}=B \hat{z}$.
Using cylindrical coordinates and the symmetric gauge for the electromagnetic vector potential,
$\vec{A}=\frac{1}{2} (\vec{B} \times \vec{r})$, the effective Hamiltonian for the semiconducting core can be written as (henceforth
$\hbar=1$)
\begin{align}
  H_0=\frac{(\vec{p}+e A_{\varphi}\hat{\varphi})^2}{2m^*}-\mu+\alpha\, \hat{r}\cdot \left[\vec{\sigma} \times  (\vec{p}+e A_{\varphi}\hat{\varphi})\right].
\end{align}
Here $\vec{p}$ is the electron momentum operator, $e>0$ the electric charge, $m^*$ the effective mass, $\mu$ is the
chemical potential, $\alpha$ the strength of the Rashba spin-orbit coupling, and finally $\sigma_i$ are spin-$\frac{1}{2}$ Pauli
matrices. $\hat{r}$, $\hat{\varphi}$ and $\hat{z}$ are the cylindrical unit vectors.
For ease of presentation, we consider $r$-independent $\mu$ and $\alpha$ in our model, which may be viewed as averaged versions of the corresponding $r$-dependent quantities.
The vector potential $A_{\varphi}= \Phi(r)/2\pi r$, where $\Phi(r)=\pi B r^2$ is the flux threading the cross-section at radius $r$.
For simplicity, we neglect the Zeeman term due to the small magnetic fields required in the experiment.

The superconducting shell (Al) induces superconducting correlations in the nanowire due to Andreev processes at the semiconductor-superconductor interface.
If the tunnel coupling to the superconductor is weak, the induced pairing in the nanowire can be expressed as a local potential $\Delta(\vec{r})$ (see Ref.~\cite{Supplementary}).
In the Nambu basis $\Psi=(\psi_{\uparrow}, \psi_{\downarrow},\psi^\dag_{\downarrow}, -
\psi^\dag_{\uparrow})$, the Bogoliubov-de-Gennes (BdG) Hamiltonian for the proximitized nanowire is then given by
\begin{align}
  H_{\rm BdG}=\left[\begin{array}{cc}
                      H_0(\vec{A}) & \Delta(\vec{r}) \\
                      \Delta^*(\vec{r}) & -\sigma_y H_0(-\vec{A})^* \sigma_y
                    \end{array}
                                          \right].\label{eq:HBdG}
\end{align}

We assume that the thickness of the shell is smaller than London penetration depth: $R_3-R_2\ll\lambda_L$.
Therefore, the magnetic flux threading the shell is not quantized.
However, the magnetic field induces a winding of the superconducting phase of the order parameter $\Delta(\vec{r})=\Delta(r) e^{-i n\varphi}$ with $\varphi$ the angular coordinate and $n\,\in\,\mathbb{Z}$ the winding number determined by the external magnetic flux.

%%%%%%%%%%%%%%%%%%%%%%%%%%%%%%%%%%%%%%%%%%%%%%%%%%%%%%
%%%%%%%%%%%%%%%%%%%%%% FIG. 3 %%%%%%%%%%%%%%%%%%%%%%%%
%%%%%%%%%%%%%%%%%%%%%%%%%%%%%%%%%%%%%%%%%%%%%%%%%%%%%%
%%%%%%%%%%%%%%%%%%%%%%%%%%%%%%%%%%%%%%%%%%%%%%%%%%%%%%
%%%%%%%%%%%%%%%%%%%%%%%%%%%%%%%%%%%%%%%%%%%%%%%%%%%%%%

We notice the following rotational symmetry of the BdG Hamiltonian: $[J_z,H_{\rm BdG}]=0$ with $J_z=-i \partial_{\varphi}+\frac{1}{2}\sigma_z+\frac{1}{2}n \tau_z$, where we introduced $\tau_i$ matrices acting in Nambu space.
Eigenstates of $H_\textrm{BdG}$ can thus be labeled by a conserved quantum number $m_J$:
$  \Psi_{m_J}(r,\varphi, z)\propto e^{i \left(m_J-\frac{1}{2}\sigma_z-\frac{1}{2}n \tau_z\right)\varphi} \Psi_{m_J}(r, z)$.
The wave function has to be single-valued, which imposes the following
constraint on $m_J$:
\begin{align}
  m_J \in \begin{cases}
    \mathbb{Z} &  n\;{\rm odd}\,,\\
    \mathbb{Z}+\frac{1}{2} & n\;{\rm even}\,.
  \end{cases}
\end{align}
It is interesting to note that the particle-hole symmetry relates states with opposite energy and angular quantum number $m_J$, that is 
$\mathcal{P}\Psi_{E, m_J}=\Psi_{-E, -m_J}$ with
$\mathcal{P}=\tau_y \sigma_y \mathcal{K}$, where $\mathcal{K}$
represents complex conjugation. Thus, the $m_J=0$ sector---allowed when the winding number $n$ is odd---is special as it allows non-degenerate Majorana solutions at zero energy, as we show below.

The angular dependence of $H_{\rm BdG}$ can be eliminated via a unitary
transformation $U=\exp\left[ -i \left(m_J\!-\!\frac{1}{2}\sigma_z\!-\!\frac{1}{2}n
    \tau_z\right)\varphi \right]$, namely
$\tilde{H}_{\rm BdG}=UH_{\rm BdG}U^\dag$ where
\begin{align}\label{eq:HBdGr}
\begin{split}
  \tilde{H}_{\rm BdG}&=\left(\frac{p_{z}^2}{2m^*}+\frac{p_r^2}{2m^*}-\mu\right)\tau_z \\
  &+ \frac{1}{2m^*r^2}\left(m_J-\frac{1}{2}\sigma_z-\frac{1}{2}n\tau_z + eA_\varphi r\tau_z\right)^2\tau_z \\
  &- \frac{\alpha}{r}\sigma_z \tau_z\left(m_J-\frac{1}{2}\sigma_z-\frac{1}{2}n \tau_z + eA_\varphi r\tau_z\right) \\
  &+\alpha p_z \sigma_y \tau_z+\Delta(r)\tau_x.
  \end{split}
\end{align}
Here $p_r^2=-\frac{1}{r}\frac{\partial}{\partial r}r\frac{\partial}{\partial r}$ and $p_z=-i \frac{\partial}{\partial z}$. Note that naively one might expect spin-orbit coupling to average
out; however, the non-trivial structure of $m_J$-eigenvectors yields
finite matrix elements proportional to the Rashba spin-orbit coupling.
%We will now show that the above BdG Hamiltonian supports topological superconductivity and MZMs.

Assuming that the electrons in the core are localized at the interface we set $R_1 \approx R_2$ (Fig.~\ref{fig:3}A).
This approximation is motivated by the fact that there is an accumulation layer in certain semiconductor-superconductor heterostructures such as InAs/Al, as explained below.
In this case the electrons in semiconductor effectively form a thin-wall hollow cylinder  and one can consider only the lowest-energy radial mode in Eq.~\eqref{eq:HBdGr}. This allows for an analytical solution of the model. The effective Hamiltonian for the hollow-cylinder model reads
\begin{align}\label{eq:H_mj}
\begin{split}
\tilde{H}_{m_J}= &
\left[\frac{p_{z}^2}{2m^*}-\mu_{m_J}\right]\!\tau_z+V_{\rm Z}\sigma_z +A_{m_J}\! + C_{m_J}\sigma_z\tau_z\\
&+\alpha p_z \sigma_y \tau_z+\Delta\tau_x.
\end{split}
\end{align}
Here, $\Delta\equiv\Delta(R_2)$ and the parameters $\mu_{m_J}$ and $V_{\rm Z}$ correspond to the effective chemical potential and Zeeman energy. $A_{m_J}$ and $C_{m_J}$ represent the coupling of the generalized angular momentum $J_z$  with magnetic field and electron spin, respectively. They are defined as
\begin{align}
\mu_{m_J}&=\mu-\frac{1}{8m^*R_2^2}\left(4m_J^2+1+\phi^2\right)-\frac{\alpha}{2R_2}\,, \label{eq:effparam}\\
    V_{\rm Z}&=\phi\,\left(\frac{1}{4m^*R_2^2}+\frac{\alpha}{2R_2}\right)\,,\\
    A_{m_J}&=-\frac{\phi m_J}{2m^* R_2^2}\,,\\
    C_{m_J}&=-m_J\left(\frac{1}{2m^*R_2^2}+\frac{\alpha}{R_2}\right)\,. \label{eq:effparam4}
    \end{align}
with $\phi=n-\Phi(R_2)/\Phi_0$.

Equations~(\ref{eq:effparam})-(\ref{eq:effparam4}) allow to identify a topological phase in the $m_J=0$ sector of the first lobe where $n=1$. In this case, $A_{0}=0$ and $C_{0}=0$, and one can map
Eq.~\eqref{eq:H_mj} to the Majorana nanowire model of
Refs.~\cite{Lutchyn2010, Oreg2010}. Notice that the effective Zeeman term has an orbital origin here and is present even when the $g$ factor in the semiconductor is zero. Both $\mu_0$ and $V_Z$ can be tuned by the magnetic flux $\Phi(R_2)$, which may induce a topological phase transition. 
In the hollow cylinder approximation $V_Z=0$ when the core is penetrated by one flux quantum ($\Phi(R_2)=\Phi_0$). This regime corresponds to the trivial (s-wave) superconducting phase.
However, a small deviation of the magnetic flux can drive the system into the topological superconducting phase~\cite{FootnoteFlux}. 
Indeed, the Zeeman and spin-orbit terms in Eq.~\eqref{eq:H_mj} do not commute and thus $V_Z$ opens a gap in the spectrum at $p_z=0$.  At the topological quantum phase transition between the two phases, the gap in the $m_J=0$ sector,
\begin{align}\label{eq:topological_condition}
E_\text{gap}^{(0)}=\left||V_Z|-\sqrt{\mu_{0}^2+\Delta^2}\right|\,,
\end{align}
closes. The resulting phase diagram is shown in Fig.~\ref{fig:3},
where the gap closing at the topological transition is indicated by black dashed lines. Close to the transition, the quasiparticle spectrum in the $m_J=0$ sector is given by
$E(p_z)=\sqrt{\left(E_\text{gap}^{(0)}\right)^2+(vp_z)^2}$
with $v=\alpha \Delta/\sqrt{\Delta^2+\mu_0^2}$ and the corresponding topological coherence length $\xi_{\rm T}\sim v/E_\text{gap}^{(0)}$.

%%%%%%%%%%%%%%%%%%%%%%%%%%%%%%%%%%%%%%%%%%%%%%%%%%%%%%
%%%%%%%%%%%%%%%%%%%%%% FIG. 4 %%%%%%%%%%%%%%%%%%%%%%%%
%%%%%%%%%%%%%%%%%%%%%%%%%%%%%%%%%%%%%%%%%%%%%%%%%%%%%%
%%%%%%%%%%%%%%%%%%%%%%%%%%%%%%%%%%%%%%%%%%%%%%%%%%%%%%
%%%%%%%%%%%%%%%%%%%%%%%%%%%%%%%%%%%%%%%%%%%%%%%%%%%%%%

A well-defined topological phase requires the quasiparticle bulk gap to be finite for all values of $m_J$. Due to the angular symmetry of Eq.~\eqref{eq:H_mj}, different $m_J$ sectors do not mix and, as a result, the condition for a finite gap in the $m_J\neq 0$ sectors is $ \Delta^2+ (C_{m_J}-\mu_{m_J})^2 > (A_{m_J}+V_{\rm Z})^2$~\cite{Supplementary}. In general, the topological phase diagram can be obtained by calculating the topological index ${\cal Q}$~\cite{Kitaev01},
\begin{align}
{\cal Q}=\mbox{\rm sign}\prod_{m_J \in Z}\left[\Delta^2+(C_{m_J}-\mu_{m_J})^2-(A_{m_J}+V_{\rm Z})^2\right], \label{indexQ}
\end{align}
where ${\cal Q}=1$ and ${\cal Q}=-1$ correspond to trivial and
topological phases, respectively.
Thus, the topological phase supporting MZMs appears due to the change of ${\cal Q}$ in the $m_J=0$ sector. In Fig.~\ref{fig:3} we show the topological phase diagram and energy gap of the hollow cylinder model determined by taking into account all $m_J$ sectors.

The hollow cylinder model provides conceptual understanding for the existence of the topological phase in full shell nanowires. The model, however, is limited to small chemical potentials and a conserved angular momentum. For a direct comparison with the experiment more realistic simulations extending to the regime with multiple radial modes are needed.

\section{Realistic simulations}

Recent advances in the modelling of semiconductor-superconductor hybrid structures have led to more accurate simulations of proximitized nanowires~\cite{Vuik16, Mikkelsen2018, Antipov2018, Winkler2018}. Here, the essence of our approach is to integrate out the superconductor into self-energy boundary conditions, as discussed in~\cite{Supplementary}.
This approximation allows for three-dimensional (3D) simulations of proximitized nanowires, including important effects such as self-consistent electrostatics and orbital magnetic field contribution~\cite{Winkler17}.

We model a hexagonal InAs wire with 130\,nm corner-to-corner diameter coated by a 30\,nm thick Al shell, see Fig.~\ref{fig:4}A.
The work function difference between InAs and Al leads to a band offset between the conduction band of InAs and the Fermi level of Al resulting in an electron accumulation layer close to the interface, see Fig.~\ref{fig:4},~A~and~B. This band offset is on the order of 100 meV~\cite{Antipov2018, Mikkelsen2018, Winkler2018,Schuwalow2019}. Due to the accumulation layer, there is an intrinsic electric field for the electrons, resulting in Rashba spin-orbit coupling with the symmetry axis in approximately radial direction~\cite{Footnote1,Winkler2003}. The magnitude of $\alpha$ has been experimentally determined to be in the range of 0.02 to 0.08\,eV\,nm~\cite{Lutchyn2018}.

Given the uncertainties, we calculate the topological phase diagram as a function of band offset, $U_0$, and the Rashba spin-orbit coupling, $\alpha$~\cite{Wimmer2012}. The band offset controls the number of subbands in the nanowire as well as their population. For $U_0<40$~meV the system is in the single radial mode regime and the phase diagram appears qualitatively similar to the hollow-cylinder model, see Fig.~\ref{fig:4},~C~and~D.  Around 5~meV there is a gapped topological phase which we identify with the $m_J=0$ angular sector, analog to the hollow-cylinder model. Specifically in this regime, apart from the $m_J=0$ sector, the topological phases have very small gap. The vertical feature at $U_0\sim40$~meV band offset in Fig.~\ref{fig:4}D corresponds to a second radial subband with $m_J=0$ crossing the Fermi level.

For $U_0>40$~meV the phase diagram becomes qualitatively different. Due to the increased number of bands the different topological phases hybridize and merge into extended topological regions~\cite{Sticlet2017, LutchynFisher11}.
Furthermore, as $U_0$ increases the wave functions are pushed closer to the superconductor leading to a stronger hybridization of the wave functions with Al. In this region one finds extended topological regions with sizable gaps which are a significant fraction of the superconducting gap. 

%%%%%%%%%%%%%%%%%%%%%%%%%%%%%%%%%%%%%%%%%%%%%%%%%%%%%%
%%%%%%%%%%%%%%%%%%%%%% FIG. 5 %%%%%%%%%%%%%%%%%%%%%%%%
%%%%%%%%%%%%%%%%%%%%%%%%%%%%%%%%%%%%%%%%%%%%%%%%%%%%%%
\begin{figure}[t]
  \begin{center}
    \includegraphics[width=\columnwidth]{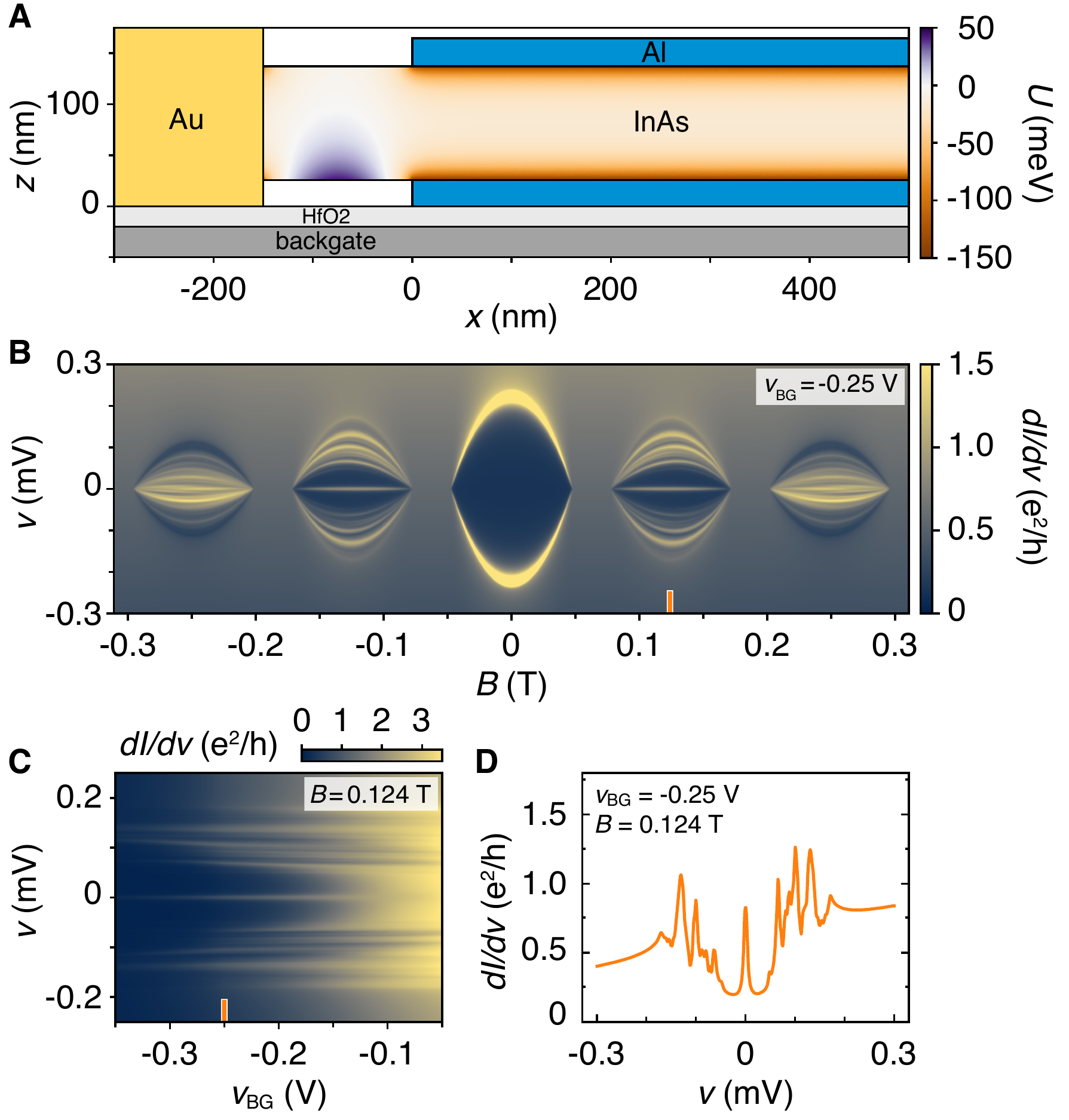}
\caption{\textbf{Simulation of tunneling transport.}
(\textbf{A}) Schematic side-view of the normal-superconducting junction device superimposed with the simulated potential energy, $U$, in the semiconductor computed for band offset $U_0 =150$\,meV and back-gate voltage $\mathrm{\textsl v}_{\rm BG}=-0.25$\,V. 
(\textbf{B}) Differential conductance $dI/d\mathrm{\textsl v}$ as a function of axial magnetic field, $B$, and bias voltage, $\mathrm{\textsl v}$, simulated at $\mathrm{\textsl v}_\mathrm{BG}=-0.3$\,V, $U_0=150$\,meV and spin-orbit coupling $\alpha=-0.1$\,eV\,nm. 
(\textbf{C}) Differential conductance as a function of $\mathrm{\textsl v}_\mathrm{BG}$ at $B=0.124$\,T for the same $U_0$ and $\alpha$ as in (B). 
(\textbf{D}) Line-cut of the conductance at $\mathrm{\textsl v}_\mathrm{BG}=-0.25$\,V and $B=0.124$\,T.
  \label{fig:5}}
\end{center}
\end{figure}
%%%%%%%%%%%%%%%%%%%%%%%%%%%%%%%%%%%%%%%%%%%%%%%%%%%%%%
%%%%%%%%%%%%%%%%%%%%%%%%%%%%%%%%%%%%%%%%%%%%%%%%%%%%%%

Mixing of different angular sectors, facilitated by the broken cylindrical symmetry due to the hexagonal cross section, lifts the restriction of gapped topological phases to the $m_J=0$ sector. In this context, we note that when angular symmetry is broken (due to disorder in superconductor or geometrical effects) and $m_J$ is not a good quantum number, the topological superconducting phase may also appear at even winding numbers (see~\cite{Supplementary} for the topological phase diagram in the second lobe).

In addition to the gap size we also compute the topological coherence length, $\xi_{\rm T}$ (Fig.~\ref{fig:4}E), from the eigenvalue decomposition of the translation operator at zero energy~\cite{Nijholt16}. As expected, regions with large gap also have a short coherence length. Note that due to the smaller Fermi velocity in the semiconductor, the topological coherence length can be smaller than the s-wave coherence length. We find that the shortest $\xi_{\rm T}\sim 120$~nm, whereas the typical values for realistic spin-orbit coupling strength and band offset range between 140 and 200~nm.

Having established bulk properties, we numerically compute a three-dimensional full shell wire in a transport geometry. The corresponding longitudinal cross section of the simulated device is shown in Fig.~\ref{fig:5}A. After calculating the electrostatic potential of the 3D structure we simulate the quantum transport using the package Kwant and adaptive~\cite{kwant, adaptive}. In the main text we focus on a single point in the phase diagram with band offset of 150~meV and $\alpha=-0.1$\,eV\,nm (see the white circle in Fig.~\ref{fig:4},~D~and~E). Results for other representative points can be found in~\cite{Supplementary}. 

Computed conductance, $dI/d\mathrm{\textsl v}$, as a function of bias voltage, $\mathrm{\textsl v}$, and magnetic field, $B$, is shown in Fig.~\ref{fig:5}B. The simulated back-gate voltage, $\mathrm{\textsl v}_\mathrm{BG}$, is chosen such that there is good visibility of states in the wire. 
Similar to the experiment, the zeroth lobe shows a hard gap with no subgap states. The first and second lobes on the other hand show multiple subgap states~\cite{FootnoteAsymmetry}.
The first lobe has a gap with a zero-bias peak due to Majorana end states. The size of the gap is consistent with the bulk phase diagram in Fig.~\ref{fig:4}D. The second lobe has more subgap states and appears to be gapless.

The evolution of the simulated spectrum with the back-gate voltage in the topological phase is displayed in Fig.~\ref{fig:5}C.
As expected, the bias voltages at which zero-bias peak and subgap states are visible is independent of $\mathrm{\textsl v}_\mathrm{BG}$ but the intensities of the states change.
Since the wire is fully covered by a superconducting shell the effect of the back gate is completely screened inside the bulk of the wire and does not influence the topological phase or bulk states.
When the junction becomes very open at $\mathrm{\textsl v}_\mathrm{BG}>-0.1$\,V the zero-bias peak transforms into a zero-bias dip, as expected in this regime.

\section{Coulomb blockade spectroscopy}

%%%%%%%%%%%%%%%%%%%%%%%%%%%%%%%%%%%%%%%%%%%%%%%%%%%%%%
%%%%%%%%%%%%%%%%%%%%%% FIG. 6 %%%%%%%%%%%%%%%%%%%%%%%%
%%%%%%%%%%%%%%%%%%%%%%%%%%%%%%%%%%%%%%%%%%%%%%%%%%%%%%
\begin{figure}[h!t]
\includegraphics[width=0.975\linewidth]{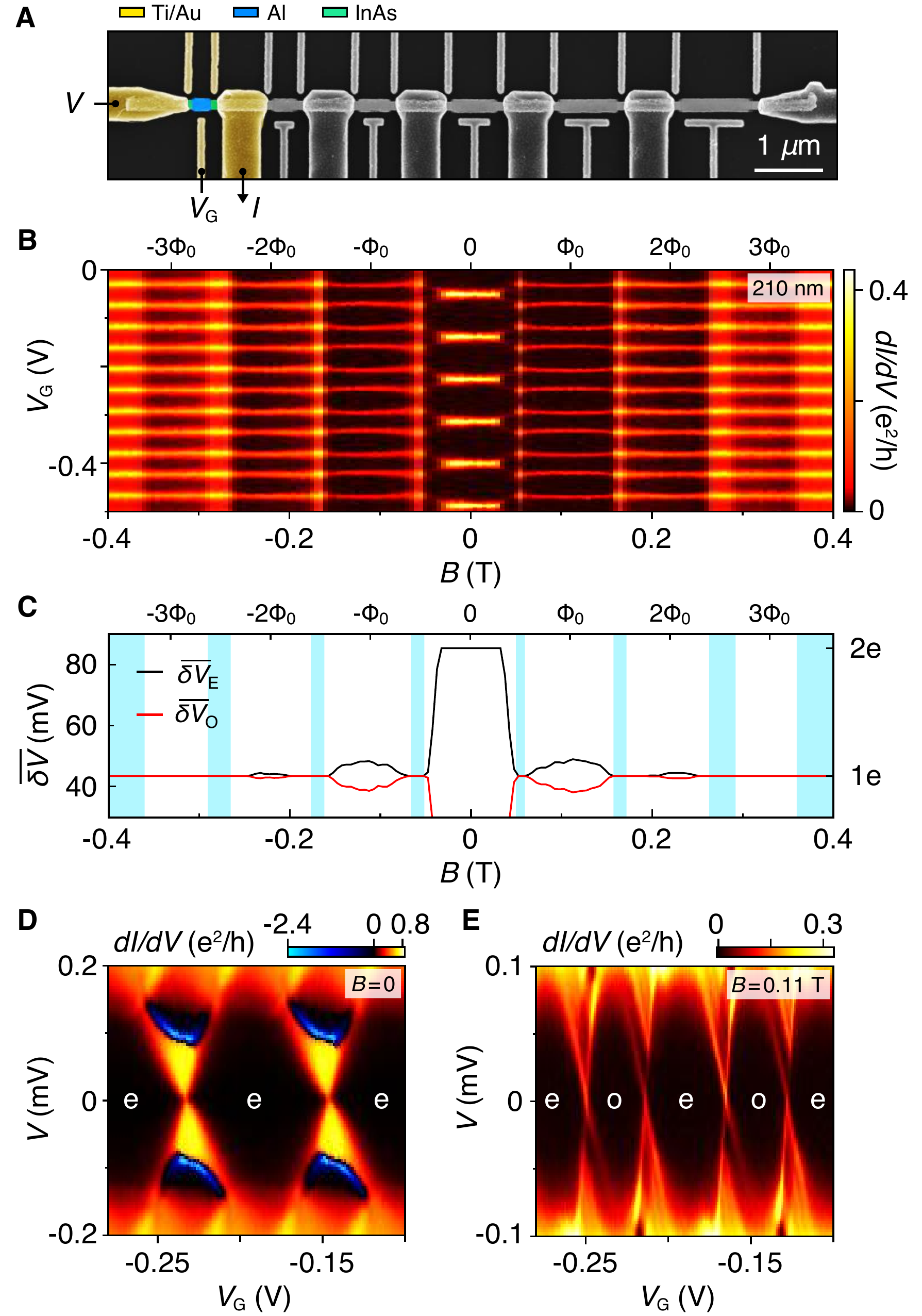}
\caption{\label{fig:6} \textbf{Coulomb blockade: $\bf 2e$ peaks in the zeroth lobe, even-odd peaks in the first lobe.} (\textbf{A}) Micrograph of device 2 comprising six islands with individual gates and leads, spanning a range of lengths from $210$~nm to $970$~nm. The measurement setup for 210~nm segment is highlighted in colors. (\textbf{B}) Zero-bias conductance for the 210~nm segment showing Coulomb blockade evolution as a function of plunger gate voltage, $V_{\rm G}$, and axial magnetic field, $B$. (\textbf{C}) Average peak spacings for even (black) and odd (red) Coulomb valleys, $\overline{\delta V}$, from the data in (A) as a function of $B$, with destructive regimes shown in blue. Coulomb peaks spaced by $2e$ split in field and become $1e$-periodic around $55$~mT. At higher field, odd Coulomb valleys shrink, reaching a minimum around $120$~mT. In the second destructive regime around 165~mT peaks are $1e$-periodic again. (\textbf{D}) Zero-field conductance as a function of $V$ and $V_{\rm G}$, showing $2e$ Coulomb diamonds with even (e) valleys only. The negative differential conductance is associated with quasiparticle trapping on the island (see text). (\textbf{E}) Similar to (D) but measured in the first lobe at $B = 110$~mT, reveals discrete, near-zero-energy state, even (e) and odd (o) valleys of different sizes, and alternating excited state structure.
}
\end{figure}
%%%%%%%%%%%%%%%%%%%%%%%%%%%%%%%%%%%%%%%%%%%%%%%%%%%%%%
%%%%%%%%%%%%%%%%%%%%%%%%%%%%%%%%%%%%%%%%%%%%%%%%%%%%%%

Returning to experiment, we next investigate MZM hybridization, which can be measured using the spacing of Coulomb blockade conductance peaks in Coulomb  islands as a function of island length \cite{Albrecht2016,vanHeck2016,OFarell2018,Shen2018}. The exponential length dependence of hybridization energy is a signature of MZMs localized at the opposite ends of the nanowire~\cite{Cheng2009,Fu2010,Chiu2017}. We investigated full-shell islands over a range of device lengths from $210$~nm to $970$~nm, fabricated on a single nanowire, as shown in Fig.~\ref{fig:6}. 

Zero-bias conductance as a function of plunger-gate voltage, $V_{\rm G}$, and $B$ for device 2 yielded series of Coulomb blockade peaks for each segment, examples of which are shown in Fig.~\ref{fig:6}B. The corresponding average peak spacings, $\overline{\delta V}$, for even and odd Coulomb valleys as a function of $B$ are shown in Fig.~\ref{fig:6}C. Around zero field, Coulomb-blockade peaks with $2e$ periodicity were found. These peaks split at $\sim$~40~mT toward the high-field end of the zeroth superconducting lobe, as the superconducting gap decreased below the charging energy of the island. The peaks then became $1e$-periodic (within experimental sensitivity) around $55$~mT and throughout the first destructive regime (see also Fig.~\ref{fig:1} for the onset of destructive regime). When superconductivity reappeared in the first lobe, the Coulomb peaks did not become spaced by $2e$ again, but instead showed nearly $1e$ spacing with even-odd modulation. The $210$~nm island showed a qualitatively similar even-odd also in the second lobe. Unlike device 1 described in Fig.~\ref{fig:2}, the shortest island in device 2 also showed a third superconducting lobe, which can be identified from the peak height contrast in Fig.~\ref{fig:6}B. Coulomb blockade peaks were $1e$-periodic within experimental sensitivity throughout the third lobe.

Tunneling spectra at finite source-drain bias showed $2e$ Coulomb diamonds around zero field (Fig.~\ref{fig:6}D) and nearly $1e$ diamonds at $B = 110$~mT, near the middle of the first lobe (Fig.~\ref{fig:6}E). The zero-field diamonds are indistinguishable from each other, showing a region of negative differential conductance associated with the onset of quasiparticle transport \cite{Hekking1993,Hergenrother1994,Higginbotham2015}. In the first lobe (Fig.~\ref{fig:6}E), Coulomb diamonds alternate in size and symmetry, with degeneracy points showing sharp, gapped structure, indicating that the near-zero-energy state is discrete. Additional resonances at finite bias reflect excited discrete subgap states away from zero energy.

%%%%%%%%%%%%%%%%%%%%%%%%%%%%%%%%%%%%%%%%%%%%%%%%%%%%%%
%%%%%%%%%%%%%%%%%%%%%% FIG. 7 %%%%%%%%%%%%%%%%%%%%%%%%
%%%%%%%%%%%%%%%%%%%%%%%%%%%%%%%%%%%%%%%%%%%%%%%%%%%%%%
\begin{figure}[t]
\includegraphics[width=0.975\linewidth]{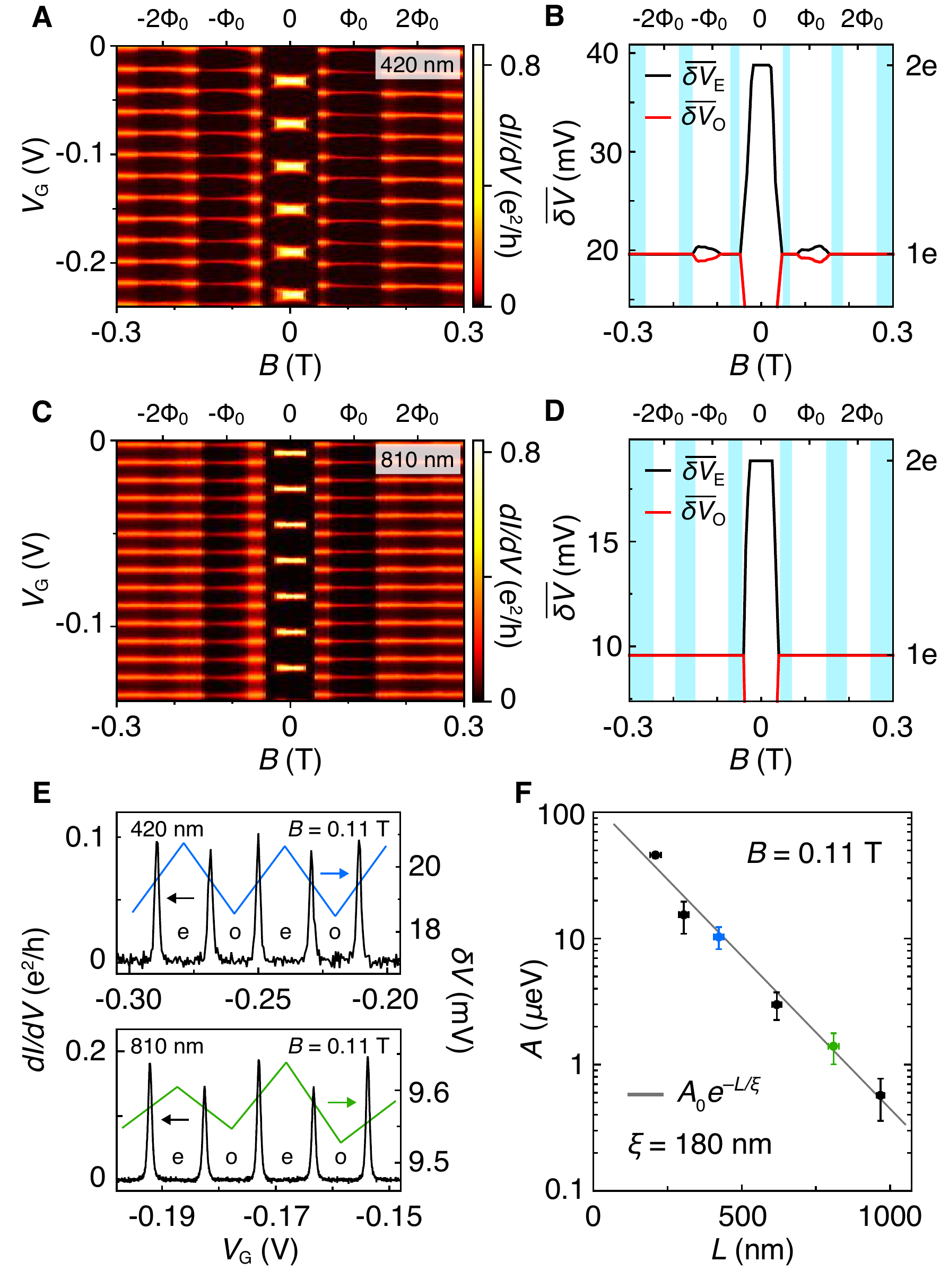}
\caption{\label{fig:7} \textbf{Length dependence of even-odd peak spacing.} (\textbf{A}) Zero-bias conductance showing Coulomb blockade evolution with $V_{\rm G}$ and $B$ for $420$~nm island. (\textbf{B}) Average peak spacing for data in (A). Even-odd pattern is evident in the first lobe, around $B = 110$~mT. (\textbf{C} and \textbf{D}) Similar to (A) and (B) for $810$~nm island. Even-odd spacing in the first lobe is not visible on this scale. (\textbf{E}) Fine-scale Coulomb peak conductance (black, left axis) and spacing (colored, right axis) as a function of plunger gate voltage, $V_{\rm G}$ at $B=110$~mT for $420$~nm and $810$~nm islands. (\textbf{F}) Average even-odd peak spacing difference converted to energy, $A$, using separately measured level arms for each segment, at $B=110$~mT as a function of island length, $L$, along with the best fit to the exponential form $A = A_{\rm 0} e^{-L/\xi}$, giving the best fit parameters $A_{\rm 0}=105~\mu$eV and $\xi=180$~nm. Vertical error bars indicate uncertainties from standard deviation of $\overline{\delta V}$ and lever arms. Experimental noise floor, $\sigma_{\rm A}<0.1\,\mu{\rm eV} \ll k_{\rm B}T$, measured using $1e$ spacing in destructive regime. Horizontal error bars indicate uncertainties in lengths estimated from the micrograph.
}
\end{figure}
%%%%%%%%%%%%%%%%%%%%%%%%%%%%%%%%%%%%%%%%%%%%%%%%%%%%%%
%%%%%%%%%%%%%%%%%%%%%%%%%%%%%%%%%%%%%%%%%%%%%%%%%%%%%%

Coulomb peaks for two longer islands are shown in Fig.~\ref{fig:7},~A~to~E, with full data sets for other lengths reported in \cite{Supplementary}. All islands showed $2e$-periodic Coulomb peaks in the zeroth lobe and nearly $1e$ spacing in the first lobe. Examining the 420~nm and 810~nm data in Fig.~\ref{fig:7},~A,~C~and~E already reveals that the mean difference between even and odd peak spacings in the first lobe decreased with increasing island length. To address this question quantitatively, we determine the lever arm, $\eta$, for each island independently in order to convert plunger gate voltages to chemical potentials on the islands, using the slopes of the Coulomb diamonds \cite{Thijssen2008, Albrecht2016}.  This allows the peak spacing differences (Fig.~\ref{fig:7},~B~and~D) to be converted to island-energy differences, $A(L)$, between even and odd occupations, as a function of device length, $L$ (a detailed exemplar peak spacing analysis is presented in \cite{Supplementary}). Within a Majorana picture, the energy scale $A(L)$ reflects the length dependent hybridization energy of MZMs. Values for $A(L)$ at $B = 110$~mT, in the middle of the first lobe, spanning over two orders of magnitude are shown in Fig.~\ref{fig:7}F. A fit to an exponential $A = A_{\rm 0} e^{-L/\xi}$ yields fit parameters $A_{\rm 0}=105~\mu$eV and $\xi=180$~nm. The data are well described by an exponential length dependence, implying that the low-energy modes are located at the ends of the wire, not bound to impurities or local potential fluctuations as expected for overlapping Majorana modes. The comparison of exponential and power-law fits as well as the calculated length dependence that shows exponential decay only in topological regimes are provided in \cite{Supplementary}. The measured $\xi$ is consistent with the calculated $\xi_{\rm T}$ using realistic parameters.

Along with length dependent even-odd peak spacing difference, we observe even-odd modulation in peak heights, see Fig.~\ref{fig:7}E. Possible explanation of these phenomena was proposed in Ref.~\cite{Hansen2018}. Additionally, we find a complex alternating peak-height structure depending on magnetic field within the first lobe. Peak height modulation accompanying peak spacing modulation was observed previously \cite{Albrecht2016,OFarell2018,Shen2018}.

To investigate how coherence length $\xi$, extracted from the exponential decrease of even-odd peak spacing with length, depends on the superconducting gap, $\Delta$, we examine peak spacing near the high-field edge of the first lobe, $B=140$~mT, where the gap is reduced to $\Delta_{140} = 40\,\mu$eV, and shows no subgap features besides the zero-bias peak \cite{Supplementary}. At this reduced gap we again find an exponential dependence on length, and incompatibility with a power law, now with $\xi = 230$~nm. We observe that $\xi_{140}/\xi_{110}=230~\mbox{nm}/180~\mbox{nm}\sim 1.3$ is consistent with simple scaling, $\xi\propto\Delta^{-1}$ (not accounting for a field-dependent velocity). From Fig.~\ref{fig:2}B and Ref.~\cite{Supplementary}, $\delta_{110}/\Delta_{140}=50~\mu$eV$/40~\mu$eV $\sim 1.2$, where $\delta_{110}$ is the lowest non-zero subgap state, and $\delta_{140}=\Delta_{140}$. We also note that both $\xi_{110}$ and $\xi_{140}$ are slightly smaller than the coherence length in the superconducting shell at corresponding B-field values: $\xi_{\rm S}(110~\mbox{mT}) \sim 190$~nm and $\xi_{\rm S}(140~\mbox{mT}) \sim 250$~nm, extracted from data in Fig.~\ref{fig:1}D using Eq.~\eqref{eq:xis} and the corresponding values of $T_C(B)$. This discrepancy may be interpreted as resulting from the velocity renormalization in the semiconductor in the strong coupling limit~\cite{Stanescu2011, nadj2013, Peng2015}. 

\section{Conclusions and Outlook}

In this Article, we demonstrated experimentally and theoretically that threading magnetic flux through a semiconducting nanowire fully covered by a superconducting shell can induce a topological phase with Majorana zero modes at the nanowire ends.
While being of similar simplicity and practical feasibility~\textbf{}\cite{Krogstrup2015} as the original nanowire proposals with a partial shell coverage~\cite{Lutchyn2010,Oreg2010}, full-shell
nanowires provide several key advantages.
First, the topological transition in a full-shell wire is driven by the field-induced winding of the superconducting order parameter, rather than by the Zeeman effect so that, as demonstrated in the reported measurements, the required magnetic fields can be very low ($\sim 0.1$~T).
Therefore, the present proposal is compatible with conventional superconducting electronics and removes the need for a large $g$ factor semiconductor, potentially expanding the landscape of candidate materials.
Moreover, the full shell naturally protects the semiconductor from impurities and random surface doping, thus enabling a reproducible way of growing many wires with essentially identical electrostatic environments.
Although full-shell wires do not allow for direct gating of the electron density in the semiconducting core, we demonstrated that via a careful design of the wire properties, for example by choosing the appropriate radius, it is possible to obtain wires that harbor MZMs at a predictable magnetic field.
The modest magnetic field requirements, protection of the semiconducting core from surface defects, and locked phase winding in discrete lobes together suggest a new and relatively easy route to creating and controlling Majorana zero modes in hybrid materials. Our findings open a possibility to study an interplay of mesoscopic and topological physics in this novel system.

\section{Acknowledgments}

We thank A.~Antipov, L.~Casparis, M.~Freedman, A.~Higginbotham, E.~Martinez for valuable discussions, J.~Gamble and J.~Gukelberger for contributions to the simulation code, as well as C.~S\o rensen, R.~Tanta and S.~Upadhyay for contributions to material growth and device fabrication. Research was supported by Microsoft, the Danish National Research Foundation, and the European Commission. This work was performed in part at Aspen Center for Physics, which is supported by National Science Foundation grant PHY-1607611. M.T.D.~acknowledges support from State Key Laboratory of High Performance Computing, China. L.I.G. acknowledges support from NSF DMR Grant No. 1603243.  

\bibliography{references}

\begin{thebibliography}{10}
\bibitem{FootnoteFlux}
Note that magnetic flux piercing the finite-thickness superconducting shell can
  be significantly different from that penetrating the core.
\bibitem{Footnote1}
With our sign convention this electric field results in an $\alpha$ of negative
  sign. However, there are also other contributions to $\alpha$ from strain and
  microscopic details of the InAs-Al interface. The sign of the spin-orbit
  coupling is difficult to predict due to this.
\bibitem{FootnoteRadius}
$R$ is measured from the center to a corner of an effective hexagon.
\bibitem{FootnoteXi}
While the $\xi_d$ chosen in the simulation is slightly different from the
  experimentally measured $\xi_S=180$\,nm, the gap-field dependence is still
  very close in simulation and experiment.
\end{thebibliography}

\bibliographystyle{Science}

%\nocite{Dao2009,Lee2013, Larkin1965, Koshnick2007, Beenakker1991, Cochran1958, DasSarma2012}

\onecolumngrid
\clearpage
\onecolumngrid
\setcounter{figure}{0}
\setcounter{equation}{0}
\section{\large{S\texorpdfstring{\MakeLowercase{upplemental} M\MakeLowercase{aterial}}{}}}
\renewcommand\thefigure{S\arabic{figure}}
\renewcommand{\tablename}{Table.~S}
\renewcommand{\thetable}{\arabic{table}}
\twocolumngrid

\section{Nanowire growth}

The hybrid nanowires used in this work were grown by molecular beam epitaxy on InAs (111)B substrate at $420~^\circ$C. The growth was catalyzed by Au via the vapor-liquid-solid method. The nanowire growth was initiated with an axial growth of InAs along the $[0001]$ direction with wurtzite crystal structure, using an In flux corresponding to a planar InAs growth rate of $0.5~\mu$m/hr and a calibrated As$_{4}$/In flux ratio of 14. The InAs nanowires with core diameter of ${\sim}130$~nm were grown to a length of ${\sim}10~\mu$m. Subsequently, an Al shell with thickness of ${\sim}30$~nm (or ${\sim}10$~nm for the nanowire used in device 4) was grown at $-30~^\circ$C on all six facets by continuously rotating the growth substrate with respect to the metal source. The resulting full shell had an epitaxial, oxide-free interface between the Al and InAs~\cite{Krogstrup2015}.
 
\section{Device fabrication}
 
The devices were fabricated on a degenerately n-doped Si substrate capped with a $200$~nm thermal oxide. Prior to the wire deposition, the fabrication substrate was pre-fabricated with a set of alignment marks as well as bonding pads. Individual hybrid nanowires were transferred from the growth substrate onto the fabrication substrate using a manipulator station with a tungsten needle. Standard electron beam lithography techniques were used to pattern etching windows, contacts and gates.

The quality of the Al etching was found to improve when using a thin layer of AR 300-80 (new) adhesion promoter. Double layer of EL6 copolymer resists was used to define the etching windows. The Al was then selectively removed by submerging the fabrication substrate for ${\sim}70$~s into MF-321 photoresist developer.

As the native InAs and Al oxides have different work functions, different cleaning processes had to be applied before contacting the wires. To contact the Al shell in devices 1, 3, 4 and 5, a stack of A4 and A6 PMMA resists was used. Normal Ti/Al ($5/210$~nm) ohmic contacts to Al shell were deposited after \textit{in-situ} Ar-ion milling (RF ion source, $25$~W, $18$~mTorr, $9$~min). To contact the InAs core in all seven devices, a single layer of A6 PMMA resist was used. A gentler Ar-ion milling (RF ion source, $15$~W, $18$~mTorr, $6.5$~min) was used to clean the InAs core followed by metalization of the normal Ti/Al ($5/180$~nm) ohmic contacts to InAs core.

A single layer of A6 PMMA resist was used to form normal Ti/Al ($5/150$~nm) side-gate electrodes in devices 2, 6 and 7, and top-gate electrode in device 4, separated from the wire by ${\sim}8$~nm layer of atomic layer deposited dielectric HfO$_2$.

\section{Measurements}

Each of the dc lines used to measure and gate the devices was equipped with RF and RC filters (QDevil \cite{QDevil}), adding a line resistance $R_{\rm Line}=6.7$~k$\Omega$. This has a negligible effect on the data in a weak tunneling regime, where the device resistance is much greater than line resistance. In the strong tunneling regime, however, a significant fraction of the applied voltage drops over the line resistance (dominated by the filters), resulting in smaller measured conductance values. A comparison between measured two-terminal and spectra and numerically corrected spectra is presented in Fig.~\ref{fig:S4}. The 4-probe differential resistance measurements were carried out using an ac excitation of $I_{\rm ac}=200$~nA. The 2-probe tunneling conductance measurements were conducted using ac excitation of $V_{\rm ac}= 5~\mu$V.

\section{Numerical simulations}

The normal-state Hamiltonian used in the numerical simulations is given by
\begin{align}
\begin{split}
  &H_0 =\\ &\left[(\vec{p}+e A_{\varphi}\hat{\varphi})^T/(2m(\vec{r}))(\vec{p}+e A_{\varphi}\hat{\varphi})-E_\mathrm{F}(\vec{r})+U(\vec{r}) \right]\sigma_0+\\
  &\frac{1}{2}\left[\alpha(\vec{r})\, \hat{r} \left(\vec{\sigma} \times  (\vec{p}+e A_{\varphi}\hat{\varphi})\right)+\hat{r} \left(\vec{\sigma} \times  (\vec{p}+e A_{\varphi}\hat{\varphi})\right)\,\alpha(\vec{r})\right]+\\
  & Bg(\vec{r})\frac{\mu_B}{2}\sigma_z,
  \end{split}
\end{align}
where $E_\mathrm{F}$ is the Fermi level, $U$ is the potential energy and $\alpha$ is the radial spin-orbit coupling. We solve for the electrostatic potential in a separate step using the Thomas-Fermi approximation analog to Ref.~\cite{Winkler2018}. In the semiconductor (InAs) we take $m_\mathrm{semi}=0.026\, m_0$, $E_\mathrm{F, semi}=0$ and $g_\mathrm{semi}=14.7$~\cite{Vurgaftman2001}. In the superconductor (Al) we take $m_\mathrm{super}=m_0$, $E_\mathrm{F, super}=11.7\,$eV and $\alpha_\mathrm{super}=g_\mathrm{super}=0$~\cite{AshcroftMermin}. For simplicity we set $g_\mathrm{super}=0$. The vector potential $A_\varphi = B r/2$ corresponds to a spatially homogeneous magnetic field.

The Bogoliubov-de-Gennes Hamiltonian is given by
\begin{align}
  H_{\rm BdG}=\left[\begin{array}{cc}
                      H_0(\vec{r}, \vec{A})-i\eta \sigma_0& \Delta(\vec{r}) \\
                      \Delta^*(\vec{r}) & -\sigma_y H_0(\vec{r}, -\vec{A})^* \sigma_y-i\eta \sigma_0
                    \end{array}
                    \right],\label{eq:HBdG}
\end{align}
where we introduce a small dissipative term $\eta$. It is numerically advantageous to introduce a small level broadening $\eta=2\,\mu$eV in the transport simulations in order to avoid sharp features. In all other simulations we set $\eta=0$. A side-effect of nonzero $\eta$ is that the conductance becomes particle-hole-asymmetric for bias voltages below the Al gap~\cite{Liu2017}. Non-zero $\eta$ can also correspond to a soft-gap in the superconductor or result from coupling to an additional lead.

The superconductor is integrated-out into a self-energy boundary condition, see also Eq.~\ref{eq:Greens}.
The effective mass in Al is taken to be infinite parallel to the interface, and finite perpendicular to the interface \cite{FootnoteEffectiveMass}. This means that in the discretized Hamiltonian, every lattice site adjacent to the superconductor is attached to a semi-infinite, one-dimensional Al chain. The idea behind this arrangement is to effectively simulate the fact that in the strong coupling limit there is nearly perfect Andreev reflection from the superconductor~\cite{Kiendl2019}, as discussed in the Supplementary Text. In the semiconductor we use a lattice spacing of 5\,nm. Due to the small $\lambda_\mathrm{F}$, significantly smaller lattice spacing of 0.1\,nm is required in Al. The non-equidistant discretization across the interface is described by the method of Ref.~\cite{Antipov2018}. For the InAs-Al bonds we choose a length of 0.1\,nm -- the same as the discretization in Al -- to ensure strong coupling.

We assume the following gap dependence within a lobe [see also Eq.~\eqref{eq:gap}]
\begin{align}
\begin{split}
     \Delta_0(B, n) = \Delta_0(0, 0)\, \max\left(0, 1-(n B_0/B_\mathrm{max})^2\right)&\\ \times\max\left(0, 1-\xi_d^2/R^2 (B/B_0-n)^2\right)&,   
\end{split}
\end{align}
with an effective radius $R=80$\,nm and $B_0=\phi_0/(R^2 3 \sqrt{3}/2)$ for a hexagonal cross-section \cite{FootnoteRadius}. The full pairing $\Delta$ is then $\Delta(\vec{r}, B, n=\lfloor\frac{B}{B_0}+0.5\rfloor) = \Delta_0(B, n) \Theta(\vec{r}\ \mathrm{in}\ \mathrm{Al}) e^{in\varphi}$. We take $\Delta(0, 0)=0.24$\,meV, $\xi_d=210$\,nm and $B_\mathrm{max}=0.8$\,T which results in a similar gap-field dependence as in the experiment~\cite{FootnoteXi}.

\begin{figure}[h!]
  \begin{center}
    \includegraphics[width=0.9\columnwidth]{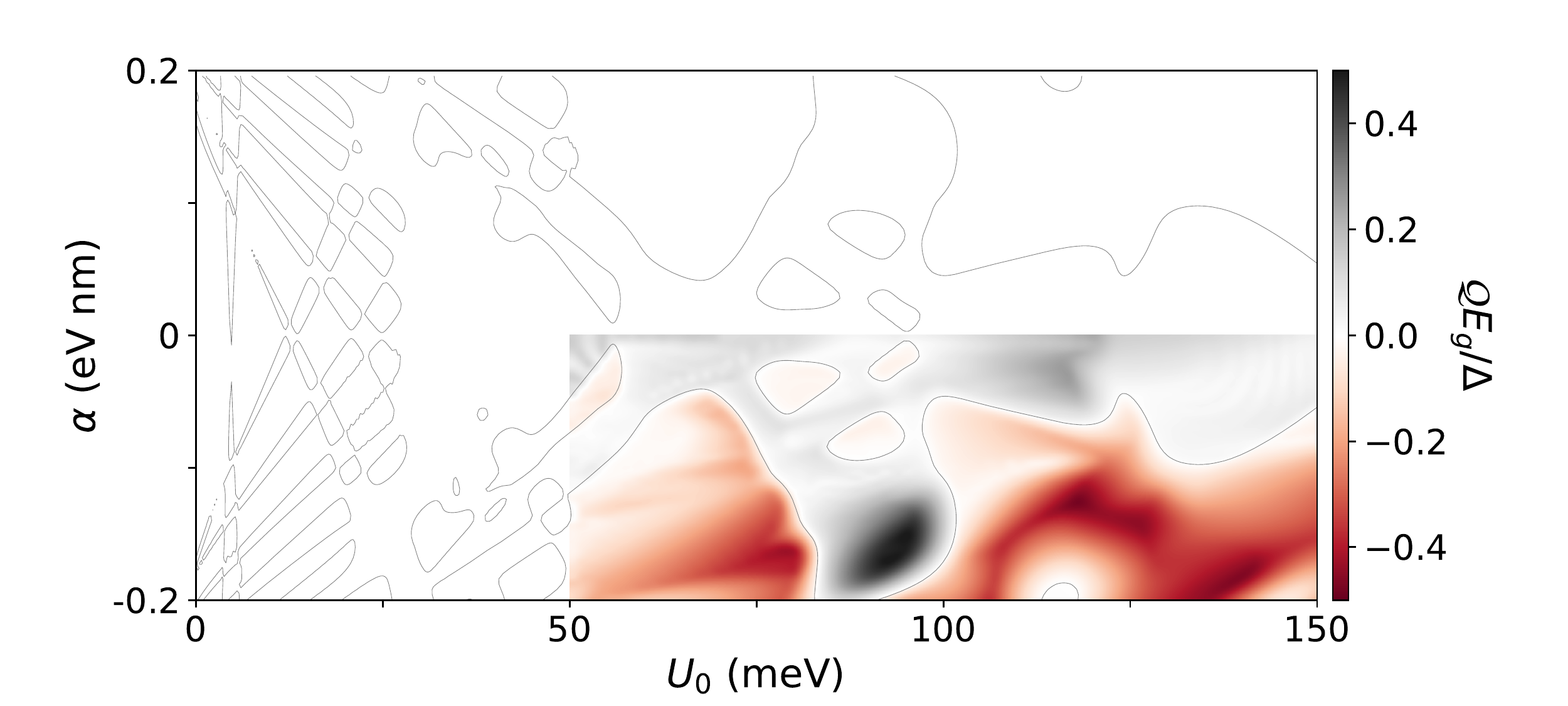}
\caption{\textbf{Topological phase diagram of a hexagonal full shell wire with disordered Al}. The thickness of Al is 20 nm. A random on-site disorder potential sampled from $\delta U
  \in [-U, U]$ with $U=3$\,eV is applied in the outer Al layer of 2 nm thickness. A discretization of $a=0.1$ nm is used throughout the system. Other parameters are chosen identical to the simulations with self-energy in the main text. Gray lines indicate a change of sign in the Pfaffian. In the lower right corner (band offset $> 50$ meV and $\alpha<0$) the energy gap of the system is indicated with color (due to the large computational complexity the energy gap is calculated only for part of the phase diagram). \label{fig:S1}}
\end{center}
\end{figure}

We validate our approach of integrating the superconductor out by performing simulations with a disordered Al shell, see Fig.~\ref{fig:S1}. We find good qualitative agreement with the phase diagram of Fig.~\ref{fig:4} of the main text.

\section{Destructive regime}\label{sec:S2.1}
As a result of fluxoid quantization, the critical temperature, $T_{\rm C}$, of a cylindrical shell is periodically modulated by an axial magnetic field \cite{Tinkham1966,Little1962}. For cylinders with radius smaller than the superconducting coherence length, $T_{\rm C}$ is expected to vanish whenever the applied flux (axial magnetic field component times shell cross-sectional area) is close to an odd half-integer multiple of the superconducting flux quantum, $n\Phi_0/2$ ($\Phi_0 = h/2e = 2.07$~mT$\,\mu$m$^{2}$, $n=1,3,5,\ldots$) \cite{deGennes1981, Schwiete2009,Dao2009}. Throughout the extended range where $T_{\rm C}$ vanishes, superconductivity is destroyed \cite{Liu2001,Sternfeld2011}.

\begin{figure}[h!]
\centering
\includegraphics[width=0.67\linewidth]{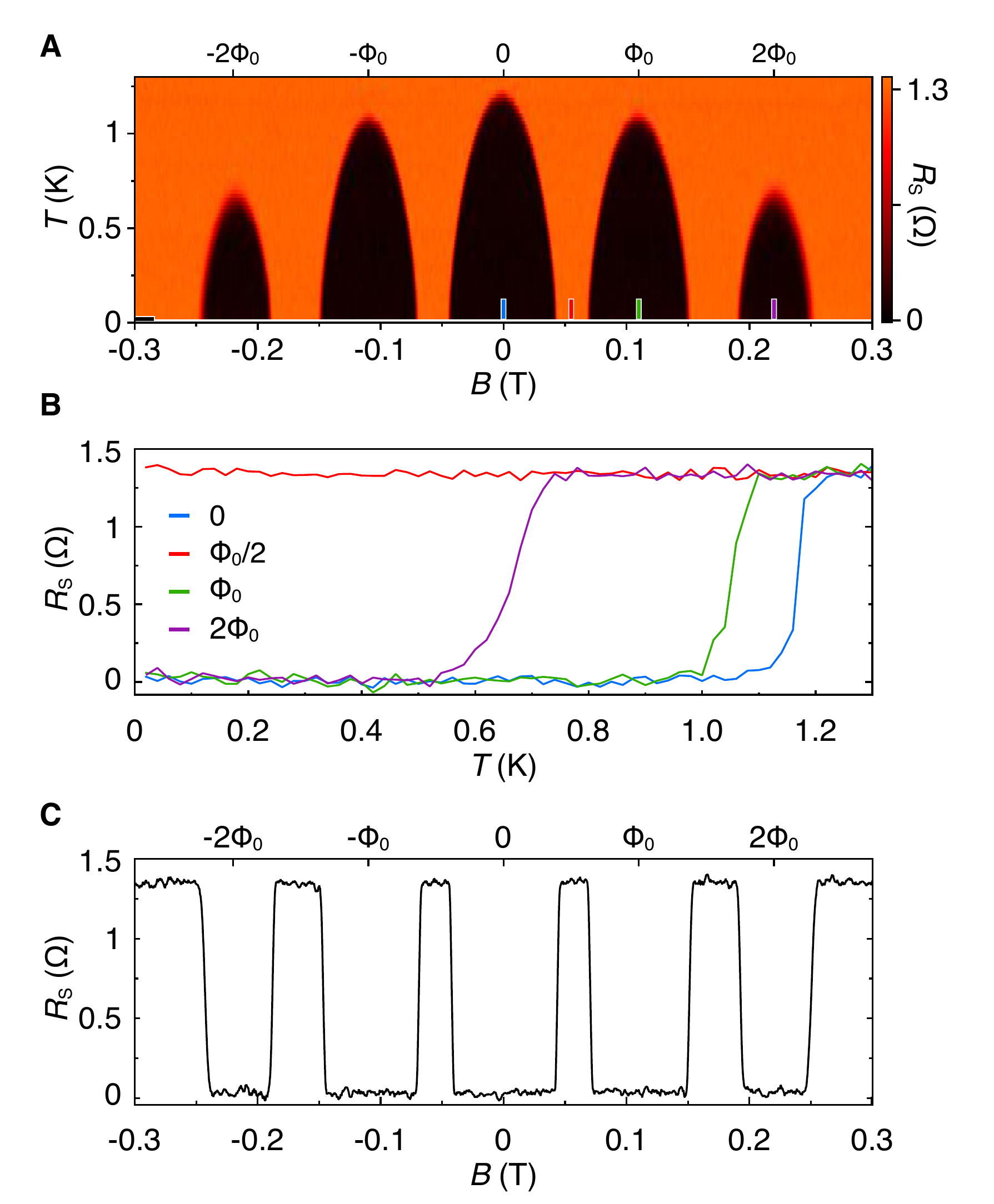}
\caption{\textbf{Shell resistance versus temperature and magnetic field (device 1).} (\textbf{A}) Same data as in the main-text Fig.~1D: Differential resistance of the Al shell, $R_{\rm s}$, measured for device 1 as a function of temperature, $T$, and axial magnetic field, $B$. (\textbf{B}) Line-cuts from (A) taken at $0$, $1/2$, $1$ and $2$ flux quanta, $\Phi_0$. At half of the flux quantum, $R_{\rm s}$ stays at the normal state resistance down to the lowest measured temperature $T=20$~mK. (\textbf{C}) Line-cut from (A) taken at $T=20$~mK. Two destructive regimes surrounded by fully superconducting phase can be  seen around $\vert B\vert = 55$ and $165$~mT.\label{fig:S2}}
\end{figure}

The nanowire used in device 1 has a hexagonal InAs core with diameter of $130$~nm and Al shell with thickness of $30$~nm, giving a mean diameter of ${\sim}160$~nm. The dirty-limit coherence length is given by $\xi_{\rm S}=\sqrt[]{\pi\hbar v_{\rm F} l/24 k_{\rm B} T_{\rm C}}$ \cite{Tinkham1966,Gordon1984}. The measured normal state resistance is $R_{\rm N}=1.3~\Omega$. The distance between the voltage probes is $L~{\sim}~940$~nm. This yields shell resistivity of $\rho_{\rm S} = R_{\rm N} A_{\rm S}/L = 21$~nm~$\Omega$, with the shell cross-sectional area $A_{\rm S}\,\sim\,3\sqrt{3}/2\,(100^2 - 65^2)$~nm$^2$.
The Fermi velocity in Al is $v_{\rm F}=2\times 10^6$~m/s \cite{Kittel2005}, giving a Drude mean free path of $l_{\rm e} = m_{\rm e}{ }v_{\rm F}/ e^2 n \rho=19$~nm, where $m_{\rm e}$ is electron mass, $e$ is electron charge and $n = k_{\rm F}^3/ 3 \pi^2$ is charge carrier density, with Fermi wave vector $k_{\rm F}$.
The measured critical temperature is $T_{\rm C}=1.2$~K. This gives $\xi_{\rm S}=180$~nm, greater than the mean nanowire radius (${\sim}80$~nm), hence the measured nanowires are expected to exhibit a destructive regime. This is consistent with the measurements, see Fig.~\ref{fig:S2}A.

At integer flux quanta ($0, \pm\Phi_0$ and $\pm2\Phi_0$), normal-to-superconducting transitions appear as the temperature is lowered, with the critical temperature decreasing as the flux number increases. Around $\pm\Phi_0/2$ and $\pm3\Phi_0/2$\ the resistance of the shell, $R_{\rm s}$, stays at the normal value down to the lowest measured temperature, $\sim20$~mK, as shown in Fig.~\ref{fig:S2}B. At the base temperature, the two destructive regimes can be identified by abrupt changes of $R_{\rm s}$ from $0$ to $R_{\rm N}$ and then back to $0$ when the flux passes $\pm\Phi_0/2$ and $\pm 3\Phi_0/2$, see Fig.~\ref{fig:S2}C.

\section{Penetration depth}\label{sec:S2.2} 
An applied magnetic field penetrates thin-film superconductors with thickness much less than penetration depth, $\lambda$, uniformly. In dirty limit, the effective penetration depth $\lambda_{\rm eff} = \lambda_{\rm L}~\sqrt[]{\xi_0/(1.33~l)}$, where $\xi_0$ and $\xi_{\rm S}$ at zero temperature are related by $\xi_{\rm S} = 0.855~\sqrt[]{\xi_0 l}$ \cite{Tinkham1966}. Taking $\lambda_{\rm L}=16~nm$ as the London penetration depth for Al \cite{Kittel2005}, yields $\lambda_{\rm eff}=150$~nm greater than Al thickness (30~nm). As a result, the flux in the wire is not quantized. Note, however, that the fluxoid is still quantized \cite{Little1962,Tinkham1966}.

\section{Tunneling spectroscopy}\label{sec:S2.3}
\begin{figure}[h!]
\centering
\includegraphics[width=0.8\linewidth]{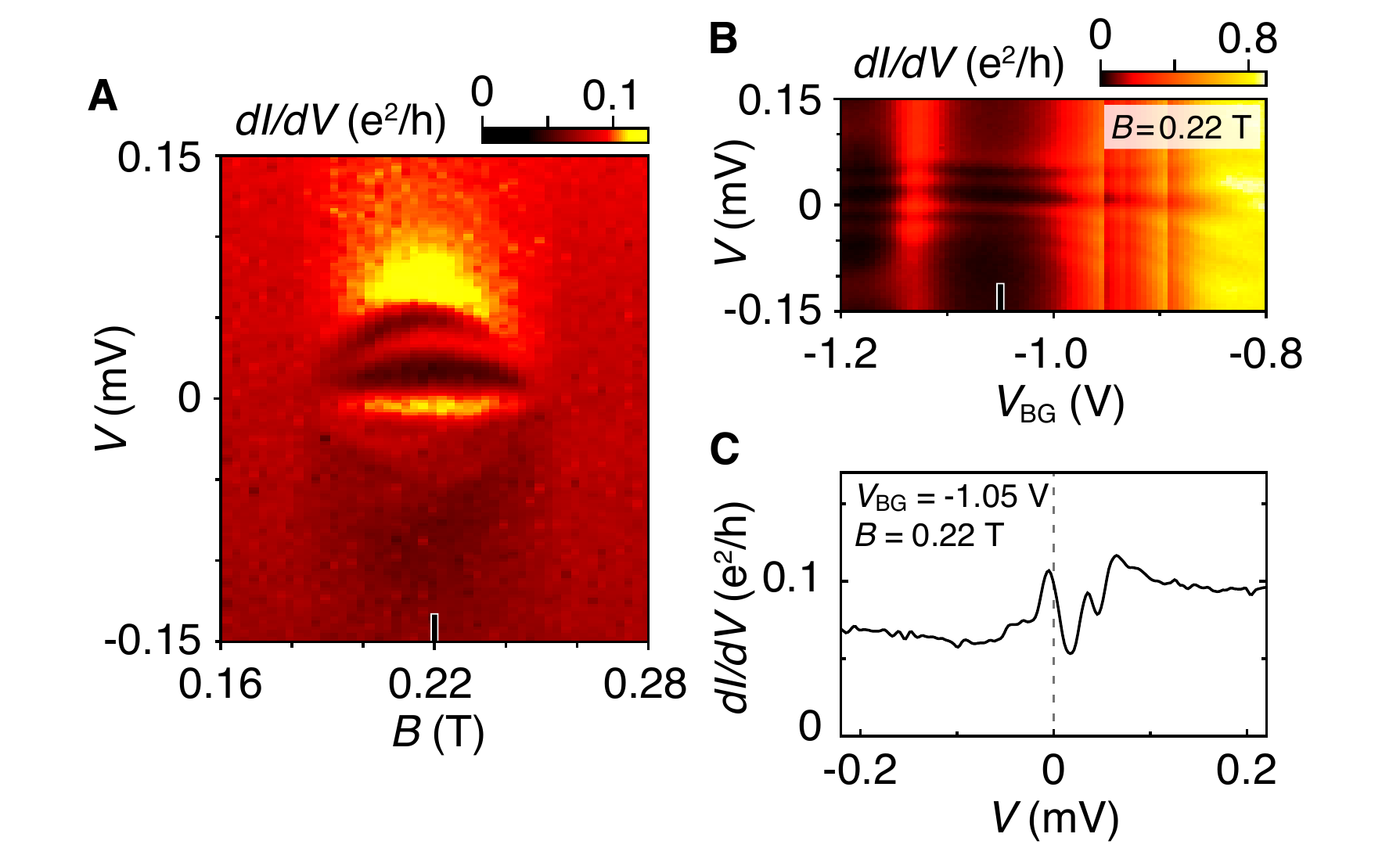}
\caption{\textbf{Tunneling spectroscopy in the second lobe (device 1).} (\textbf{A}) Zoom-in around the second superconducting lobe of the data shown in the main-text Fig.~2B. (\textbf{B}) Differential conductance as a function of source-drain voltage, $V$, and back-gate voltage, $V_{\rm BG}$. (\textbf{E}) Line-cut of the conductance taken at $B=0.22$~T and $V_{\rm BG}=-1.05$~V. The spectrum shows subgap states away from zero energy.\label{fig:S3}
}
\end{figure}

\begin{figure}[h!]
\centering
\includegraphics[width=0.8\linewidth]{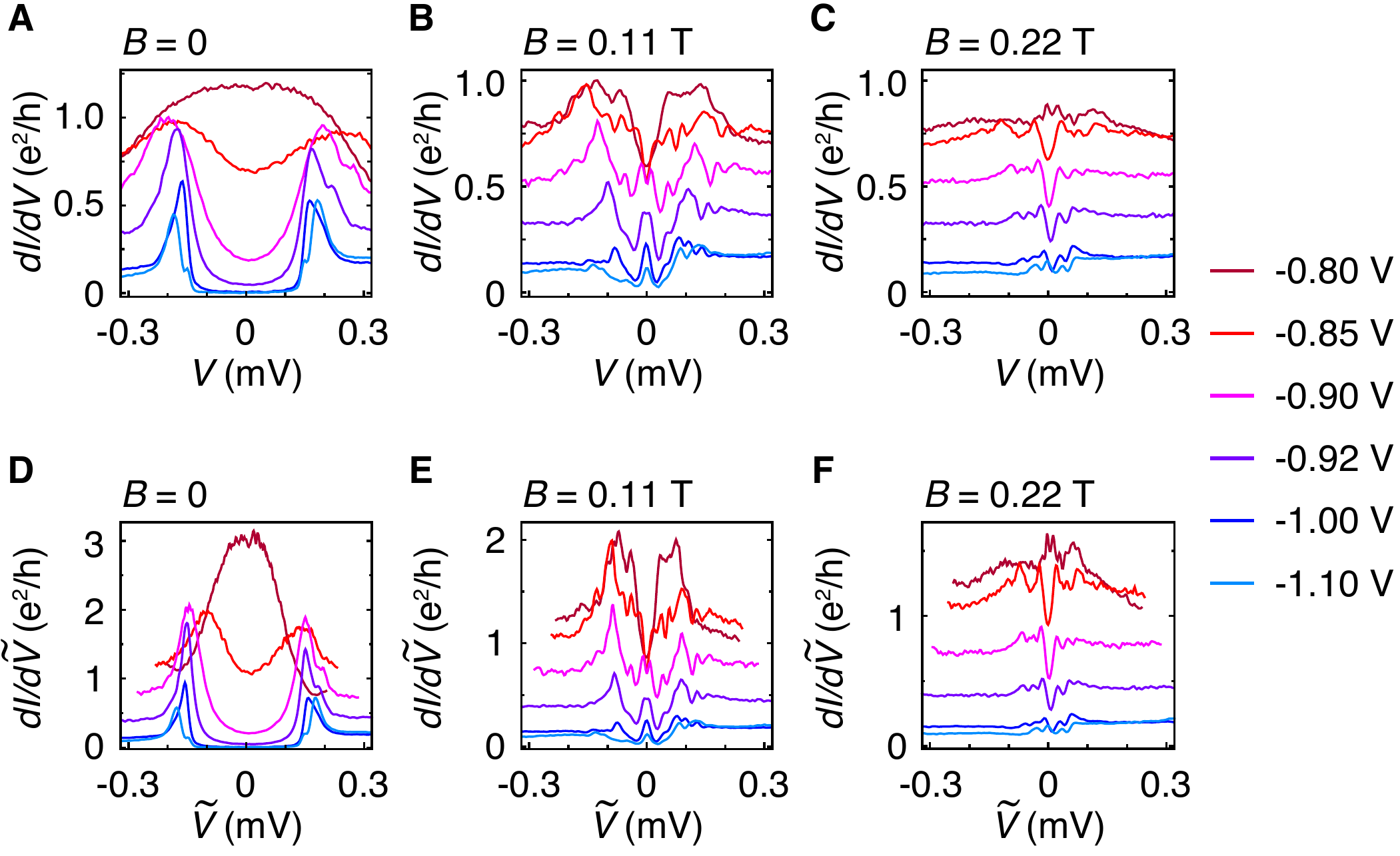}
\caption{\textbf{Spectrum evolution with barrier strength (device 1).} Line-cuts of the 2-terminal conductance at different back-gate voltages, $V_{\rm BG}$, measured for device 1 at (\textbf{A}) $B=0$, around zero flux, (\textbf{B}) $B=0.11$~T, around one flux quantum, and (\textbf{C}) $B=0.22$~T, around two flux quanta. (\textbf{D-F}) Similar to (A-C) but with removed line resistance, which was done by numerically integrating the data, subtracting the voltage drop over the filters (see Materials and methods in main text) and numerically differentiating the data.
\label{fig:S4}
}
\end{figure}

\begin{figure}[h!]
\centering
\includegraphics[width=0.67\linewidth]{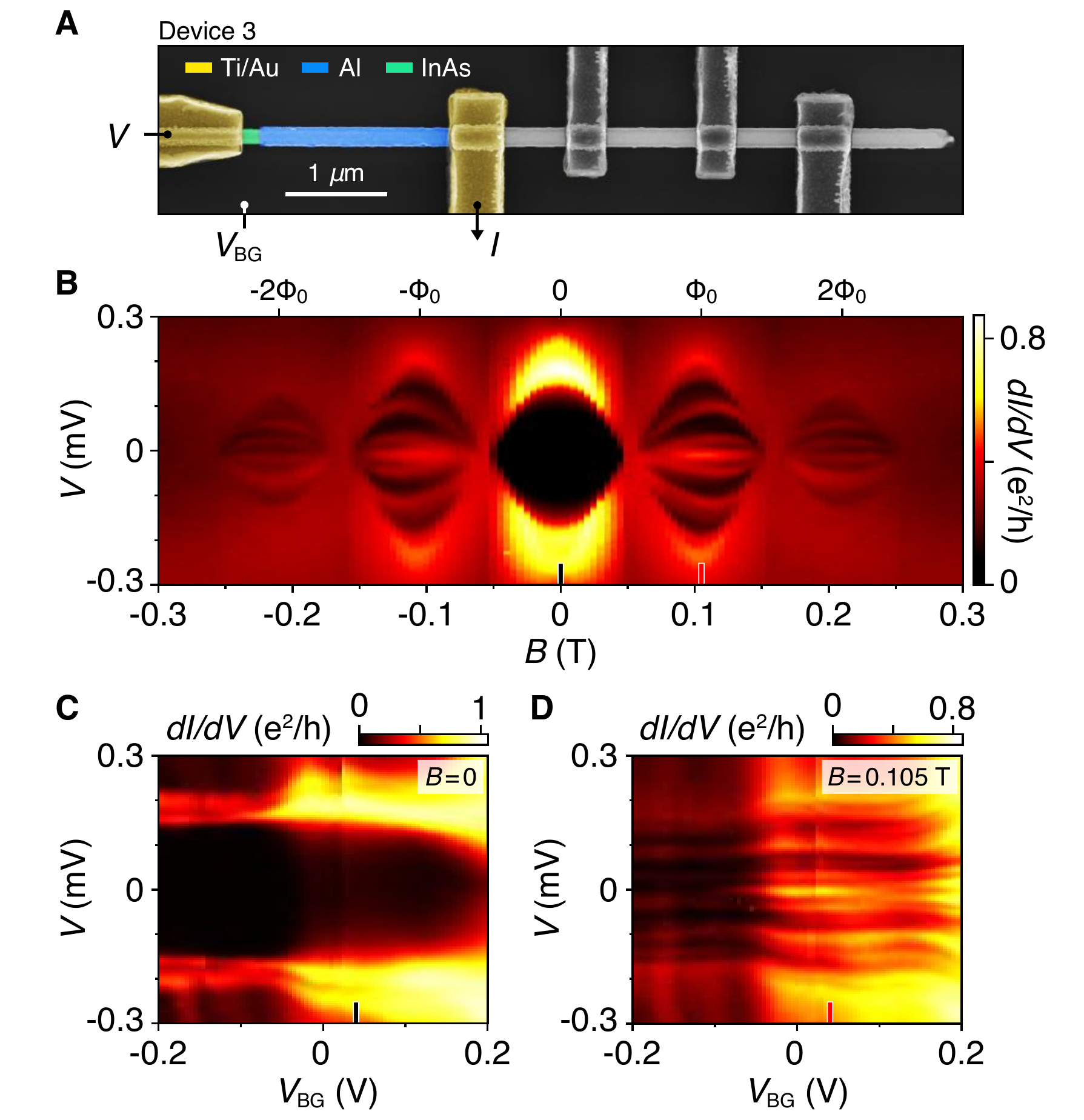}
\caption{\textbf{Tunneling spectroscopy from an additional device (device 3).}
(\textbf{A}) Micrograph of device 3, lithographically equivalent to device 1, colorized to highlight tunneling spectroscopy set-up. (\textbf{B}) Differential conductance, $dI/dV$, as a function of source-drain bias voltage, $V$, and axial field, $B$. The zeroth lobe shows a hard superconducting gap, the first lobe shows zero-bias peak, the second lobe shows non-zero energy subgap states. The lobes are separated by featureless normal-state spectra. (\textbf{D}) Zero-field conductance as a function of $V$ and back-gate voltage, $V_{\rm BG}$. (\textbf{E}) Similar to (D), measured in the first lobe at $B=110$~mT.
\label{fig:S5}}
\end{figure}

\begin{figure}[h!]
\centering
\includegraphics[width=0.65\linewidth]{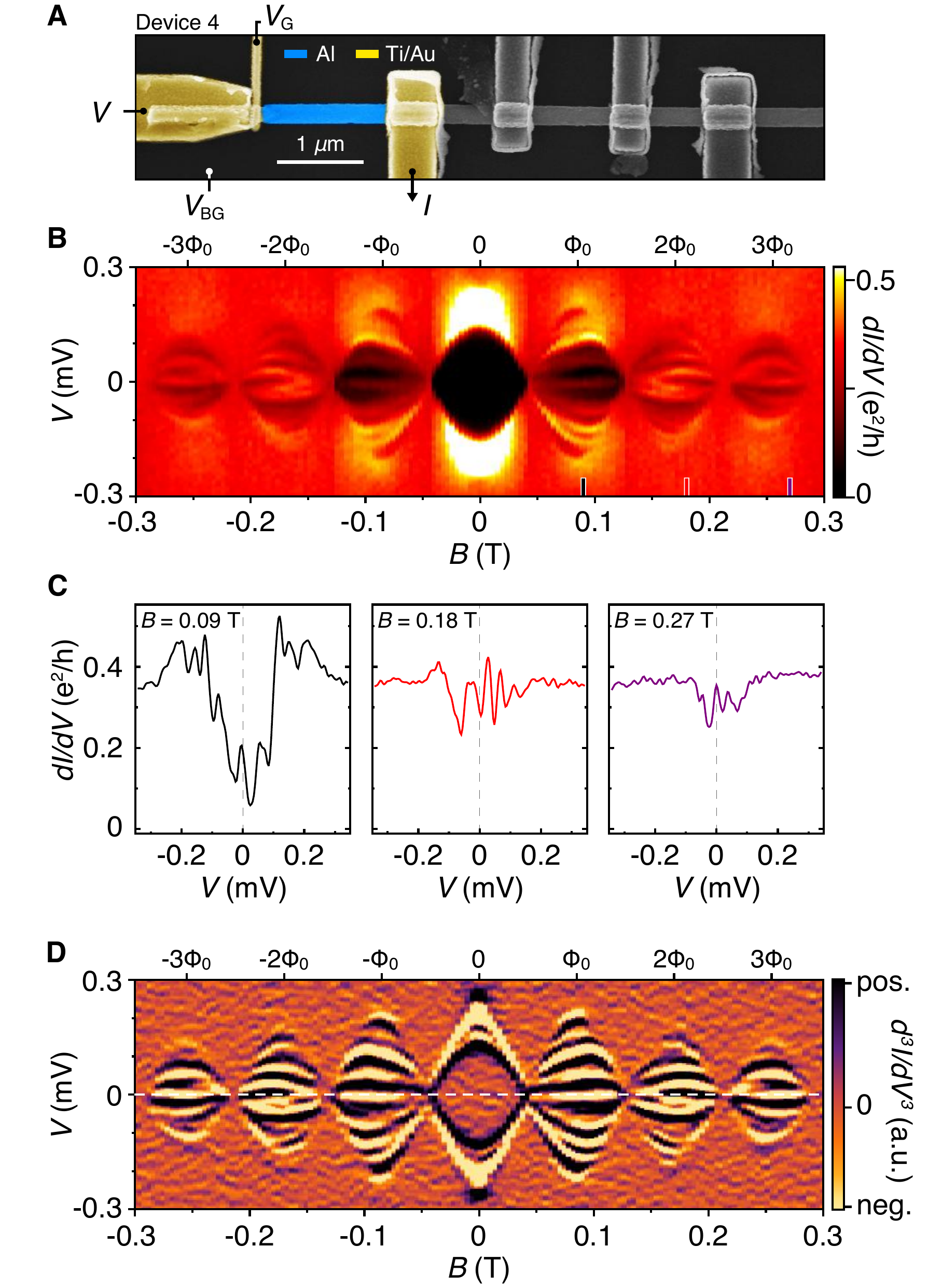}
\caption{\textbf{Zero-bias peak in the first and third lobes only (device 4).}
(\textbf{A}) Micrograph of device 4, lithographically similar to device 1, but with an additional top-gate and slightly bigger InAs diameter of ${\sim}160$~nm. (\textbf{B}) Differential conductance, $dI/dV$, as a function of source-drain bias voltage, $V$, and axial field, $B$, measured at back-gate voltage $V_{\rm BG}=-5$~V and top-gate voltage $V_{\rm BG}=-2$~V. (\textbf{C}) Line-cut of the conductance taken at (left) $B=0.09$~T, around one flux quantum, (middle) $B=0.18$~T, around two flux quanta, and (left) $B=0.27$~T, around three flux quanta. (\textbf{D}) The curvature (numerical second derivative) of the conductance, $d^3I/dV^3$, for the data shown in (B). Negative (positive) curvature corresponds to a peak (dip) in conductance. The zeroth lobe shows a hard superconducting gap, the first lobes show subgap states including a zero-bias peak, the second lobes show non-zero energy subgap states. The third lobes show subgap states again with a peak at zero bias.\label{fig:S6}}
\end{figure}

\begin{figure}[h!]
\centering
\includegraphics[width=0.67\linewidth]{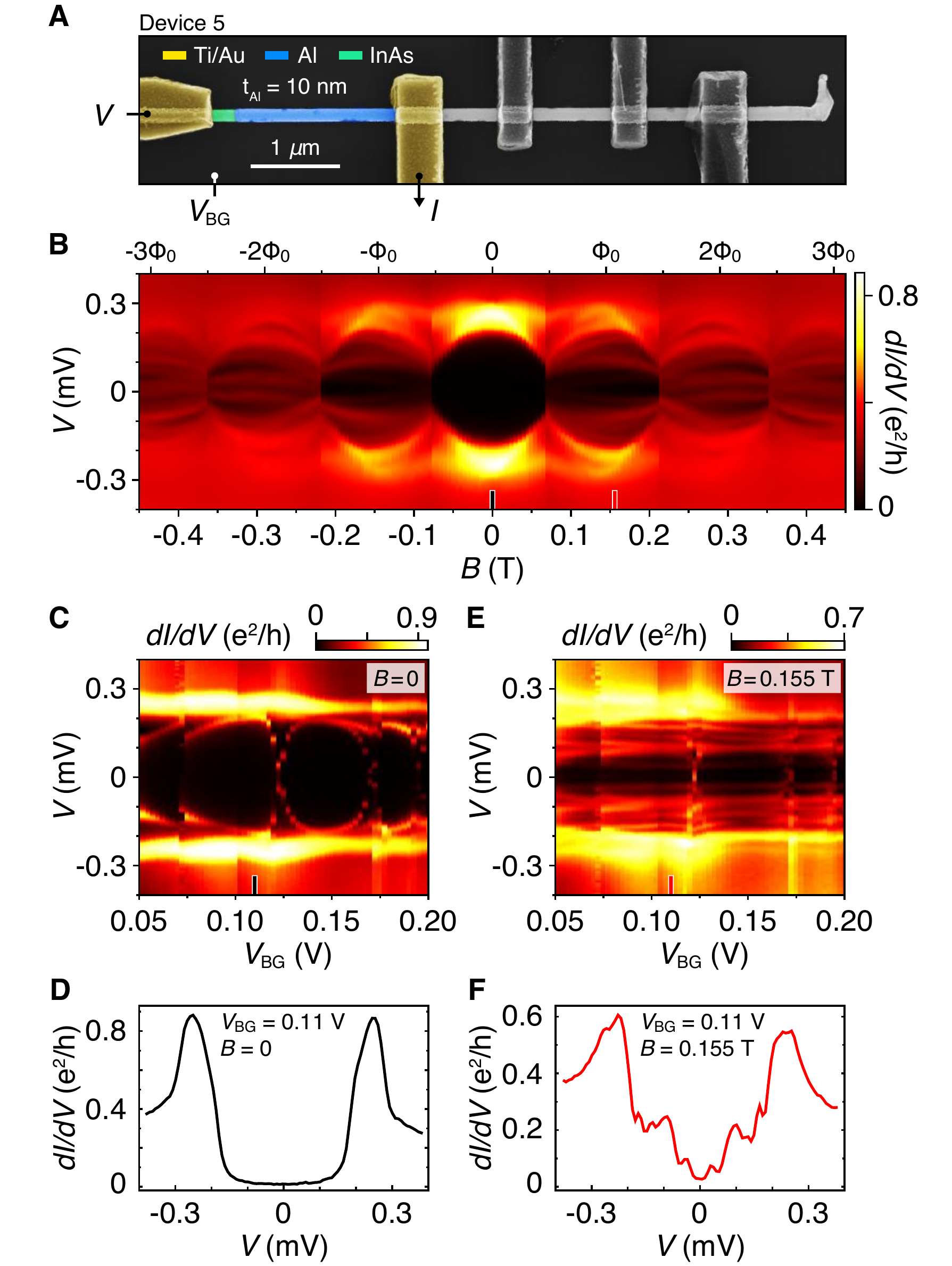}
\caption{\textbf{Tunneling spectroscopy without zero-bias peaks in device with thinner Al shell (device 5).} (\textbf{A}) Micrograph of device 5, comprised of a nanowire with a $10$~nm Al full-shell, colorized to highlight tunneling spectroscopy set-up. (\textbf{B}) Differential conductance, $dI/dV$, as a function of source-drain bias voltage, $V$, and axial field, $B$. The zeroth lobe shows a hard superconducting gap, the higher-order lobes show multiple discrete states away from zero energy. No destructive regime is present in the thinner-shell device. (\textbf{C}) Zero-field conductance as a function of $V$ and back-gate voltage, $V_{\rm BG}$. (\textbf{D}) Line-cut of the conductance taken at $B=0$ and $V_{\rm BG}=0.11$~V. (\textbf{E} and \textbf{F}) Similar to (C) and (D), measured in the first lobe at $B=110$~mT.\label{fig:S7}
}
\end{figure}

For all four tunneling-spectroscopy devices (1, 3, 4 and 5) the zeroth lobe, where the winding number is 0, shows a hard gap and no subgap states are visible. In the first lobe, with the phase winding of $2\pi$, the spectrum for devices 1, 3 and 4 (all with $30$~nm Al shell) displays a discrete, zero-energy state (see main-text Fig.~2, and Figs.~\ref{fig:S5}~and~\ref{fig:S6}), whereas for device 5 (with 10 nm Al shell) the spectrum consists of multiple discrete, but finite energy subgap states, see Fig.~\ref{fig:S7}. In the second lobe, with even number of phase windings, the spectrum for device 1 features an asymmetric superconducting density of states with the lowest energy subgap state centered around $-5~\mu$eV, see Fig.~\ref{fig:S3}; For devices 3 and 4, multiple subgap states can be identified at finite voltage, but no zero-bias peak, see Figs.~\ref{fig:S5}~and~\ref{fig:S6}; For device 5, a qualitatively similar to the first lobe spectrum with several finite-energy states is observed, see Fig.~\ref{fig:S7}. Device 4 with slightly bigger diameter, displays the third lobe, with odd number of phase windings. The spectrum features subgap states and a peak at zero bias, see Fig.~\ref{fig:S6}.

For device 5, a discrete state crosses zero-energy around $V_{\rm BG}=0.12$~V and then again at $0.17$~V, resembling a proximitized quantum dot state, similar to the one previously studied in Ref.~\cite{Lee2013}, see Fig.~\ref{fig:S7}. We usually associate such state with a resonant level in the barrier and if possible avoid it in the measurements.

\section{Model for the disordered superconducting shell}
\label{app:shell}

\begin{figure}[h!]
  \begin{center}
    \includegraphics[width=0.9\columnwidth]{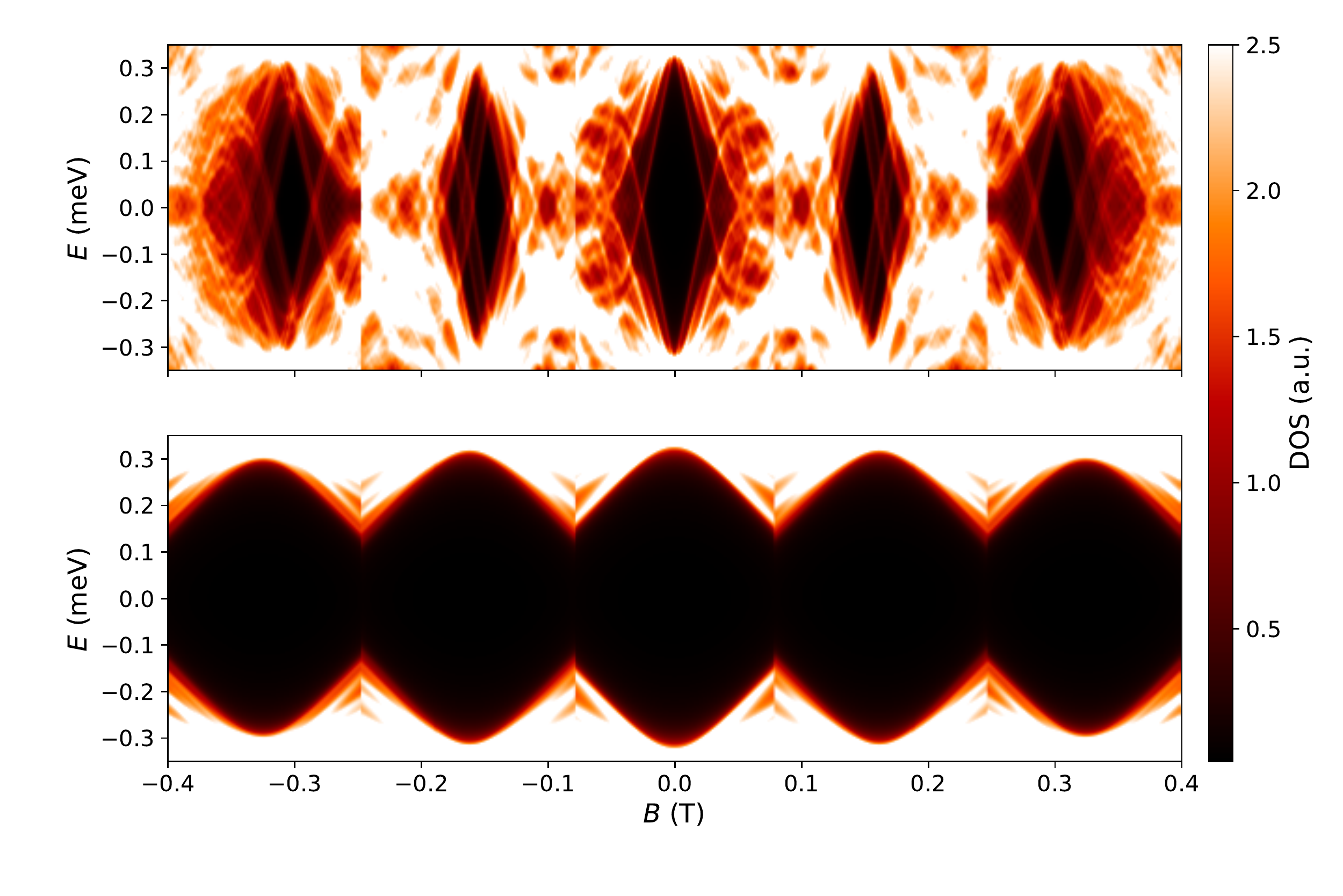}
\caption{\textbf{Simulation of disordered Al shell.} We simulate a superconducting shell, without the semiconducting core, with $R_1 = R_2=60$\,nm and $R_3=70$\,nm. Realistic parameters corresponding to Al are used: $m^*=m_e$,  $\mu=10$\,eV and $\Delta_0=0.34\,$meV~\cite{Cochran1958}. The Hamiltonian Eq.~\eqref{eq:shamiltonian} is discretized on a square lattice with $a=0.1$\,nm using the Kwant package~\cite{kwant}. {\it Top:} We show the clean case, where the superconductivity is rapidly destroyed by the magnetic field when the radius $R_3$ is smaller than the coherence length $\xi_0$ in a clean superconductor. {\it Bottom:} We show the disordered case using the on-site disorder potential $\delta U$ which is randomly sampled from $\delta U
  \in [-U, U]$ with $U=2$\,eV. The disorder is applied in an outer layer of 5\,nm thickness, with the purpose of modelling an oxidized Al$_2$O$_3$ layer. The amplitude of Little-Parks oscillations is small because $R_3 > \sqrt{\xi_0 l}$ with $l$ being the mean-free path, see also Ref. \cite{Schwiete2009}. \label{fig:S8}}
\end{center}
\end{figure}

In this Section, we consider a disordered superconducting shell (e.g., Al shell) with inner and outer radii $R_2$ and $R_3$, respectively, see Fig.~\ref{fig:3} of the main text.
We assume that the thickness of the shell  $d\equiv R_3-R_2 \ll \lambda_L$, with $\lambda_L$ being the London penetration length in the bulk superconductor.
In this case, the screening of the magnetic field by the superconductor is weak and can be neglected.
The effective Hamiltonian for the SC shell in cylindrical coordinates can be written as
\begin{align}
  \!\!H^{(s)}_{\rm BdG}\!&=\left[\frac{\hat{p}_{z}^2}{2m^*}\!+\!\frac{\hat{p}_{r}^2}{2m^*}\!+\!\frac{(\hat{p}_{\varphi}\!+\!eA_{\varphi}\tau_z)^2}{2m^*}\!-\!\mu^{(s)}+V_{\rm imp} \right] \tau_z\! \nonumber\\
                   &+\!\Delta_0\left[\cos(n \varphi)\tau_x\!+\!\sin(n \varphi)\tau_y\right]
                     \label{eq:shamiltonian}
\end{align}
Here, $\hat{p}_i$ are the electron momentum operators, $e>0$ the electric charge,
$m$ the electron mass in the SC, $A_{\varphi}=\frac{1}{2}Br$, $\mu^{(s)}$ is the
chemical potential in the SC, $\tau_i$ are Pauli matrices representing Nambu space, $\Delta_0$ is bulk SC gap at $B=0$, $n$ is the winding number for the SC phase, and $V_{\rm imp}$ represents short-range impurity scattering potential. It is enlightening to perform a gauge transformation which results in a real order parameter, i.e. $\Delta_0\left[\cos(n \varphi)\tau_x\!+\!\sin(n \varphi)\tau_y\right]\rightarrow \Delta_0 \tau_x$.
The gauge transformation introduces an effective vector potential, $A_{\varphi} \rightarrow \tilde{A}_{\varphi}$ with
\begin{equation}
\tilde{A}_{\varphi}=-\frac{1}{2er} (n-2eA_{\varphi}r)=-\frac{1}{2e r}\left[n-\frac{\Phi(r)}{\Phi_0}\right]
\end{equation}
where $\Phi(r)=\pi B r^2$ and $\Phi_0=h/2e$. It follows from this argument that the solution of Eq.~\eqref{eq:shamiltonian} should be periodic with $\Phi_0$, see Fig.~\ref{fig:S8}.
Namely, the winding number adjusts itself to the value of the magnetic field so that the energy of the superconductor is minimized.
In particular, for each winding number $n$, the maxima of the quasiparticle gap occur at
\begin{equation}
B_n\approx 4n\frac{\Phi_0}{\pi (R_2+R_3)^2}\,.
\end{equation}
We neglect the Zeeman contribution since the typical magnetic fields of interest are smaller than $100$ mT for which the Zeeman splitting is negligible.

In order to understand the magnetic field dependence of the quasiparticle gap, one needs to calculate the Green's functions for the disordered SC shell as a function of $\tilde{A}_{\varphi}$.
The disordered superconductor is characterized by an elastic mean free path $l_e$ and a corresponding diffusive coherence length $\xi_d = \sqrt{l_e\,\xi_0} \gg\textbf{} l_e$, where $\xi_0=v_F/\Delta$ is the coherence length in the bulk, clean limit ($v_F$ is the Fermi velocity in the SC).
For simplicity, we assume henceforth that the thickness of the superconducting shell $d \gtrsim \xi_d$~\footnote{our results also apply to the case of $d \sim l_e$ and $\xi_d \gtrsim d$}, so that the properties of the system can be obtained by calculating the Green's function for the disordered bulk superconductor in magnetic field $B$ and $n=0$.
This problem was considered by Larkin~\cite{Larkin1965}, who showed that within the self-consistent Born approximation the normal and anomalous Matsubara Green's function are given by
\begin{align}
G^{(m_J)}(\omega_n, \varepsilon)= \frac{i\omega_n+\bar{G}+H m_J}{(\Delta+\bar{F})^2+\varepsilon^2-(i\omega_n+\bar{G}+H m_J)^2}\\
F^{(m_J)}(\omega_n,\varepsilon)=-\frac{\Delta+\bar{F}}{(\Delta+\bar{F})^2+\varepsilon^2-(i\omega_n+\bar{G}+H m_J)^2}
\end{align}
where $H=eB/4m$ and $m_J$ is the angular momentum eigenvalue and $\epsilon$ is the eigenvalue of the Hamiltonian 
\begin{align}
H^{\rm SC}_0 \phi(\vec{r})\!=\!\varepsilon \phi(\vec{r}) \mbox{ where } H^{\rm SC}_0\!=\!\frac{\hat{p}_{z}^2}{2m^*}\!+\!\frac{\hat{p}_{r}^2}{2m^*}\!+\!\frac{\hat{p}_{\varphi}^2}{2m^*}\!-\!\mu^{(s)}\nonumber
\end{align}
The functions $\bar{G}$ and $\bar{F}$ are determined by the following equations:
\begin{align}
\bar{G}=\frac{1}{2\tau \bar{m}_J}\sum_{|m_J| < \bar{m}_J} \frac{i\omega_n+\bar{G}+H m_J}{\sqrt{(\Delta+\bar{F})^2-(i\omega_n+\bar{G}+H m_J)^2}}\\
\bar{F}=\frac{1}{2\tau \bar{m}_J}\sum_{|m_J| < \bar{m}_J} \frac{\Delta+\bar{F}}{\sqrt{(\Delta+\bar{F})^2-(i\omega_n+\bar{G}+H m_J)^2}}
\end{align}
with $\tau$ being the elastic scattering time and $\bar{m}_J \sim p_F R_3$ being the angular momentum cutoff. In the limit $H \rightarrow 0$, the leading order corrections to the above equations appear in quadratic order since linear terms vanish due the averaging over $m_J$. Indeed, one can show that the self-consistent solution for $\tau \rightarrow 0$ is given by
\begin{align}
\bar{G}&=\frac{i}{2\tau}\sin z \\
\bar{F}&=\frac{i}{2\tau}\cos z \\
\frac{\omega_n}{\Delta}& =\tan z-\kappa  \sin z
\end{align}
where $\kappa=3H^2 \tau \langle m_J^2 \rangle/\Delta$ is the characteristic scale for the magnetic field effects in the problem. Here $ \langle m_J^2 \rangle=1/ \bar{m}_J \sum_{|m_J| < \bar{m}_J} m_J^2 \sim (p_F R_3)^2$. Thus, corrections to the pairing gap are governed by the small parameter $\kappa \ll 1$. In terms of the flux quantum, this condition reads $\Phi/\Phi_0 \ll R_3/\xi_d$. Note that disorder suppresses orbital effects of the magnetic field and leads to a weaker dependence of the pairing gap on magnetic field (i.e., quadratic vs linear). In other words, the disordered superconductor can sustain much higher magnetic fields compared to the clean one, see Fig.~\ref{fig:S8}. Finally, the analysis above can be extended to $n\neq 0$. After some manipulations, one finds that \cite{Liu2001, Koshnick2007} 
\begin{align}
\frac{\Delta_0-\Delta(\Phi)}{\Delta_0} \sim \frac{\xi_d^2}{R_3^2}\left(n-\frac{\Phi}{\Phi_0}\right)^2
\label{eq:gap}
\end{align}
This estimate is consistent with the numerical calculations, see Fig. \ref{fig:S8}. 

\section{Derivation of the effective Hamiltonian}
\label{app:effective}

In the previous Section we derived the Green's function for the disordered superconducting ring. One can now use these results to study the proximity effect of the SC ring on the semiconducting core. We will focus again on the case when the SC shell is thin $d \sim l_e$ such that $\frac{\xi_d}{R_3} \ll 1$. In this case, one can neglect magnetic field dependence of the self-energy for the entire lobe. Thus, one can use zero field  Green's functions for the disordered superconductor to investigate the proximity effect which are obtained by substituting $\omega_n \rightarrow \tilde{\omega}_n = \omega_n \eta(\omega_n)$ and $\Delta_0 \rightarrow \tilde{\Delta}_0=\Delta_0 \eta(\omega_n)$ with $\eta(\omega_n)=1+1/{2\tau \sqrt{\omega_n^2+\Delta_0^2}}$ in the clean Green's functions. 

One can now integrate out the superconducting degrees of freedom and calculate the effective self-energy due to the tunneling between semiconductor and superconductor. Using the gauge convention when $\Delta_0$ is real, the tunneling Hamiltonian between SM and SC is given by~\cite{Stanescu2011}
\begin{align}
H_t=\int dr dr' T(r,r')e^{i n\varphi/2} \Psi^\dag(r)\Psi(r')+H.c.
\end{align}
where $r$ and $r'$ refer to the SM and SC domains, respectively. $T(r,r')$ is the tunneling matrix element between the two subsystems, and $\Psi$ and $\Psi^\dag$ are the fermion annihilation and creation operators in the corresponding subsystem. One can calculate the SC self-energy due to tunneling to find
\begin{align}
\Sigma^{(\rm SC)}(r,\omega_n)=\Gamma(r)\frac{i \omega_n \tau_0 -\Delta_0 \left[\cos(n \varphi)
                     \tau_x\!+\!\sin(n \varphi)\tau_y\right]}{\sqrt{\omega_n^2+\Delta_0^2}}
                     \label{eq:selfenergy}
\end{align}
where $\Gamma(r)$ is a quickly decaying function away from $r=R_2$ describing tunneling between the two subsystems. Note that the SC self-energy in this approximation is the same as for a clean superconductor because the ratio of $\tilde{\omega}_n/\tilde{\Delta}_0$ is independent of $\tau$.

The Green's function for the semiconductor can be written as
\begin{align}\label{eq:Greens}
G^{-1}(\omega_n)=-i\omega_n-H_{\rm SM}-\Sigma^{(\rm SC)}(r,\omega_n) 
\end{align}
In order to calculate quasiparticle energy spectrum one has to find the poles of above Green's function. 

In the hollow cylinder limit, $\Gamma(r=R_2)$ is a constant and one can find low energy spectrum analytically. Indeed, after expanding Eq.~\eqref{eq:Greens} in small $\omega_n$, the quasiparticle poles are determined by the spectrum of the following effective Hamiltonian:  
\begin{align}
H_{\rm eff}=\frac{H_{\rm SM}}{1+\Gamma/\Delta_0}  &- \frac{\Gamma}{1+\Gamma/\Delta_0} \left[\cos(n \varphi)
                     \tau_x\!+\!\sin(n \varphi)\tau_y\right]=0
\end{align}
By comparison with Eqs.~(6)-(10) of the main text, one can establish the correspondence between the bare parameters of the semiconductor and the renormalized parameters of the hollow cylinder model due to the proximity cooupling to the SC. We find:
\begin{eqnarray}
m^* &=& m^*_0 \frac{\Delta_0+\Gamma}{\Delta_0}  \\
\mu &=& \mu_0 \frac{\Delta_0}{\Delta_0+\Gamma} \\
\alpha &=& \alpha_0 \frac{\Delta_0}{\Delta_0+\Gamma} \\
\Delta &=&\Gamma\frac{\Delta_0}{\Delta_0+\Gamma}\,, 
\end{eqnarray}
where $m^*_0$, $\mu_0$ and $\alpha_0$ are the bare values of the effective mass, chemical potential and spin-orbit coupling strength, respectively. Specifically, note that the renormalizaiton also reduces the Fermi velocity by a factor of $1/(1+\Gamma/\Delta_0)$ which is well-known to lead to shorter coherence lengths than from naive estimates that assume bare Fermi velocities \cite{Stanescu2011,nadj2013,Peng2015}.

\section{Effect of higher $m_J$ states on the gap}\label{sec:S2.6}

As demonstrated in the main text, states with larger $m_J \neq 0$ have the potential to close the gap and thus limit the extent of the topological phase. Here we provide analytical estimates within the hollow cylinder model for the regions in parameter space that become gapless due to higher $m_J$ states. We start with the BdG Hamiltonian~(5) of the main text assuming $n=1$,
\begin{align}\label{eq:H_mj}
\begin{split}
\tilde{H}_{m_J}=
&\left[\frac{p_{z}^2}{2m^*}-\mu_{m_J}\right]\!\tau_z+V_{Z}\sigma_z +A_{m_J}\! \\ &+ C_{m_J}\sigma_z\tau_z+\alpha_2 p_z \sigma_y \tau_z+\Delta\tau_x,    
\end{split}
\end{align}
with
\begin{eqnarray}
\mu_{m_J}&=&\mu-\frac{1}{8m^*R_2^2}\left(4m_J^2+1+\phi^2\right)-\frac{\alpha_1}{2R_2}\,,\\
    V_{Z}&=&\phi\,\left(\frac{1}{4m^*R_2^2}+\frac{\alpha_1}{2R_2}\right)\,,\\
    A_{m_J}&=&-\frac{\phi m_J}{2m^* R_2^2}\,,\\
    C_{m_J}&=&-m_J\left(\frac{1}{2m^*R_2^2}+\frac{\alpha_1}{R_2}\right)\,,
    \end{eqnarray}
with $\phi=1-\Phi(R_2)/\Phi_0$. With respect to the main text, we also introduced anisotropic spin-orbit coupling with $\alpha_1$ and $\alpha_2$ representing the strength of coupling to the transversal and longitudinal ($z$) momentum direction. In the main text, we used isotropic spin-orbit $\alpha_1=\alpha_2=\alpha$ but it is convenient for the discussion below to distinguish the two contributions.

\begin{figure}[t!]
\begin{center}
\includegraphics[width=0.8\columnwidth]{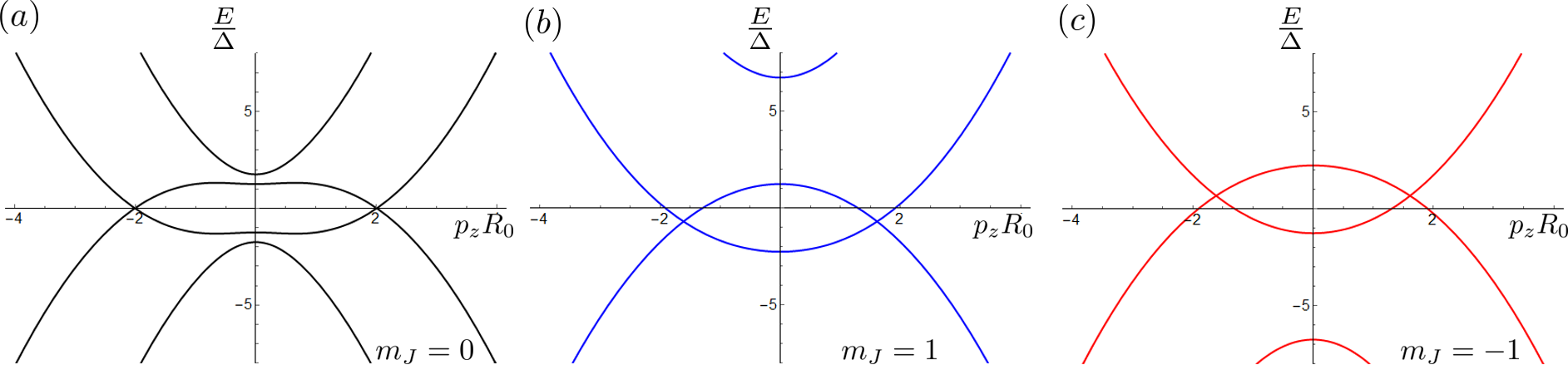}
\caption{\textbf{Energy spectrum for the lowest $m_J$ sectors for $n=1$.} Here parameters used are the same as in Fig.~3 of the main text, except $\Delta=0$. For finite $\Delta$ the intersections of particle and hole bands become avoided crossings. Note that for $m_J\neq 0$ these avoided crossings happen at finite energy, which leads to condition \eqref{cond_2a}.}
\label{fig:S9}
\end{center}
\end{figure}

Example energy spectra for the lowest $m_J$ sectors are shown in Fig.~\ref{fig:S9}. Particle-hole symmetry relates $m_J$ and $-m_J$ sectors. Therefore, $m_J=0$ sector is special in this sense. Note that $\alpha_2$ is crucial to estimate the topological gap in the $m_J=0$ sector, i.e., a topological gap requires $\alpha_2\neq 0$ and $\Delta\neq 0$. The conditions for a finite gap in $m_J\neq 0$ sectors are more stringent. First of all, the pairing term hybridizes states within each $m_J$ sector. Thus, the system is {\it gapless} if there is an odd number of particle and hole bands at the Fermi level, which leads to the  condition 
\begin{align}
(A_{mj}+V_Z\sigma_z)^2 & > \Delta^2+ (C_{mj}\sigma_z-\mu_{mj})^2\, , \label{cond_1}
\end{align}
which follows from the gap closing at $p_z=0$. The gapless region in the upper right corner of Figs.~\ref{fig:3}, A~and~B, of the main text is due $m_J=\pm 2$ states fulfilling condition \eqref{cond_1}.

If condition~\eqref{cond_1} is not satisfied and the number of bands at the Fermi levels is even, the system can be gapped -- see, for instance, panels (b) and (c) in Fig.~\ref{fig:S9}. This happens, for example, if the effective chemical potential for a given subband $\mu_{mj}-C_{mj}\sigma_z<0$ in which case the subband is empty and gapped. However, if the subband is filled, i.e. if $\mu_{mj}-C_{mj}\sigma_z>0$, one has to investigate the closing of the gap at finite momenta. In this case, the system is gapless when $\Delta$ is smaller than a certain critical value required to hybridize particle and hole bands with mismatched Fermi momenta, see Fig.~\ref{fig:S9}, (b) and (c). In the limit $\alpha_2\rightarrow 0$, the condition for a {\it vanishing} gap reads
\begin{align}
|A_{mj}+V_Z\sigma_z| &> \Delta \label{cond_2a}\\
\mu_{mj}-C_{mj}\sigma_z &>0. 
\label{cond_2b}
\end{align}
One may notice that the term $A_{mj}+V_Z\sigma_z$ acts as a Pauli limiting field for a given subband and leads to pair-breaking effects. 
     
We can understand the generally finite extent of the gapped regions in the $\alpha$-$\mu$ plane by considering conditions \eqref{cond_2a} and \eqref{cond_2b}. Condition \eqref{cond_2a} is either met for sufficiently large $m_J$ or sufficiently large $\alpha_1$ (when $m_J$ is kept constant). At the same time, large $m_J$ states generally violate condition \eqref{cond_2b} since the bottom of the band is shifted up $\propto m_J^2$ which needs to be compensated by sufficiently large $\mu$. We therefore expect to find gapless states for large $\mu$ (which enable large $m_J$) or very large $\alpha$ which fulfill condition $\eqref{cond_2a}$ while still being compatible with condition $\eqref{cond_2b}$.

\section{Phase diagram in the second lobe}\label{sec:S2.8}

\begin{figure}[h!]
  \begin{center}
    \includegraphics[width=0.9\columnwidth]{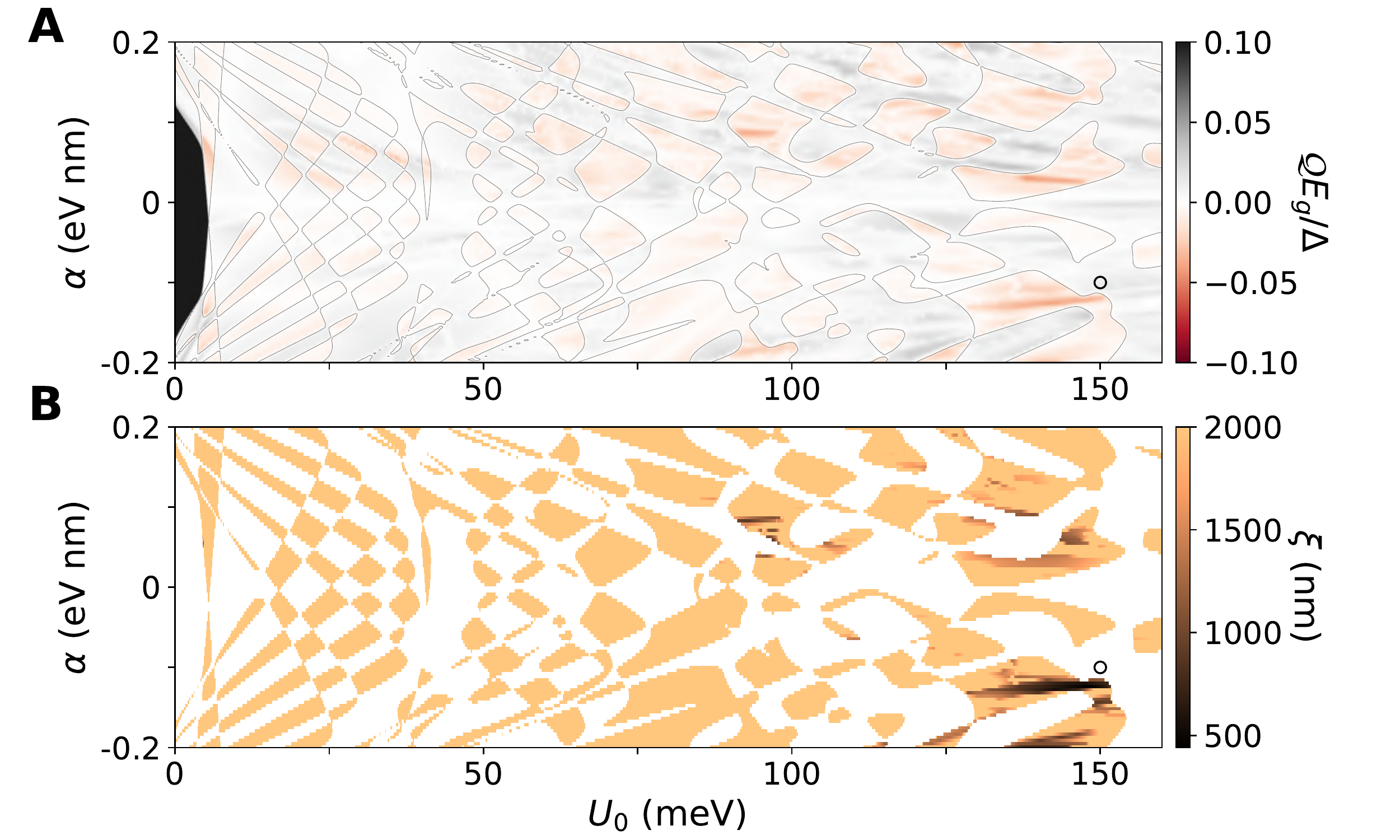}
\caption{\textbf{Topological properties in the second lobe.}
(\textbf{A}) Topological phase diagram of the full shell nanowire in the second lobe at $B=0.248$\,T as a function of band offset and SOC. The gray lines indicate a change of the sign of the Pfaffian.
(\textbf{B}) Decay length of the MZMs in the topological phase in the second lobe.
The point at which the transport simulations in the main text were performed is marked with a black circle in (A and B).
  \label{fig:S10}}
\end{center}
\end{figure}

In Fig.~\ref{fig:S10} we simulate the topological phase diagram and Majorana coherence length in the second lobe. We find that for the parameters of the transport simulations in the main text, the system is in the topologically trivial phase in the second lobe. In general, we find that topological phases in the second lobe have extremely small gap and therefore also very long Majorana coherence lengths. 

\section{Coulomb spectroscopy}\label{sec:S2.9} 
The Coulomb peak spacing is dictated by the lowest energy state at energy $E_0$, may it be a subgap state or the superconducting gap itself. The periodicity of the Coulomb peaks is determined by the ratio between $E_0$ and the charging energy, $E_{\rm C}$. The Coulomb blockade is $2e$ periodic for $E_0>E_{\rm C}$; It becomes even-odd once $E_0$ is less than $E_{\rm C}$; And it is $1e$ periodic in case $E_0 = 0$. Non-interacting Majorana modes have zero energy, hence a Coulomb island hosting Majoranas can be charged in portions of single electrons. If the wavefunctions of the opposing Majorana modes have a finite overlap, for example because of the finite island length, the energy of the corresponding modes will deviate from zero \cite{Albrecht2016,vanHeck2016}.

In the even-odd Coulomb blockade regime, the Coulomb-peak spacing, $\delta V$, is proportional to $E_{\rm C} + E_0$ for even diamonds and $E_{\rm C} - E_0$ for odd diamonds, which implies that $\delta V_{\rm E} - \delta V_{\rm O} \propto E_0$  \cite{Albrecht2016,Higginbotham2015}. This makes the Coulomb spectroscopy a powerful tool to study the interaction of Majorana modes in hybrid superconducting islands with finite size.

\begin{figure}[h!]
\centering
\includegraphics[width=0.7\linewidth]{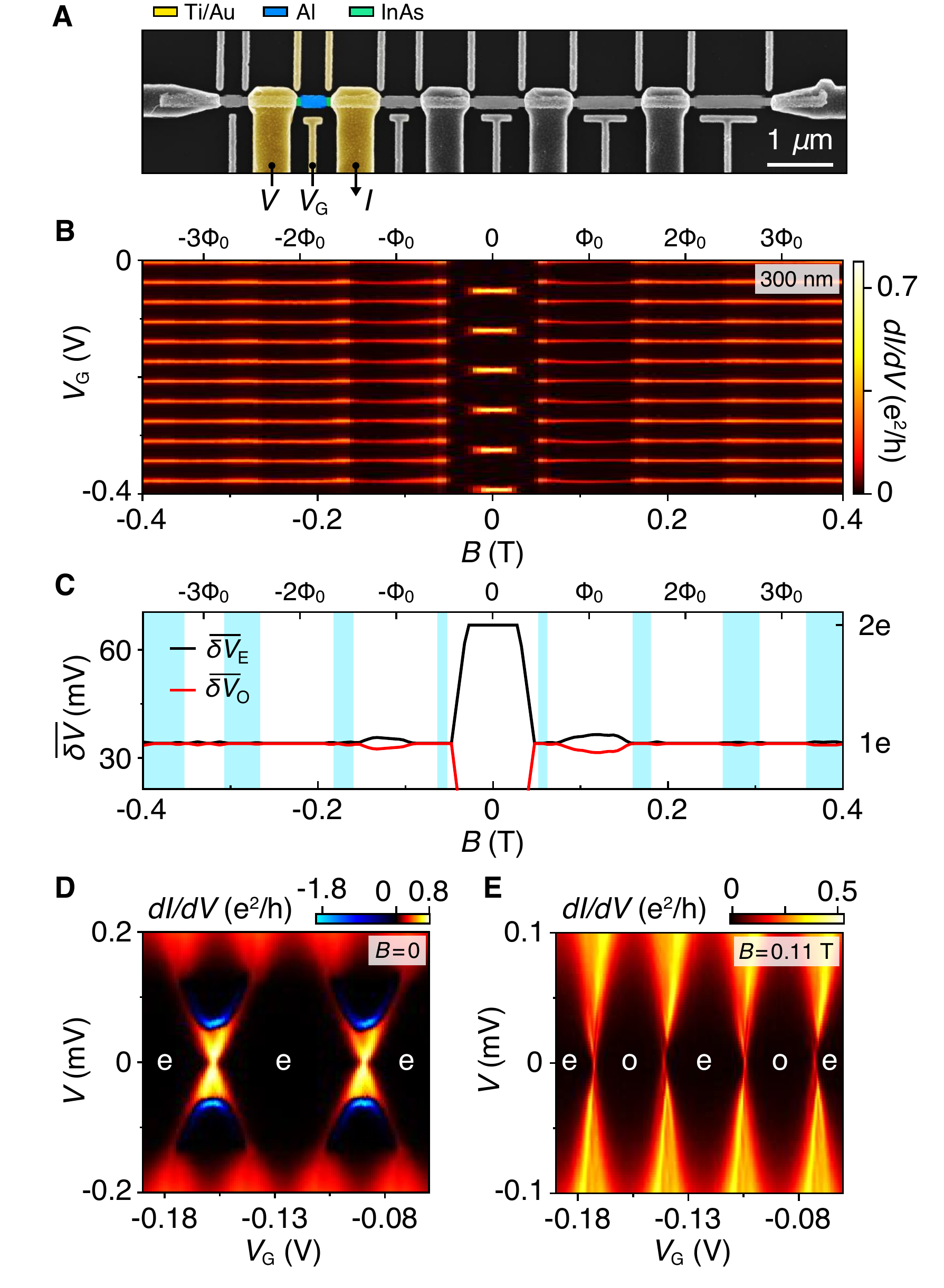}
\caption{\textbf{300 nm Coulomb island (device 2).}
(\textbf{A}) Micrograph of device 2 with the measurement setup for $300$~nm island highlighted in colors. (\textbf{B}) Zero-bias conductance showing Coulomb blockade evolution as a function of plunger gate voltage, $V_{\rm G}$, and magnetic field, $B$. (\textbf{C}) Average peak spacing for even (black) and odd (red) Coulomb valleys, $\overline{\delta V}$, from the measurements shown in (A) as a function of $B$. The blue background indicates the magnetic field ranges where superconductivity is absent. (\textbf{D}) Zero-field conductance as a function of $V$ and $V_{\rm G}$. (\textbf{E}) Similar to (D) but measured at $B = 110$~mT.\label{fig:S11}}
\end{figure}

\begin{figure}[h!]
\centering
\includegraphics[width=0.7\linewidth]{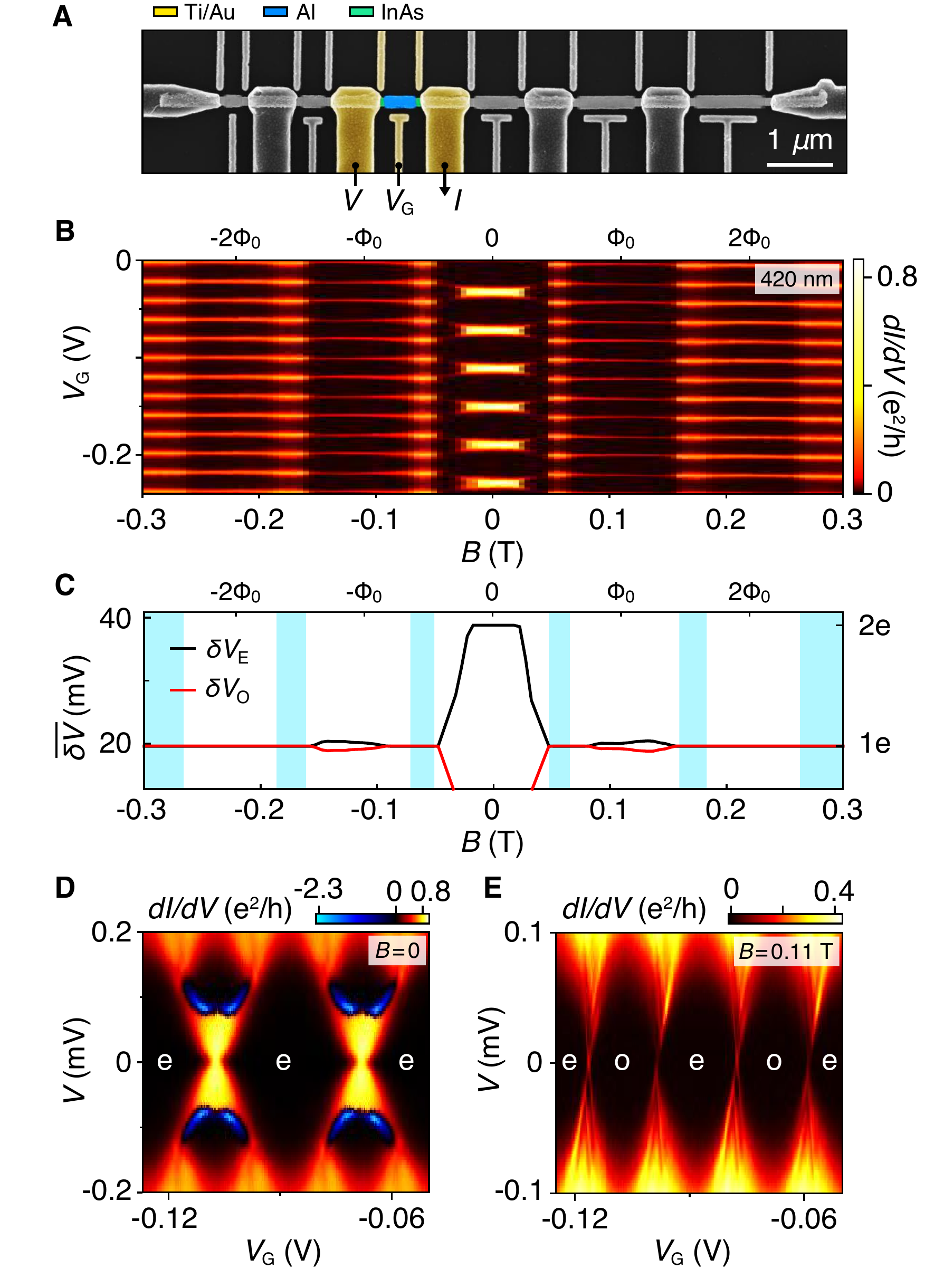}
\caption{\textbf{420 nm Coulomb island (device 2).}
(\textbf{A}) Micrograph of device 2 with the measurement setup for $420$~nm island highlighted in colors. (\textbf{B}) Zero-bias conductance showing Coulomb blockade evolution as a function of plunger gate voltage, $V_{\rm G}$, and magnetic field, $B$. (\textbf{C}) Average peak spacing for even (black) and odd (red) Coulomb valleys, $\overline{\delta V}$, from the measurements shown in (A) as a function of $B$. The blue background indicates the magnetic field ranges where superconductivity is absent. (\textbf{D}) Zero-field conductance as a function of $V$ and $V_{\rm G}$. (\textbf{E}) Similar to (D) but measured at $B = 110$~mT.\label{fig:S12}}
\end{figure}

\begin{figure}[h!]
\centering
\includegraphics[width=0.7\linewidth]{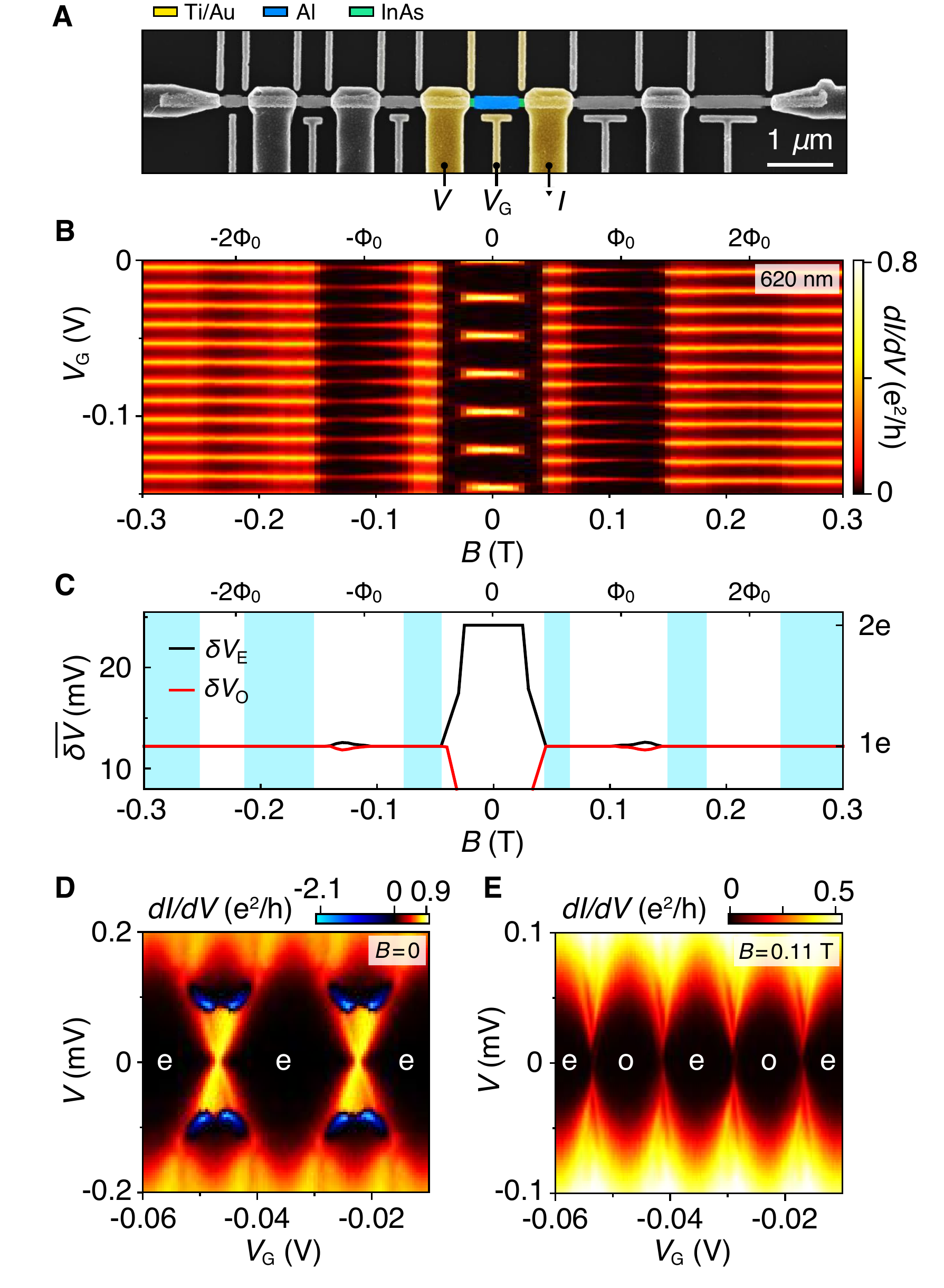}
\caption{\textbf{620 nm Coulomb island (device 2).}
(\textbf{A}) Micrograph of device 2 with the measurement setup for $620$~nm island highlighted in colors. (\textbf{B}) Zero-bias conductance showing Coulomb blockade evolution as a function of plunger gate voltage, $V_{\rm G}$, and magnetic field, $B$. (\textbf{C}) Average peak spacing for even (black) and odd (red) Coulomb valleys, $\overline{\delta V}$, from the measurements shown in (A) as a function of $B$. The blue background indicates the magnetic field ranges where superconductivity is absent. (\textbf{D}) Zero-field conductance as a function of $V$ and $V_{\rm G}$. (\textbf{E}) Similar to (D) but measured at $B = 110$~mT.\label{fig:S13}}
\end{figure}

\begin{figure}[h!]
\centering
\includegraphics[width=0.7\linewidth]{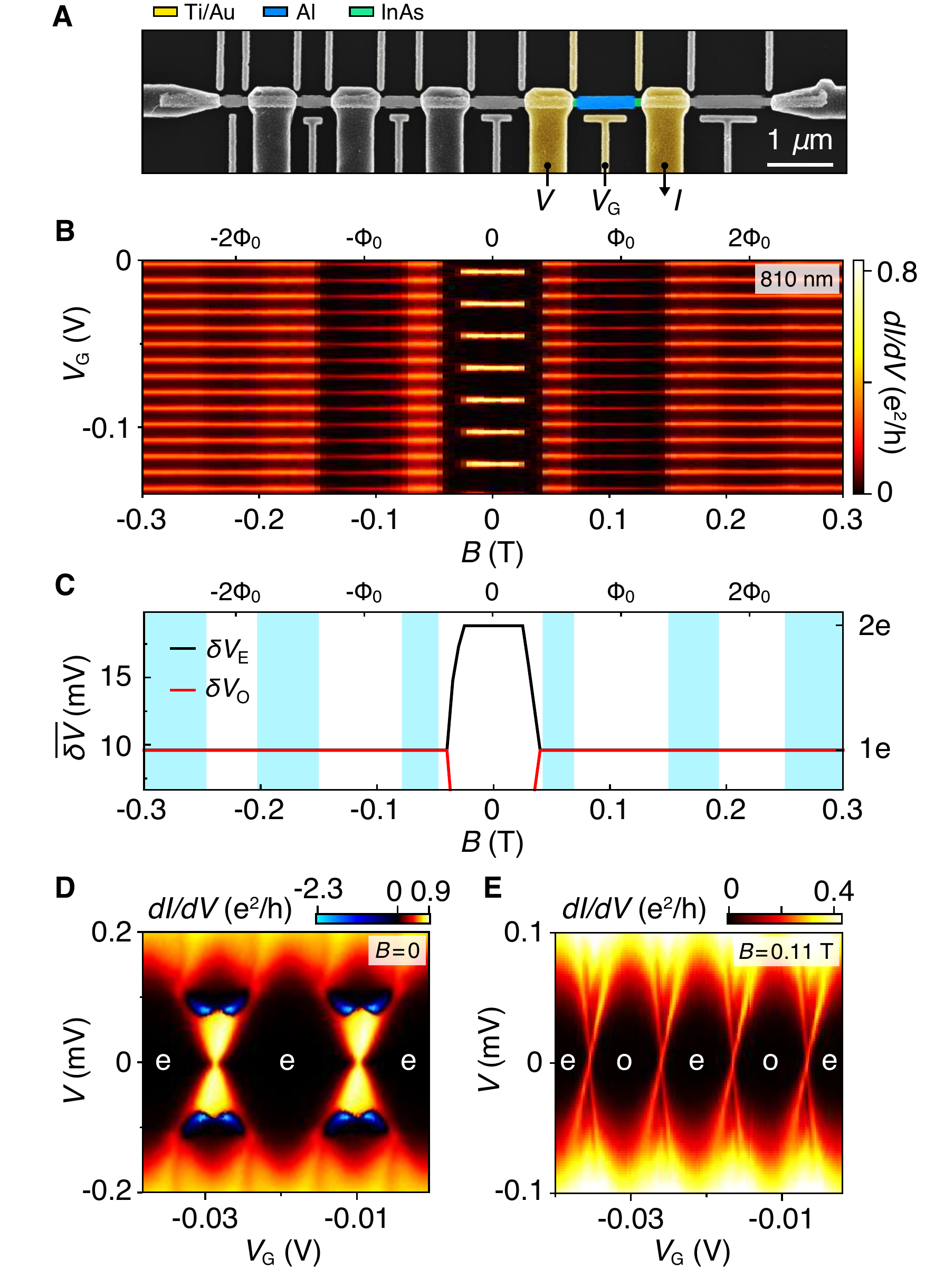}
\caption{\textbf{810 nm Coulomb island (device 2).}
(\textbf{A}) Micrograph of device 2 with the measurement setup for $810$~nm island highlighted in colors. (\textbf{B}) Zero-bias conductance showing Coulomb blockade evolution as a function of plunger gate voltage, $V_{\rm G}$, and magnetic field, $B$. (\textbf{C}) Average peak spacing for even (black) and odd (red) Coulomb valleys, $\overline{\delta V}$, from the measurements shown in (A) as a function of $B$. The blue background indicates the magnetic field ranges where superconductivity is absent. (\textbf{D}) Zero-field conductance as a function of $V$ and $V_{\rm G}$. (\textbf{E}) Similar to (D) but measured at $B = 110$~mT.\label{fig:S14}}
\end{figure}

\begin{figure}[h!]
\centering
\includegraphics[width=0.7\linewidth]{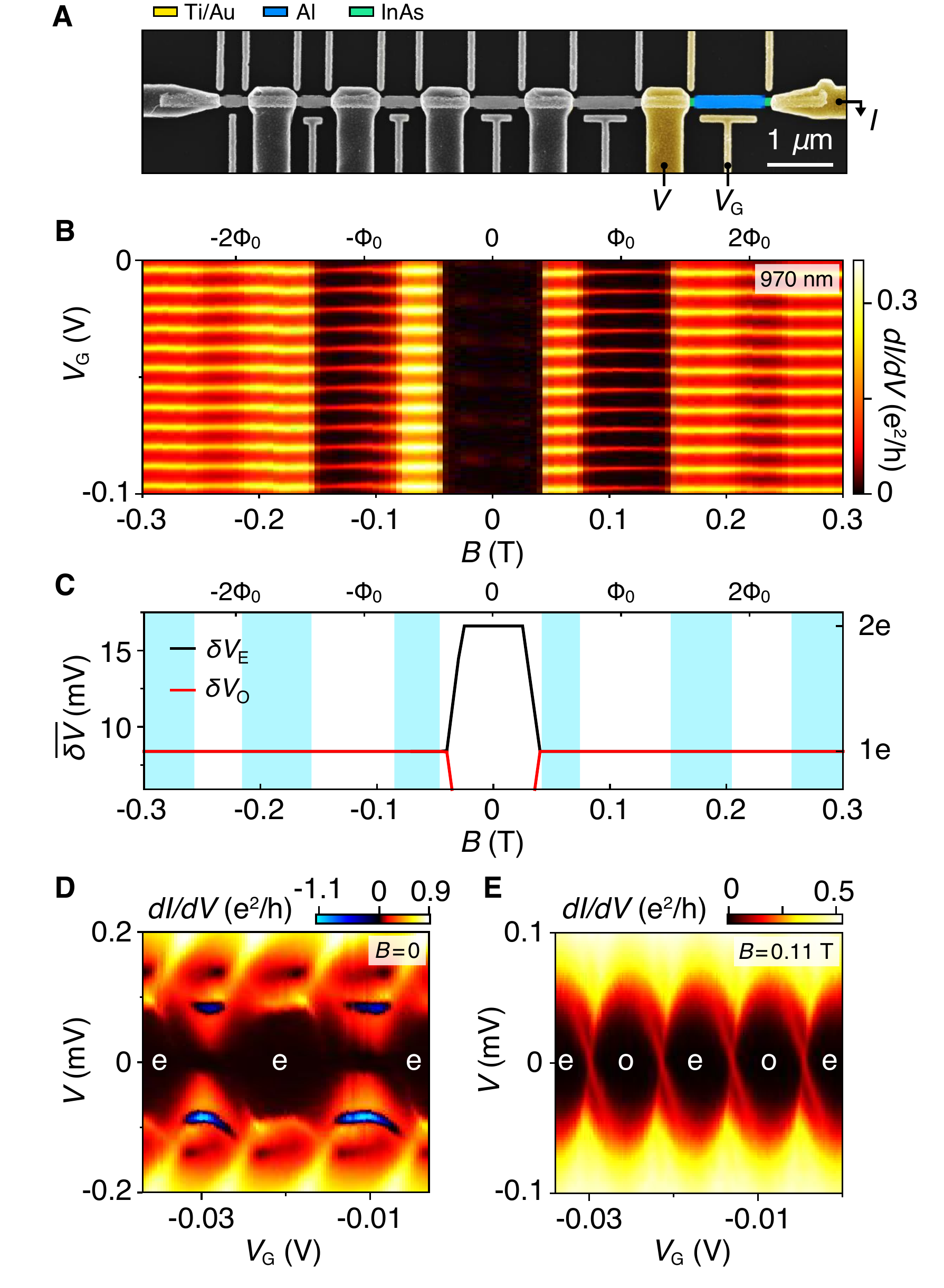}
\caption{\textbf{970 nm Coulomb island (device 2).}
(\textbf{A}) Micrograph of device 2 with the measurement setup for $970$~nm island highlighted in colors. (\textbf{B}) Zero-bias conductance showing Coulomb blockade evolution as a function of plunger gate voltage, $V_{\rm G}$, and magnetic field, $B$. (\textbf{C}) Average peak spacing for even (black) and odd (red) Coulomb valleys, $\overline{\delta V}$, from the measurements shown in (A) as a function of $B$. The blue background indicates the magnetic field ranges where superconductivity is absent. (\textbf{D}) Zero-field conductance as a function of $V$ and $V_{\rm G}$. (\textbf{E}) Similar to (D) but measured at $B = 110$~mT.\label{fig:S15}}
\end{figure}

Device 2 consists of six hybrid islands with lengths $L$ ranging from $210$~nm up to $970$~nm. Figure~\ref{fig:6} in the main-text presents measurements for the shortest island. Data for the other five islands are presented in Figs.~\ref{fig:S11}--\ref{fig:S15}. In each of the figure, panel (A) displays the scanning electron micrograph with the measurement setup for corresponding island highlighted in false colors; Panel (B) shows zero-bias conductance as a function of the axial magnetic field, $B$, and gate voltage, $V_{\rm G}$; Panel (C) depicts average even and odd peak spacing evolution in magnetic field, extracted from the data shown in panel (B); Panels (D) and (E) show Coulomb diamonds in the middle of the zeroth and first lobes, the later featuring zero-bias peaks at the degeneracy points for each island.

The same measurement routine was carried out at several different gate configurations for each island to gather more statistics. The average lever arm, $\overline{\eta}$, the difference of average even and odd peak spacings $\Delta\overline{\delta V}_{110}$ as well as the corresponding amplitude $A=\overline{\eta}~\times~\Delta\overline{\delta V}_{110}$---all measured at $110$~mT---are given in Table~\ref{tab:S1}.

\begin{figure}[h!]
\centering
\includegraphics[width=0.7\linewidth]{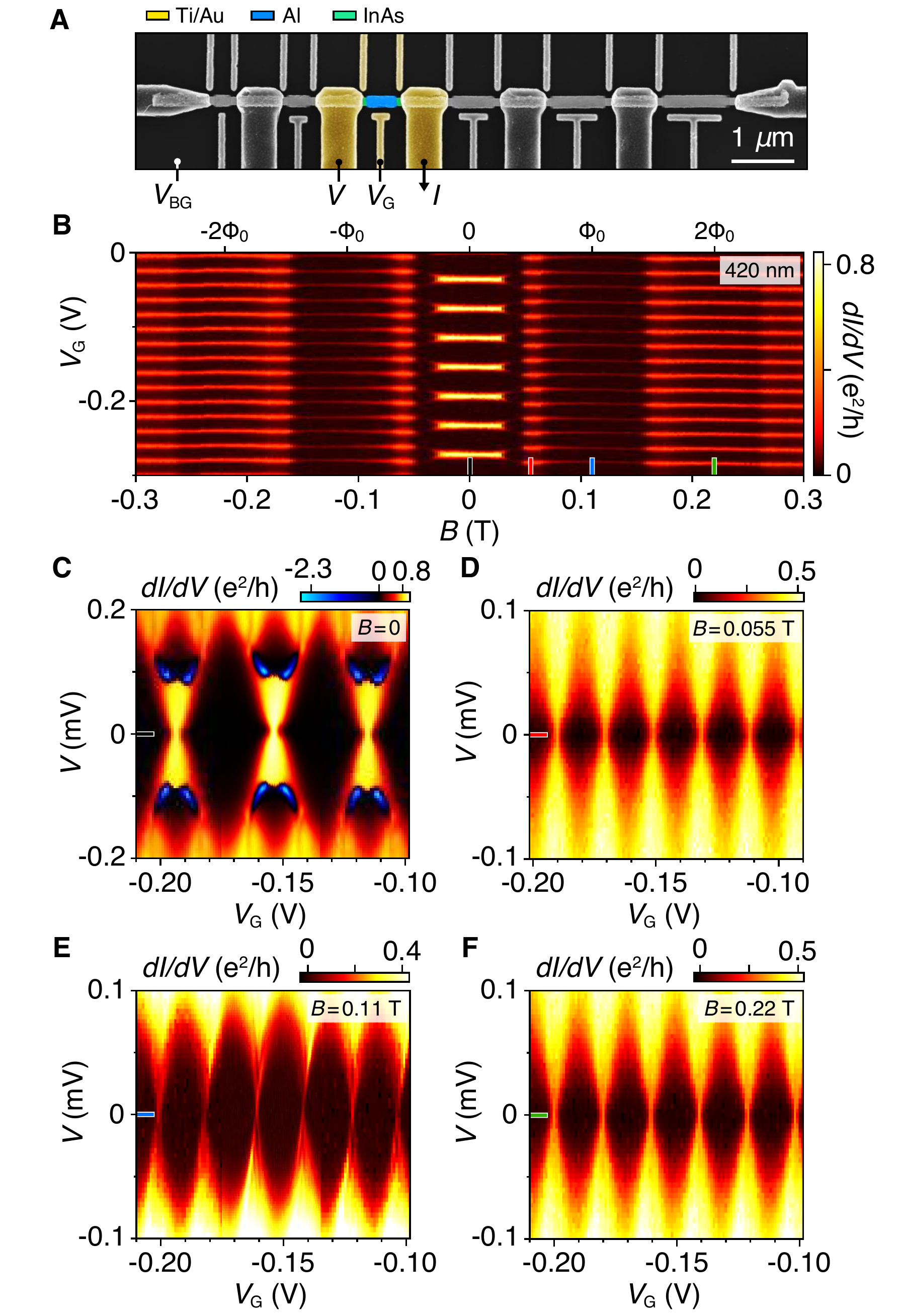}
\caption{\textbf{Additional data from 420 nm Coulomb island (device 2).}
(\textbf{A}) Micrograph of device 2 with the measurement setup for $420$~nm island highlighted in colors. (\textbf{B}) Zero-bias conductance showing Coulomb blockade evolution as a function of plunger gate voltage, $V_{\rm G}$, and magnetic field, $B$, taken at $V_{\rm BG}=-9.7$~V. For comparison, the data shown in Fig.~\ref{fig:S12} is taken at $V_{\rm BG}=-11.4$~V. (\textbf{C} to \textbf{F}) Differential conductance as a function of $V$ and $V_{\rm G}$ at various magnetic fields. The degeneracy points in the first lobe (E) exhibit discrete zero-bias peaks, whereas in the second lobe (F), a smooth spectrum similar to the one in destructive regime (D) is observed.\label{fig:S16}}
\end{figure}

A set of additional data from device 2 measured on the $420$~nm Coulomb island at $V_{\rm BG}=-9.7$~V is shown in Fig.~\ref{fig:S16}. For comparison, the data from the same island presented in Fig.~\ref{fig:S12} is taken at $V_{\rm BG}=-11.4$~V. Similar to the other island lengths, $2e$-periodic Coulomb diamonds are found in the zeroth lobe (Fig.~\ref{fig:S16}C); In the destructive regime, the superconducting gap is suppressed and the Coulomb blockade becomes $1e$ periodic with the degeneracy points displaying normal density of states (Fig.~\ref{fig:S16}D); In the first lobe, the diamonds are nearly $1e$-periodic with discrete, low energy states at the degeneracy points (Fig.~\ref{fig:S16}E); In the second lobe, the superconductivity is restored (see Fig.~\ref{fig:S2}), however, the Coulomb blockade is $1e$ periodic and no clear discrete states can be resolved (Fig.~\ref{fig:S16}F). Qualitatively similar behavior is observed for all six island lengths.

\section{Peak spacing analysis} 
An exemplar routine of the peak spacing analysis is illustrated in Fig.~\ref{fig:S17} for data from device 2 measured on the $810$~nm Coulomb island. The peak positions and spacings are deduced from a multi-Lorentzian fit to the data. A sharp distinction between the destructive regime and the first lobe is found: While the peak spacing evolution with the plunger-gate voltage is featureless at $55$~mT (blue line in Fig.~\ref{fig:S17}A), a clear zigzag-like alternating pattern between the adjacent spacings emerges at $110$~mT (green line in Fig.~\ref{fig:S17}B). The destructive regime, where the Coulomb blockade is $1e$ periodic, provides a useful tool for calibrating the analysis and determining the experimental noise floor. 

Conductance line shape of the Coulomb oscillations in the regime $\Gamma$ (tunneling rate to the leads) $< k_{\rm B}T$ (electron temperature) $< \delta$ (level spacing) $< E_{\rm C}$ (charging energy) is given by (see Ref.~\cite{Beenakker1991})

\begin{equation}
\label{eq:peak_fit}
  dI/dV =A\cosh^{-2}\left[(p-V_{\rm G})/2w\right],
\end{equation}

\noindent where $A$ is the amplitude of the peak, $p$ is the peak position and $w$ is the peak width parameter that is related to the electron temperature by $w=k_{\rm B}T/\eta$, with the lever arm $\eta$. The Full width at half maximum of the peak is given by 3.5~$w$.

Figure \ref{fig:S17}C shows three Coulomb peaks measured at $110$~mT and fit to a linear combination of Eq.~\eqref{eq:peak_fit} with the parameter estimates and their standard errors given in the figure caption. The average peak width $\overline{w}=(0.203\pm0.002)$~mV together with the lever arm $\eta=(17\pm 1)$~meV/V, deduced form the Coulomb diamonds shown in Fig.~\ref{fig:S17}E, yield electron temperature $T=(40~\pm~1)$~mK. Note that the effective electron temperature $w \eta /k_{\rm B}$ is two orders of magnitude higher than the standard fit error of the peak-position estimate.

\section{Distinguishing topological from trivial regime}

In this section we compare simulated observables in the topological and in the trivial regime. For this purpose, we perform transport simulations in different locations of the phase diagram, see Fig.~\ref{fig:S23}. From these simulations, it is clear that a ZBP is not a unique signature of the topological phase. For example, a very small bulk gap might mimic a faint ZBP as in Fig.~\ref{fig:S23}B. Furthermore, crystalline symmetries might stabilize additional topological phases with an even number of MZMs at each end, see Fig.~\ref{fig:S23}D.
Therefore, we also calculate the lowest excitation energy for different wire lengths in Fig.~\ref{fig:S24}, similar in spirit to the experimental Coulomb blockade spectroscopy. We find that only topological phases with a large gap show an exponential decay of the lowest excitation energy over a large range of wire lengths, consistent with localized end states. Fig.~\ref{fig:S24}C corresponds to a scenario with even number of MZMs at each end (two in this case) protected by a spatial mirror symmetry. Initially the behavior is exponential for short wire lengths but flattens out at longer wire lengths. Due to limited numerical accuracy the mirror symmetry is slightly broken in our system and the two MZMs are not at exactly zero energy. Analogous, any spatial symmetry will be broken in an experiment due to imperfections. As we have shown here with multiple different scenarios, it is possible to distinguish trivial from topological ZBPs with this technique.

%%Figures
%\clearpage
\onecolumngrid

%%%%%%%%%%%%%%%%%%%%%% FIG. S1 %%%%%%%%%%%%%%%%%%%%%%%%

%%%%%%%%%%%%%%%%%%%%%% FIG. S2 %%%%%%%%%%%%%%%%%%%%%%%%

%%%%%%%%%%%%%%%%%%%%%% FIG. S3 %%%%%%%%%%%%%%%%%%%%%%%%

%%%%%%%%%%%%%%%%%%%%%% FIG. S4 %%%%%%%%%%%%%%%%%%%%%%%%

%%%%%%%%%%%%%%%%%%%%%% FIG. S5 %%%%%%%%%%%%%%%%%%%%%%%%

%%%%%%%%%%%%%%%%%%%%%% FIG. S6 %%%%%%%%%%%%%%%%%%%%%%%%

%%%%%%%%%%%%%%%%%%%%%% FIG. S7 %%%%%%%%%%%%%%%%%%%%%%%%

%%%%%%%%%%%%%%%%%%%%%% FIG. S8 %%%%%%%%%%%%%%%%%%%%%%%%

%%%%%%%%%%%%%%%%%%%%%% FIG. S9 %%%%%%%%%%%%%%%%%%%%%%%%

%%%%%%%%%%%%%%%%%%%%%% FIG. S10 %%%%%%%%%%%%%%%%%%%%%%%%

%%%%%%%%%%%%%%%%%%%%%% FIG. S11 %%%%%%%%%%%%%%%%%%%%%%%%

%%%%%%%%%%%%%%%%%%%%%% FIG. S12 %%%%%%%%%%%%%%%%%%%%%%%%

%%%%%%%%%%%%%%%%%%%%%% FIG. S13 %%%%%%%%%%%%%%%%%%%%%%%%

%%%%%%%%%%%%%%%%%%%%%% FIG. S14 %%%%%%%%%%%%%%%%%%%%%%%%

%%%%%%%%%%%%%%%%%%%%%% FIG. S15 %%%%%%%%%%%%%%%%%%%%%%%%

%%%%%%%%%%%%%%%%%%%%%% FIG. S16 %%%%%%%%%%%%%%%%%%%%%%%%

%%%%%%%%%%%%%%%%%%%%%% FIG. S17 %%%%%%%%%%%%%%%%%%%%%%%%
\newpage
\begin{figure}[h!]
\centering
\includegraphics[width=0.7\linewidth]{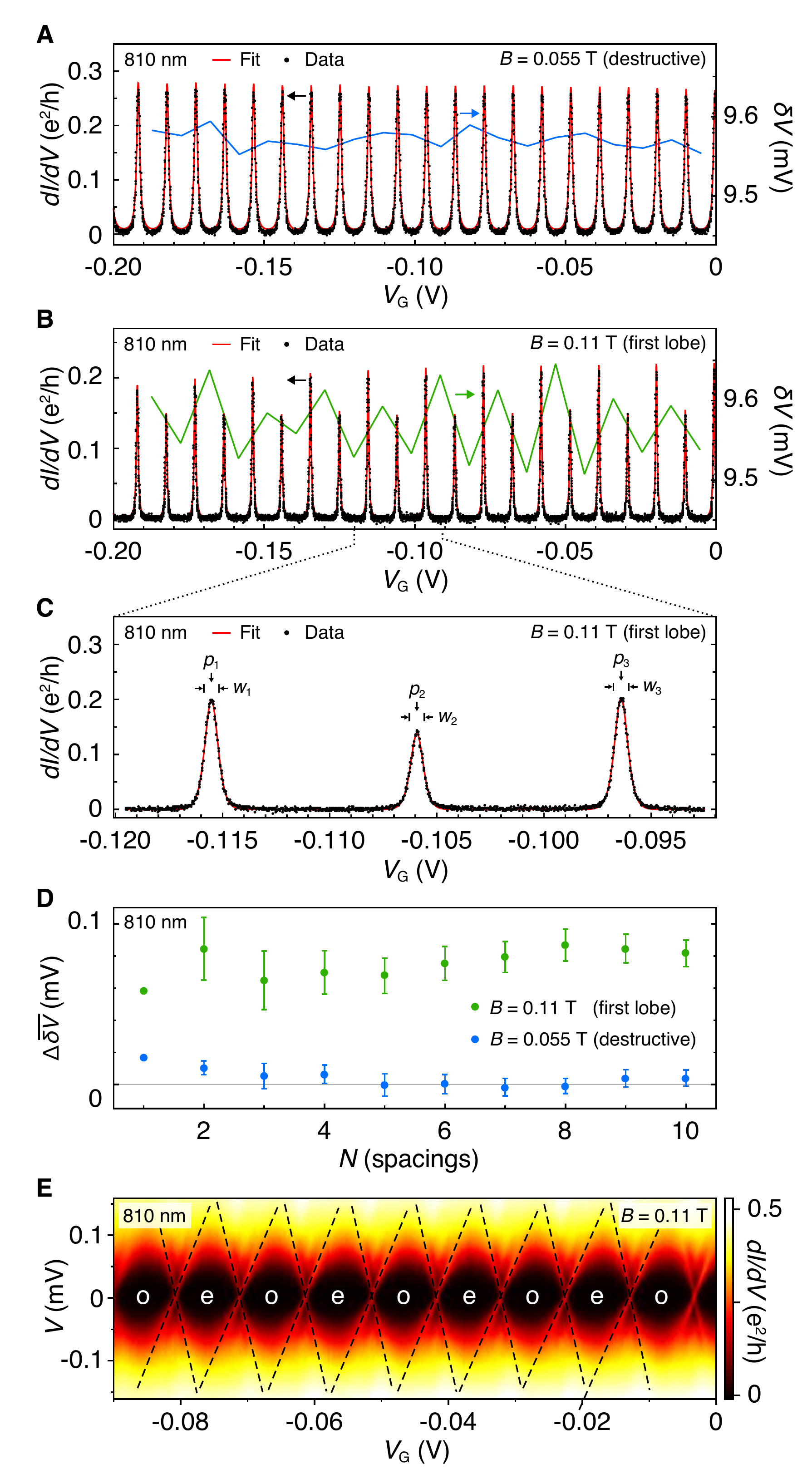}
\end{figure}
\newpage
\begin{figure}[h!]
\centering
\caption{\label{fig:S17} \textbf{Peak spacing analysis for 810 nm Coulomb island (device 2).}
(\textbf{A}) Zero-bias conductance as a function of plunger gate voltage, $V_{\rm G}$, measured over 20 Coulomb peaks in the destructive regime at $B=55$~mT (black dots, left axis). The position of each peak is deduced from multi-Lorentzian fit (red line, left axis). The corresponding individual peak spacings (blue line, right axis) do not show any clear pattern---the peaks are $1e$ periodic. (\textbf{B}) Similar to (A), measured in the first lobe at $B=110$~mT. Here, the peak spacings display a zigzag-like alternating patter indicating even-odd periodicity. Note that the right-axes in (A) and (B) have the same scale. (\textbf{C}) Zoom-in on three peaks from (B), each consisting of ${\sim}25$ data points over FWHM. The fit is described by Eq.~\ref{eq:peak_fit}, giving peak positions $p_{\rm 1}=(-115.524 \pm 0.001)$~mV, $p_{\rm 2}=(-105.931 \pm 0.002)$~mV, $p_{\rm 3}=(-96.397 \pm 0.001)$~mV and width parameters $w_{\rm 1}=(0.204 \pm 0.001)$~mV, $w_{\rm 2}=(0.198 \pm 0.001)$~mV, $w_{\rm 3}=(0.206 \pm 0.001)$~mV, with the FWHM of each peak given by 3.5~$w$.  (\textbf{D}) Difference of the average even and odd peak spacings, $\Delta\overline{\delta V}$, as a function of number of the spacings taken to determine the average, $N$, measured at $B=55$~mT (blue) and $B=110$~mT (green). The error bars illustrate the standard error of the mean given by $\sigma/\sqrt{N}$, where $\sigma$ is the standard deviation of the spacings. Using $N=$~10 gives $\Delta\overline{\delta V}_{\rm 55}=(0.004\pm 0.005)$~mV, which sets the experimental noise floor, and $\Delta\overline{\delta V}_{\rm 110}=(0.082\pm 0.008)$~mV. (\textbf{E}) Tunneling conductance as a function of source-drain bias voltage, $V$, and $V_{\rm G}$ measured at $B = 110$~mT, reveals nearly $1e$-periodic Coulomb diamonds with even (e) and odd (o) valleys and discrete zero-bias peaks at the degeneracy points. The black, dashed lines are fits to the resonant energy levels used to infer the average lever arm, $\overline{\eta}=(17\pm 1)$~meV/V, yielding electron temperature $T=(40\pm 1)$~mK and $\Delta\overline{\delta V}_{110} = (1.4\pm 0.2)~\mu$eV for this data set.
}
\end{figure}

%%%%%%%%%%%%%%%%%%%%%% FIG. S18 %%%%%%%%%%%%%%%%%%%%%%%%
\begin{figure}[h!]
\centering
\includegraphics[width=0.9\linewidth]{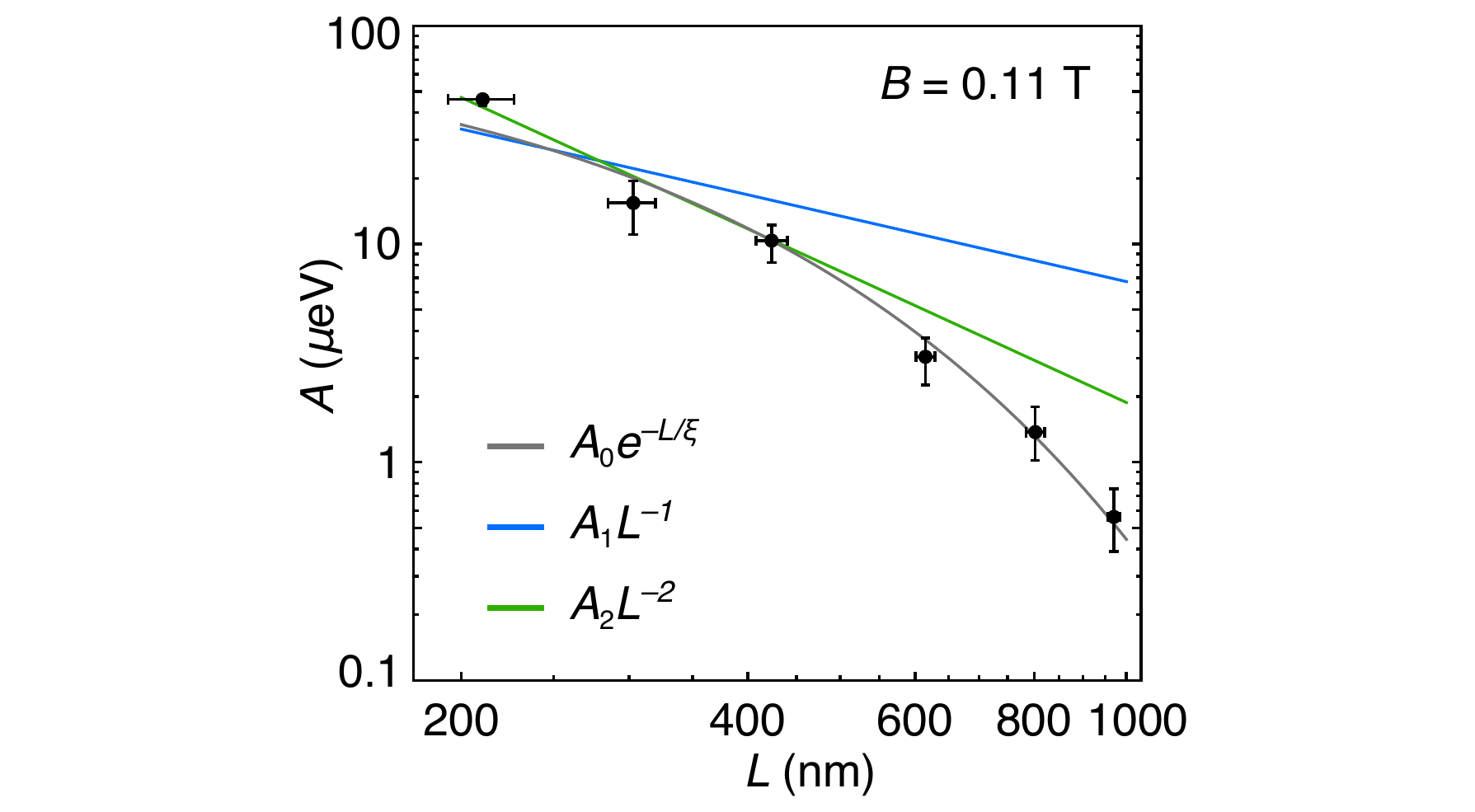}
\caption{\textbf{Exponential and power-law fits of peak spacing difference (device 2).} 
Same data as in main-text Fig.~7F: Average even-odd peak spacing difference, $A$, as a function of island length, $L$, measured at $B=110$~mT. In case of topological bound states, the hybridization energy is expected to decay exponentially with $L$, whereas for trivial boundstates, the amplitude is expected to follow a power-law dependence. The best fit to the exponential form $A = A_0 e^{-L/\xi}$, gives $A_0=(105\pm 1)~\mu$eV and $\xi=(180\pm 10)$~nm. For a particle in a box, energy level spacing scales as $L^{-1}$, and for a quantum harmonic oscillator---as $L^{-2}$. The best fit to $A=A_1 L^{-1}$ gives $A_{\rm 1}=(7\pm 1)$~meV~nm, and to $A=A_2 L^{-2}$ gives $A_{\rm 2}=(1.9\pm 0.1)$~eV~nm$^2$.\label{fig:S18}}
\end{figure}

%%%%%%%%%%%%%%%%%%%%%% FIG. S19 %%%%%%%%%%%%%%%%%%%%%%%%
\begin{figure}[h!]
\centering
\includegraphics[width=0.9\linewidth]{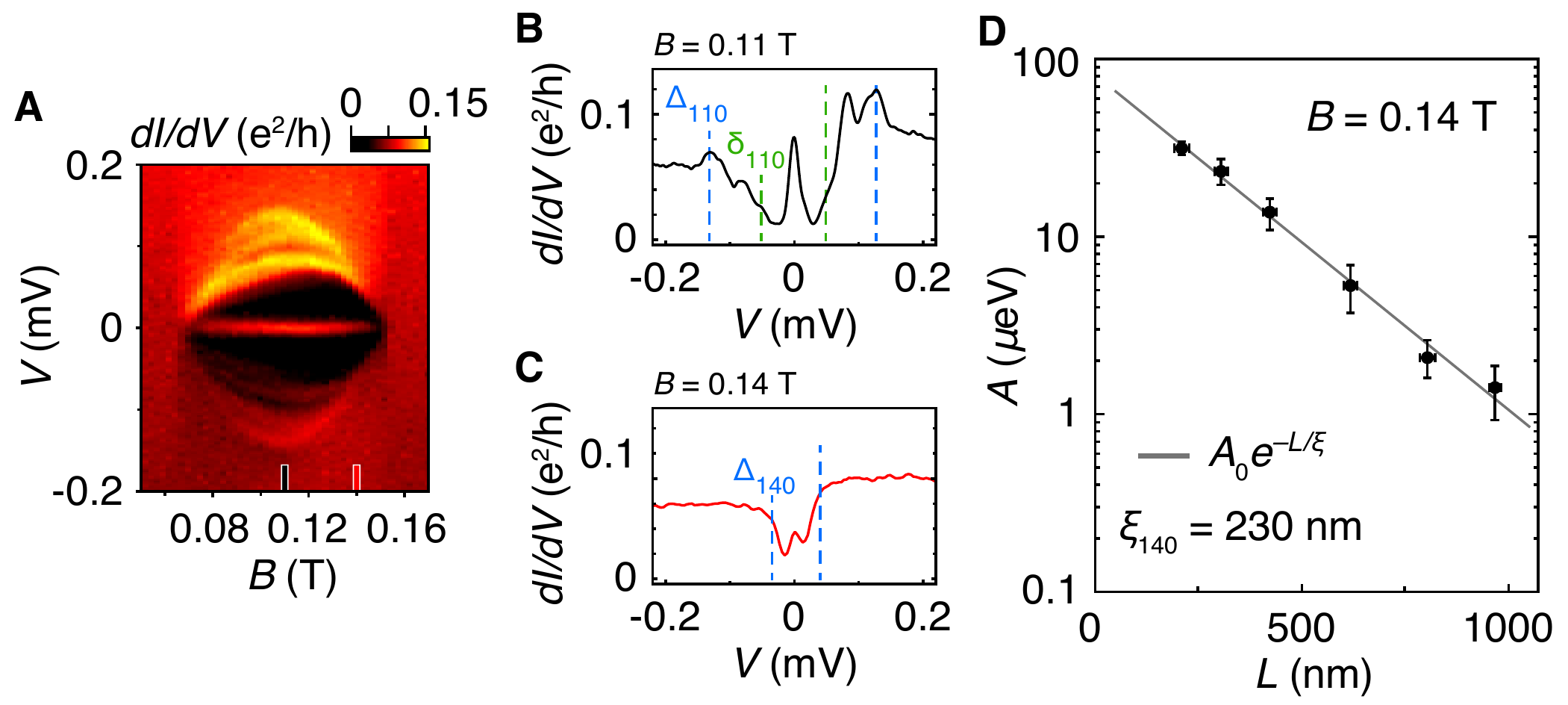}
\caption{\textbf{Superconducting gap and coherence length in the first lobe (device 1 and 2).}
(\textbf{A}) Zoom-in around the first superconducting lobe of the data shown in the main-text Fig.~2B. (\textbf{B}) Line-cut of the conductance taken from (A) at $B = 110$~mT. The dashed lines indicate the main superconducting gap at $\Delta_{110}=\pm 130~\mu$eV (blue) and the lowest excited state at $\delta_{110}=\pm 50~\mu$eV (green). (\textbf{C}) Line-cut of the conductance taken from (A) at $B = 140$~mT. The blue dashed lines indicate the gap at 140~mT,  $\Delta_{140}=\pm 40~\mu$eV. No subgap states are observed at 140~mT. (\textbf{D}) Difference of average even and odd Coulomb peak spacings, $A$, measured at $B = 140$~mT as a function of island length, $L$. The gray line is the best fit to the exponential $A = A_0 e^{-L/\xi}$, yielding $A_0 = 81~\mu$eV and $\xi_{140} = 230$~nm. Vertical error bars indicate uncertainties from standard deviation of $\overline{\delta V}$ and lever-arm measured at different gate configurations. Horizontal error bars indicate uncertainties in lengths estimated from the electron micrograph.\label{fig:S19}}
\end{figure}

%%%%%%%%%%%%%%%%%%%%%% FIG. S20 %%%%%%%%%%%%%%%%%%%%%%%%
\begin{figure}[h!]
\centering
\includegraphics[width=0.9\linewidth]{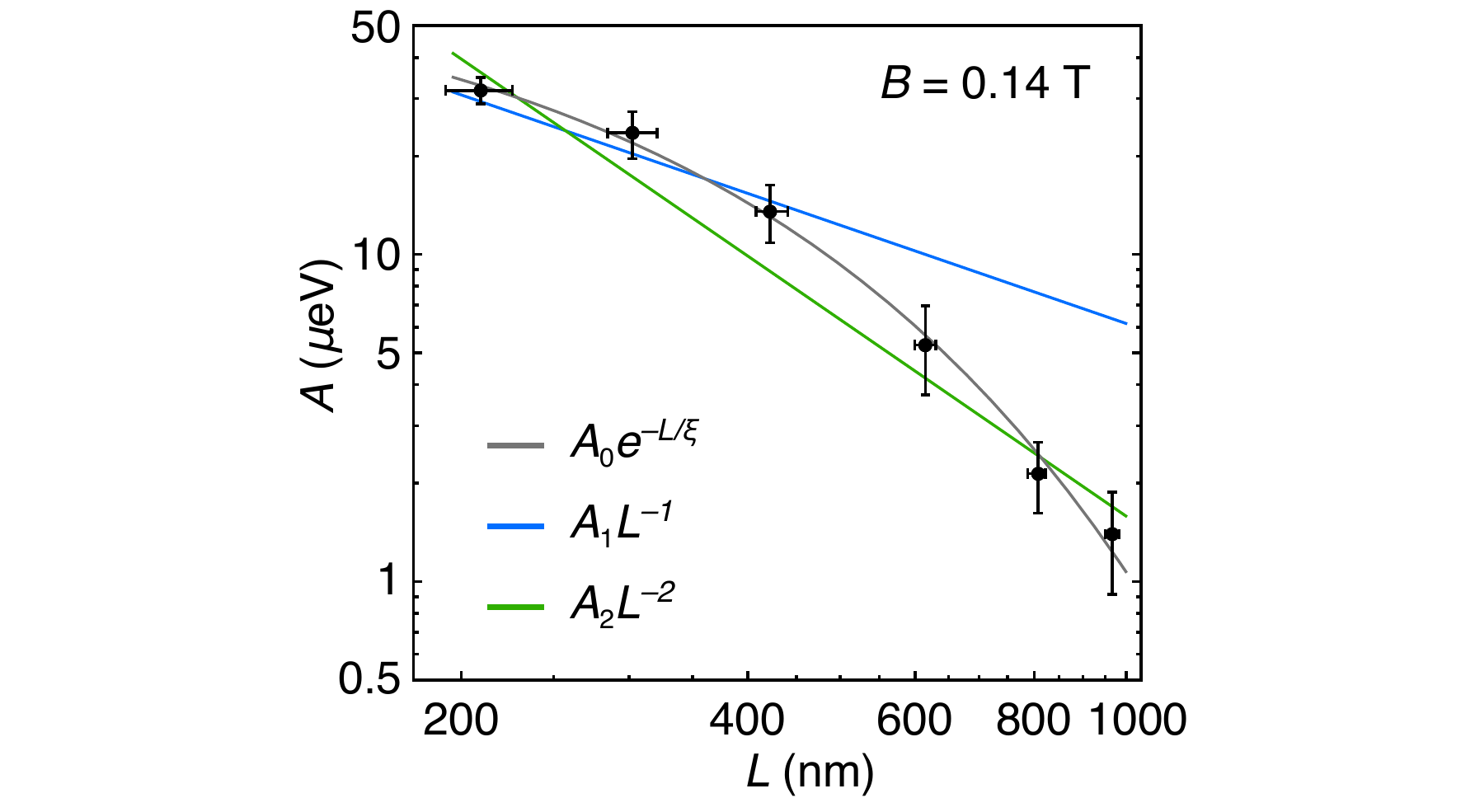}
\caption{\textbf{Exponential and power-law fits of peak spacing difference at $\bf B=0.14$~T (device 2).} 
Same data as in Fig.~\ref{fig:S19}D: Difference of average even-odd peak spacings, $A$, as a function of island length, $L$, measured at $B=140$~mT. The best fit to the exponential form $A = A_0 e^{-L/\xi}$, gives $A_0=(81 \pm 1)~\mu$eV and $\xi_{140}=(230\pm 10)$~nm. The best fit to $A=A_1 L^{-1}$ gives $A_{\rm 1}=(6.2\pm 0.7)$~meV~nm, and to $A=A_2 L^{-2}$ gives $A_{\rm 2}=(1.6\pm 0.2)$~eV~nm$^2$.\label{fig:S20}}
\end{figure}

%%%%%%%%%%%%%%%%%%%%%% FIG. S21 %%%%%%%%%%%%%%%%%%%%%%%%
\begin{figure}[h!]
\centering
\includegraphics[width=0.7\linewidth]{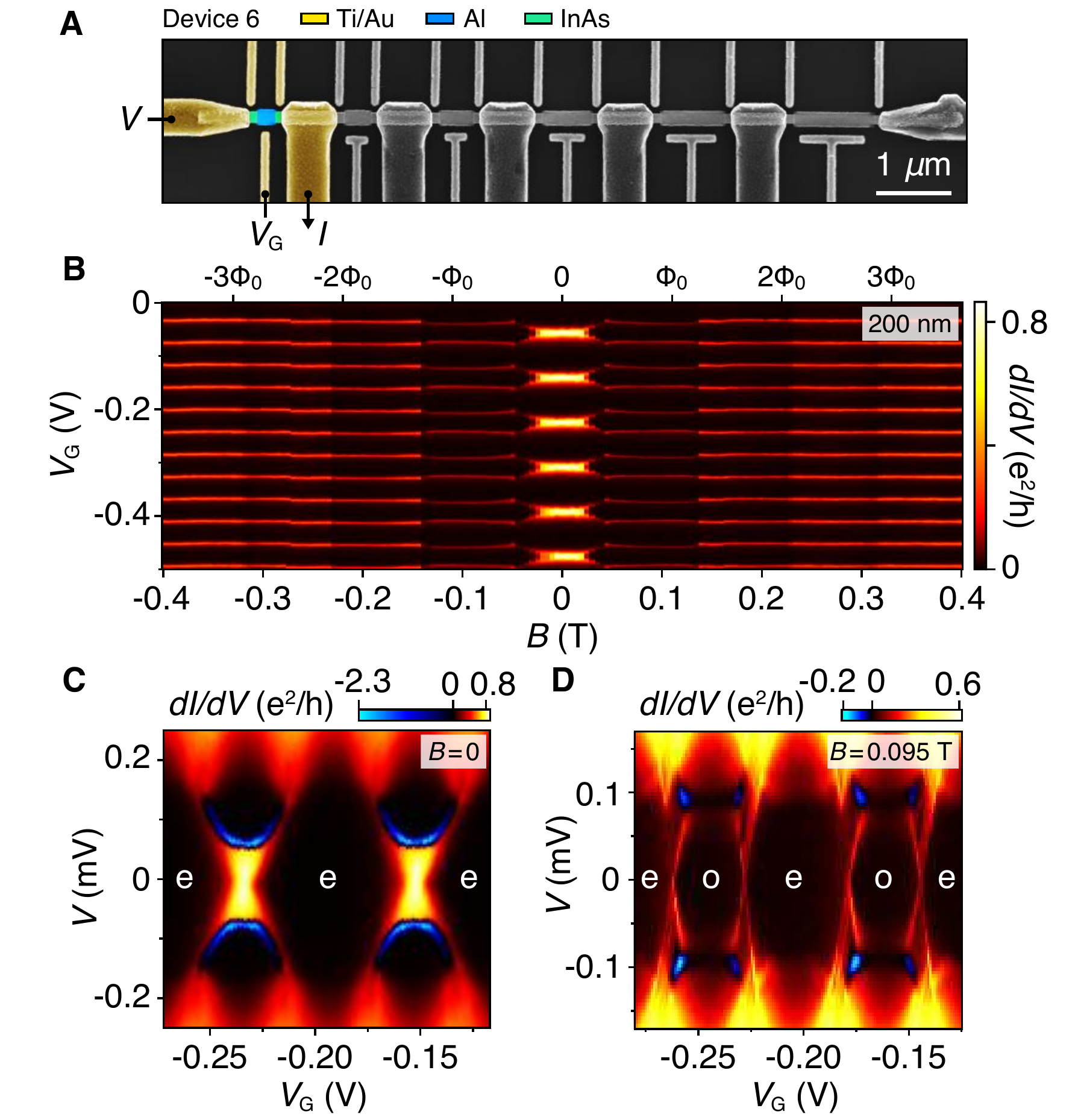}
\caption{\textbf{200 nm Coulomb island (device 6).}
(\textbf{A}) Micrograph of device 6 with the measurement setup for $200$~nm island highlighted in colors. (\textbf{B}) Zero-bias conductance showing Coulomb blockade evolution as a function of plunger gate voltage, $V_{\rm G}$, and magnetic field, $B$. (\textbf{C}) Zero-field conductance as a function of $V$ and $V_{\rm G}$. (\textbf{D}) Similar to (C) but measured at $B = 95$~mT.\label{fig:S21}}
\end{figure}

%%%%%%%%%%%%%%%%%%%%%% FIG. S22 %%%%%%%%%%%%%%%%%%%%%%%%
\begin{figure}[h!]
\centering
\includegraphics[width=0.7\linewidth]{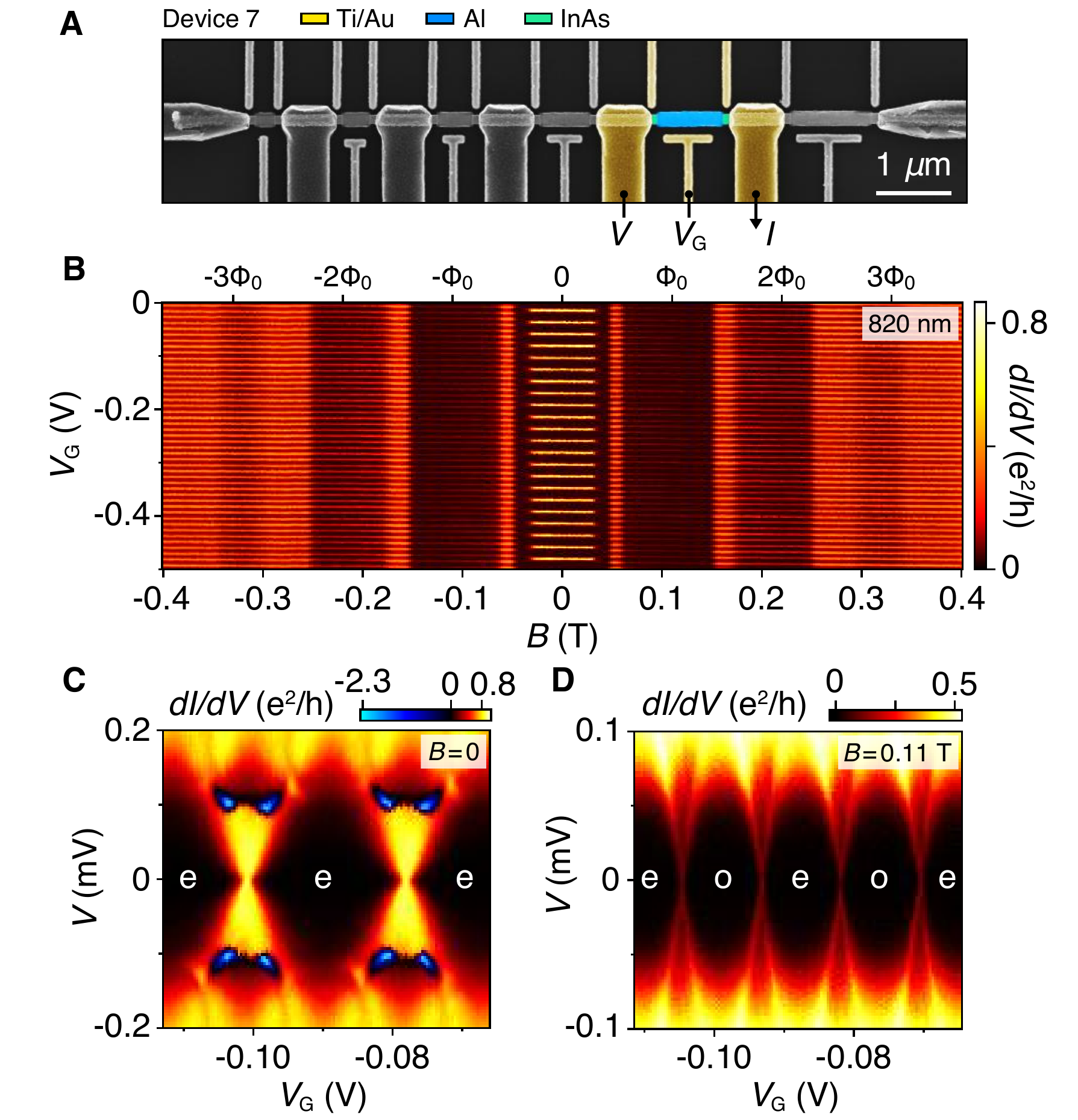}
\caption{\textbf{820 nm Coulomb island  (device 7).}
(\textbf{A}) Micrograph of device 7 with the measurement setup for $820$~nm island highlighted in colors. (\textbf{B}) Zero-bias conductance showing Coulomb blockade evolution as a function of plunger gate voltage, $V_{\rm G}$, and magnetic field, $B$. (\textbf{C}) Zero-field conductance as a function of $V$ and $V_{\rm G}$. (\textbf{D}) Similar to (C) but measured at $B = 110$~mT.\label{fig:S22}}
\end{figure}

%%%%%%%%%%%%%%%%%%%%%% FIG. S23 %%%%%%%%%%%%%%%%%%%%%%%%
\begin{figure}
  \begin{center}
    \includegraphics[width=0.8\columnwidth]{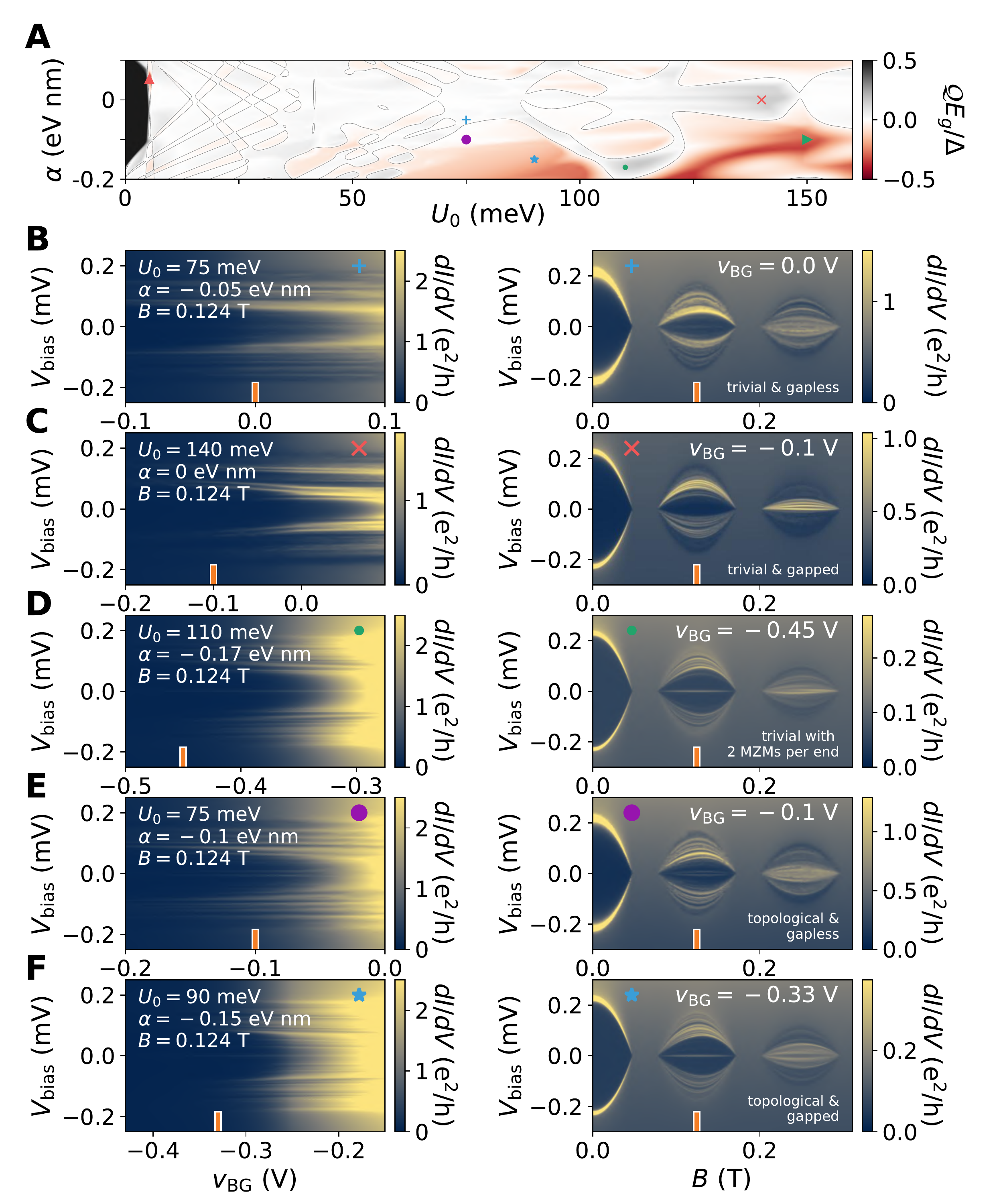}
\caption{\textbf{Conductance for different points in the phase diagram.} (\textbf{A}) Topological phase diagram with the points marked where the conductance was calculated. The green triangle corresponds to the point where the transport was calculated in Fig.~5 of the main text. The rows below show the $\mathrm{\textsl v}_\mathrm{BG}$-dependence in the topological phase (left column) and the $B$-dependence for a selected $\mathrm{\textsl v}_\mathrm{BG}$ with good visibility (right column). 
(\textbf{B}) Trivial regime with very small gap. A faint ZBP is visible.
(\textbf{C}) Trivial regime with a sizable gap.
(\textbf{D}) Trivial regime with two MZMs per end of the wire, protected by a mirror symmetry~\cite{Winkler2018}. A clear ZBP is visible.
(\textbf{E}) Topological regime with very small gap. A faint split ZBP is visible.
(\textbf{F}) Topological regime with sizable gap. A clear ZBP is visible similar to Fig.~5 in the main text.
\label{fig:S23}}
\end{center}
\end{figure}

%%%%%%%%%%%%%%%%%%%%%% FIG. S24 %%%%%%%%%%%%%%%%%%%%%%%%
\begin{figure}
  \begin{center}
    \includegraphics[width=0.7\columnwidth]{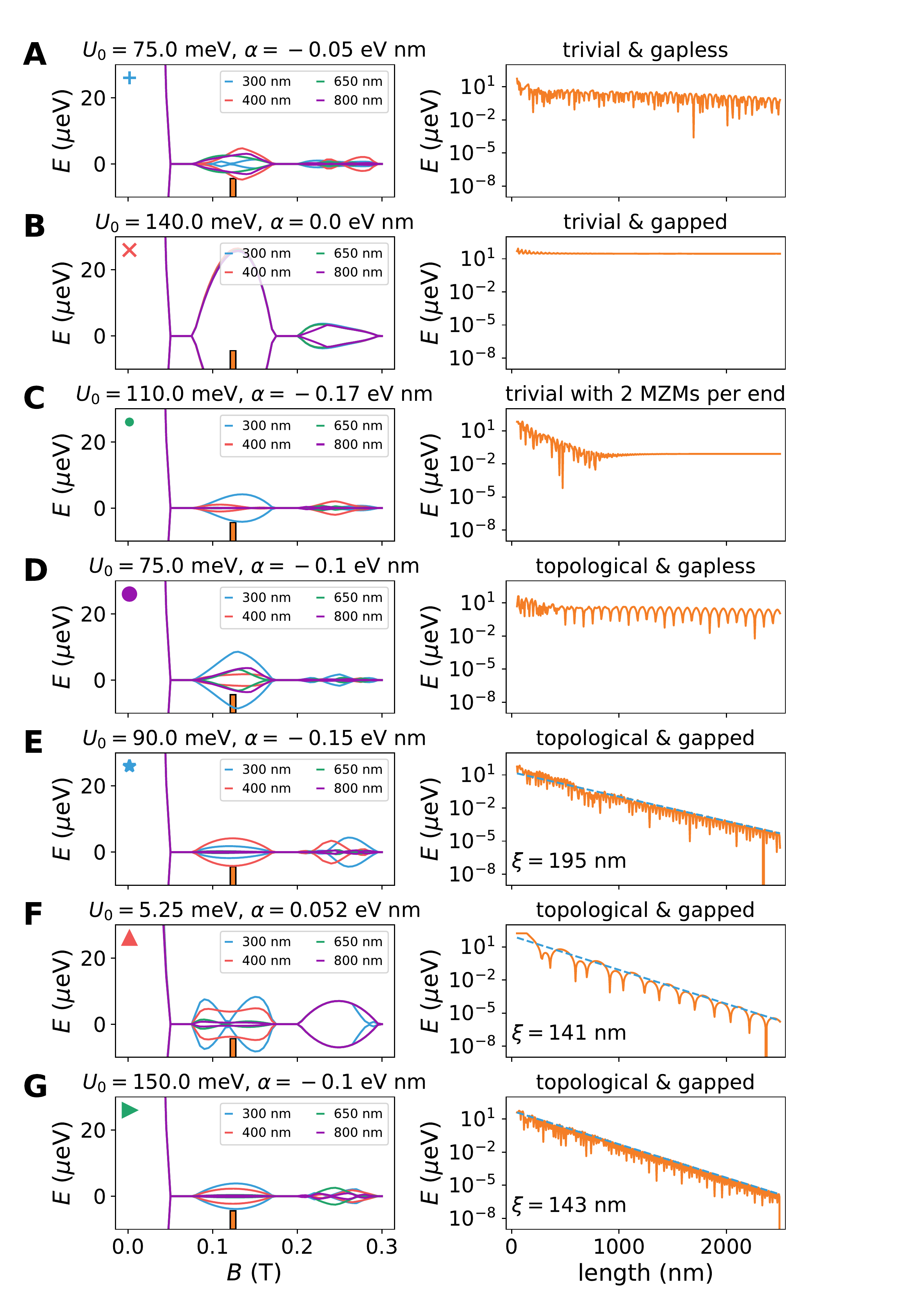}
\caption{\textbf{Lowest excitation energy in finite length wires.} The points where the lowest excitation energies where calculated are also marked in Fig.~\ref{fig:S23}A. The left column shows the $B$-dependence and the right column the length-dependence for $B=0.124$\,T. (\textbf{A}--\textbf{C}) Correspond to trivial regimes. (\textbf{D}) Topological regime with very small gap. (\textbf{E}--\textbf{G}) Correspond to the topological regime with sizable gap. They show an exponential decay of the lowest excitation energy with system length, consistent with end-localized MZMs. The oscillations in wire length are Majorana oscillations~\cite{DasSarma2012}. The dashed blue lines are exponential fits and the corresponding coherence lengths $\xi$ are inset. The extracted values of $\xi$ are consistent with Fig.~4 in the main text. \label{fig:S24}}
\end{center}
\end{figure}

\clearpage
\begin{table}[!h]
\centering
\begin{tabular*}{0.55\linewidth}{@{\extracolsep{\fill}}cccc}
\hline\hline
$L$ (nm)&$\overline{\eta}$ (meV/V)&$\Delta\overline{\delta V}_{110}$ (mV)&$A$ ($\mu$eV)\\
\hline
210 & 4.9 & 9.3 & 45\\
300 & 6.1 & 2.5 & 15\\
420 & 11 & 0.91 & 10\\
620 & 17 & 0.17 & 3\\
810 & 17 & 0.08 & 1.4\\
970 & 15 & 0.04 & 0.6\\
\hline\hline
\end{tabular*}
\caption{\textbf{Parameters for device 2.} $L$ is the length of the island. $\overline{\eta}$ is the average lever arm extracted from slopes of the Coulomb diamonds measured at $110$~mT. $\Delta(\overline{\delta V}_{110})$ are the differences of even and odd peak spacings measured at $110$~mT. $A=\overline{\eta}~\times~\Delta\overline{\delta V}_{110}$ is the corresponding amplitude in energy.\label{tab:S1}}
\end{table}

\end{document}